\documentclass[12pt]{article}
\usepackage{tikz}
\usetikzlibrary{decorations.pathreplacing,angles,quotes}
\usepackage{etex}
\usepackage[T1]{fontenc}
\usepackage{amssymb,amsmath,geometry}
\usepackage{mathtools}
\usepackage{graphicx}
\usepackage{fancyhdr}
\usepackage{fancybox}
\usepackage{shadow}
\usepackage{empheq}
\usepackage{titlesec}
\usepackage{titletoc}
\usepackage{soul, color}
\usepackage{float}
\usepackage{shorttoc}
\usepackage{xcolor}
\usepackage{fancybox}
\usepackage{natbib}
\usepackage{array}
\usepackage{tabularx}
\usepackage{booktabs}
\usepackage{rotating}
\usepackage{bigstrut}
\usepackage{natbib}
\usepackage{aeguill}

\usepackage{changepage}
\newenvironment{changemargin}[2]{\begin{list}{}{%
\setlength{\topsep}{0pt}%
\setlength{\leftmargin}{0pt}%
\setlength{\rightmargin}{0pt}%
\setlength{\listparindent}{\parindent}%
\setlength{\itemindent}{\parindent}%
\setlength{\parsep}{0pt plus 1pt}%
\addtolength{\leftmargin}{#1}%
\addtolength{\rightmargin}{#2}%
}\item }{\end{list}}

\usepackage{hyperref}
\def\sym#1{\ifmmode^{#1}\else\(^{#1}\)\fi}
\usepackage{caption}
\usepackage{blindtext}

\usepackage{caption}
\usepackage{subcaption}
\usepackage{mathtools}

\usepackage{graphicx}

\usepackage{enumerate}

\usepackage{hyperref}
\hypersetup{backref,
colorlinks=true, citecolor=blue!50!black}
\usepackage{graphicx}

\renewcommand{\thetable}{\Roman{table}}
\renewcommand{\thefigure}{\Roman{figure}}

\usepackage{titlesec}
\usepackage{setspace}
\usepackage{indentfirst}

\makeatletter
  \newcommand\smalls{\@setfontsize\smalls{10.3pt}{6}}
\makeatother

\makeatletter
  \newcommand\footnotesizes{\@setfontsize\footnotesizes{9.6pt}{6}}
\makeatother

\newsavebox\tmpbox

\usepackage{caption}
\usepackage{subcaption}

\begin{document}

\title{Fighting for Not-So-Religious Souls: \\ The Role of Religious Competition in \\ Secular Conflicts\thanks{We thank seminar participants at several seminars and Jeff Brown, Elisa Cavatorta, Irma Clots-Figueras, Guillermo Diaz, Leopoldo Fergusson, Jorge Gallego, David Karp,  Anirban Mitra, Maxime Liegey, Nicolas Lillo, Marie C Mabeu, Karen Macours, Ted Miguel, Mounu Prem, Jorge Restrepo and Nelson Ruiz. We also thank Juliana Salazar for outstanding research assistance.  All errors or omissions are ours.}}
\author{Hector Galindo-Silva\thanks{%
Department of Economics, Pontificia Universidad Javeriana, E-mail: galindoh@javeriana.edu.co
} \\
Pontificia Universidad Javeriana\\
 \and Guy Tchuente \thanks{%
School of Economics, University of Kent, E-mail: guytchuente@gmail.com} \\
University of Kent
}
\date{First Draft: October, 2019 \\This Draft: Jun 2021}

 \maketitle
\begin{abstract}
Many countries embroiled in non-religious civil conflicts have experienced a dramatic increase in religious competition in recent years. This study examines whether increasing competition between religions affects violence in non-religious or secular conflicts. The study focuses on Colombia, a deeply Catholic country that has suffered one of the world's longest-running internal conflicts and, in the last few decades, has witnessed an intense increase in religious competition between the Catholic Church and new non-Catholic churches.  The estimation of a dynamic treatment effect model shows that establishing the first non-Catholic church in a municipality substantially increases the probability of conflict-related violence. The effect is larger for violence by guerrilla groups, and is concentrated on municipalities where  the establishment of the first non-Catholic church leads to more intense religious competition.  Further analysis suggests that the increase in guerrilla violence is associated with an expectation among guerrilla groups that their membership will decline as a consequence of more intense competition with religious groups for followers.


\end{abstract}

\newpage

\section{Introduction}

Domestic armed conflicts are a common feature of the modern world.\footnote{Our calculations using the UCDP/PRIO Armed Conflict Dataset show that between 2000 and 2018, there were 113 internal armed conflicts.} While the majority of these conflicts are non-religious,\footnote{See \cite{SvenssonNilsson2017}, who classify a internal conflict as religious if there is a religious dimension to the original  disagreement as explicitly stated at the onset of the conflict by the primary parties. They show that for conflicts between 1975 and 2015, this condition is satisfied in only 31.2\% of cases; however, the percentage is noticeably higher from 2010 to 2015.} most of them occur in deeply religious countries,\footnote{Our calculations using the UCDP/PRIO Dataset and the Correlates of War (COW) World Religion Data (v1.1) show that,  between 2000 and 2010, the average percentage of religious adherents in countries that experienced at least one internal armed conflict was larger than the average percentage of religious adherents in countries that did not experienced a internal conflict: 97.5\% for the first group of countries versus 91.7\% for the second group.} where it seems plausible that the religious dimension of social structure matters. Think, for instance, about the role that religious leaders can play in a conflict given their spiritual leadership and influence over their community: if religious leaders can influence regional and local actors, they can help build (or block) peace.\footnote{\label{footnoteliterature}For practitioner-based evidence on the role that religious leaders have played in conflict scenarios, see \cite{Peacedirect2019}, and see \cite{Galtung1996} and \cite{Appleby2000} for a general perspective. For case study evidence, see for instance \cite{Gill1998} for Latin America, \cite{Isaacs2017} for Northern Ireland, \cite{Shore2009} and \cite{Wilson2001} for South Africa, and \cite{Haynes2009} for Nigeria, Cambodia and Mozambique. For more systematic correlations between religious adherence and conflict, see \cite{Fox1999, Fox2001}, \cite{Toft2007}, \cite{Basedau2011}, \cite{Svensson2013} and \cite{BormannCedermanVogt2015}. For econometric evidence, previous studies have examined the role that various dimensions of social structure play in economic development \citep{Giuliano2007, FernandezFogli2009, SpolaoreWacziarg2009, Nunn2012, YanagizawaDrotCampante2015, Cantoni2015}, and in armed conflict \citep{SpolaoreWacziarg2016, MosconaNunnRobinson2018}.}

In this paper, we empirically investigate the role of religious organizations in non-religious conflicts. We focus on scenarios in which religious denominations that have traditionally enjoyed a monopoly or near-monopoly in the religious marketplace give ground to other denominations.\footnote{The religious landscapes in many countries have been undergoing dramatic transformations, with religions that have traditionally enjoyed a monopoly or near-monopoly giving ground to other denominations. Our calculations using COW World Religion Data show  that for a sample of 191 countries, in the 1950s the average percentage of adherents of the most common religious denomination was 77\%, decreasing to 71\% in the 1960s, 68\% in the 1970s, 66\% in the 1980s, and 65\% in both the 1990s and 2000s.}
 In these increasingly common scenarios, we study the effect of an increase in competition  for adherents between religious organizations on armed conflict.

The study focuses on Colombia, a deeply religious country  with serious deficiencies in the state's capacity to control secular violence, and which in the last few decades has also experienced a significant increase in religious competition between the Catholic Church, which held a near-monopoly on religion for four centuries, and new evangelical Pentecostal churches. This makes Colombia an ideal case study for our investigation.

We estimate the effect of religious competition on armed conflict using a  two-way (unit and time) flexible fixed effects (FE) framework that compares conflict-related violence before and after the establishment of the first non-Catholic church in a municipality to violence in municipalities without a non-Catholic church, or where a non-Catholic church was established a long time ago. Since Colombia is still primarily a Catholic country, the establishment of the first non-Catholic church in a municipality can be interpreted as an increase in religious competition.

We find that the probability of conflict-related violence substantially increases after the establishment of the first non-Catholic church in a municipality. The effect is larger for violence by a guerrilla group, and is concentrated on municipalities where  the establishment of the first non-Catholic church  leads to more intense religious competition.  These results are robust to alternative specifications and samples, and to the use of two alternative measures of conflict-related violence. 

Several mechanisms might explain these results. We focus on the community-forming aspect of religion: the increase in religious competition, insofar as it implies more religious denominations competing for adherents, entails an increase in religious participation and a decline in participation in other secular activities. Crucially, the reduction in participation in secular activities may be due in part to a decline in collaboration with non-state armed groups --- in particular, with those whose ideology is more distinct from that of the religious organizations. In the Colombian context, we argue that these non-state armed groups are mostly left-wing guerrillas. Thus, guerrilla groups, insofar as they are aware of the negative effect of an increase in religious competition on their own recruitment efforts, have an incentive to use violence to prevent, or stop, that loss of guerrilla recruits from happening. We model this ``collaboration-motivated violence'' mechanism formally in a Hotelling-like framework.

This mechanism is not only consistent with existing anecdotal evidence, but also with quantitative evidence that uses data on armed groups' forced recruitment and on the presence of coca crops, which are labor-intensive to cultivate. We find evidence that the effect of religious competition on the probability of an attack by a guerrilla group is substantially larger in municipalities that have experienced at least one case of forced recruitment and in municipalities that have traditionally been coca-leaf producers.  Since in these municipalities the guerrillas plausibly expect that an increase in religious competition is more threatening (e.g. to its ability to recruit and retain members), we can interpret these additional results as evidence in favor of recruitment-motivated violence. Additional results that account for the effect of religious competition on other outcomes are consistent with our preferred collaboration-motivated violence mechanism.

We also examine the plausibility of four alternative explanations. First, ideological differences between Marxist guerrilla groups and religious organizations might explain our main result. However, the fact that the effect is concentrated in municipalities that experience more intense religious competition suggests that this explanation is incomplete at best; if our results were only due to ideological differences, religious competition would not be so crucial. In addition, we find that the effect of religious competition on the probability of violence by right-wing paramilitaries is also positive (although  smaller). 

A second alternative explanation is that the increase in violence by guerrilla groups is only due to a specific interest of the guerrilla groups to prevent newly established non-Catholic groups from gaining political representation. Although this explanation is consistent with our preferred mechanism --- and may complement it --- we do not find any evidence that suggests guerrilla groups are particularly concerned about political representation. In addition, we find that the effect is the same regardless of whether the first non-Catholic church is part of a larger religious organization (which may be seen by armed groups as more politically threatening). Thus, we argue that an explanation based on competition for political representation alone is unlikely.

A third alternative explanation is that religious competition exacerbates ethnic tensions, which may result in more violence if armed groups are instrumental to achieving the goals of some of the ethnic communities involved in the conflict. Using the available data on ethnic diversity in Colombia, we do not find any evidence that a change in the size or composition of the ethnic minority population is correlated with changes in religious competition, nor that the effect of religious competition on conflict is greater in regions with more pre-existing ethnic diversity. Thus, we argue that an explanation based on ethnic identity competition is unlikely.

A fourth alternative explanation is that religious competition increases the predisposition of individuals to commit crimes, which, by exhausting police resources, may weaken local institutions, incentivizing rebel groups to take advantage of the situation by increasing their attacks. We argue that this is unlikely because we find no evidence that religious competition impacts two key types of violence (homicides and robberies) that are not directly related to armed conflicts.

\medskip

Our identification strategy relies on four main assumptions. First, conditional on covariates, fixed effects and municipality-year  trends, the time-varying changes in municipalities without a non-Catholic church or municipalities with a longstanding non-Catholic church provide valid counterfactuals for the changes that would have occurred in other municipalities if they hadn't experienced the arrival of a non-Catholic church. Second, we assume there is no reverse causality. Third, we assume treatment effect homogeneity. Fourth, we assume no anticipatory behavior.

We examine the empirical validity of the first two assumptions in a variety of ways, including demonstrating that the presence of conflict-related violence (and related factors) prior to the establishment of the first non-Catholic church is not correlated with this establishment, and that there is no ``effect'' prior to the establishment of the first non-Catholic church.  To examine the validity of the assumptions about homogenous treatment effects and the absence of anticipatory behavior, we first show that our main results are robust to the use of \citeauthor{AbrahamSun2018}'s interaction-weighted estimator \cite[see][]{AbrahamSun2018}. Second, we show that  our main results are robust to using an alternative identification framework that does not require assuming no anticipation, and that is not based on (other) assumptions that are specific to quasi-experimental designs (such as difference-in-differences). Specifically, we follow \cite{Tchuente_Wind2019}, who propose an estimator for dynamic treatment effects that does not require the selection of time-varying unobservables. 

\medskip

This study contributes to the literature in several ways. Besides the aforementioned studies examining the relationship between various dimensions of social structure on armed conflict (see footnote \ref{footnoteliterature}), this paper adds to the vast conflict literature \cite[for a review, see][]{BlattmanMiguel2010} by providing new evidence on the relationship between religious competition and armed conflict. 

Insofar as an increase  in  religious competition may imply a change in people's ethnic identities, this paper  relates to the literature on the role of ethnicity in conflict \citep{FearonLaitin2003, MontalvoReynalQuerol2005, EstebanMayoralRay2012AER}, and more specifically, on the relation between group identity competition and conflict  \citep{MitraRay2014, JedwabJohnsonKoyama2019, BeckerPascali2019, GrosfeldSakalliSeyhunZhuravskaya2019}. Consistent with our study, these papers also find that an increase in identity-based competition results in more violence, and that this effect is best explained by its politico-economic origins. However,  the underlying mechanism and main focus of our paper are different: while in these papers violence is between ethnic or religious groups, in our study violence is perpetrated by secular groups against religious groups, and the main driver of this violence is that secular armed groups are negatively affected by an increase in competition between religious organizations.\footnote{The fact that the perpetrators of violence in Colombia are secular organizations is crucial for characterizing this conflict as non-religious. However, this does not rule out that group identity competition plays a role in our results, even though our preferred explanation is political-economic in nature (like in \citealt{MitraRay2014, JedwabJohnsonKoyama2019, BeckerPascali2019, GrosfeldSakalliSeyhunZhuravskaya2019}) and that religious competition in Colombia does not occur within ethnic group boundaries (unlike \citealt{Isaacs2017}). In fact, insofar as our preferred mechanism is based on the community-forming aspect of religion,  group identity matters. In particular, our theory is consistent with what \cite{Seul1999} says about the role of religious identity in conflict: although religion is not the \emph{cause} of conflict, it can still supply the fault line along which resource competition occurs. 
}  In this respect,  our study has parallels with \cite{CantoniDittmarYuchtman2018}, who study the effect of religious competition on the allocation of resources benefiting  secular rulers. Unlike \cite{CantoniDittmarYuchtman2018}, we study  how religious competition affects the allocation of resources for armed secular organizations. 

\medskip
Since our empirical evidence is exclusively from Colombia, we use caution in making claims about external validity. Nevertheless, we believe that the mechanisms and empirical evidence presented in this paper can be generalized to other countries. In particular, other countries with non-religious internal armed conflicts may also experience an increase in violence from non-state armed actors when more religious denominations start competing for adherents and where religious leaders also have spiritual leadership and influence over their communities.\footnote{Guatemala and El Salvador are two examples of deeply religious countries that also suffered non-religious civil conflicts, where religious identities have changed significantly, and in which religious leaders played a key role in promoting (or blocking) peace \cite[see][chs. 9 and 11]{Klaiber1998}. Other examples include the civil wars in Mozambique and Sierra Leone, where religious  organizations appear to have played a key role as mediators between the parties engaged in conflict \cite[see][]{Perchoc2016}. 
}

The outline of the paper is as follows: Section \ref{Background} provides a brief overview of the Colombian armed conflict and the evolution of the country's religious identity. Section \ref{Theory} presents a simple model that outlines the possible effect of religious competition on armed conflicts. The data and empirical strategy are discussed in Section \ref{DataandEmpiricalStrategy}. The main results are presented in Section \ref{MainResults}. Alternative explanatory mechanisms are considered in Section \ref{Othermechanismsandadditionalevidence}. Section \ref{Robustness} explores robustness, and Section \ref{Conclusion} concludes.


\section{Background} \label{Background}

In this section, we provide a brief historical overview of the armed conflict and of the evolution of Colombia's religious identity over the last two decades.


\subsection{Colombian armed conflict}

Colombia has suffered one of the world's longest-running internal conflicts.   The conflict has its roots in struggles for land rights and ownership, political exclusion, and weak institutions \citep{Sanchez2001}. Its persistence has been explained as the result of international influences and drug trafficking \citep{Deas2015}, as well as the decentralization of local politics and public spending \citep{SanchezPalau2006}. The start of the conflict coincided with the founding of the FARC,  Colombia's largest and best-equipped rebel group, which was originally comprised militant communists and peasant self-defense groups. The FARC's main stated aim always has been to redistribute land to the poor.

 In addition to the FARC, other armed groups have  participated in Colombia's conflict, including smaller left-wing insurgents and several right-wing paramilitary groups. The most important left-wing insurgent other than the FARC is the National Liberation Army (ELN), which initially consisted of students, Catholic radicals  inspired by the liberation theology movement, and left-wing intellectuals who hoped to replicate Fidel Castro's communist revolution.\footnote{Even though several former leaders of the ELN were Catholic priests, there has never been a close relationship or alliance between the Catholic Church and the ELN. To the contrary,  the Colombian Catholic Church harshly criticized those priests who took up arms and joined the ELN \cite[][pp. 34-35]{Gonzalez2005}, and consistently supported governmental structures and priorities of elites \cite[][p. 258]{LaRosa2000}.} 
 
 As for paramilitary groups, some authors associate their origin with local elites and drug cartels that faced threats of kidnapping and extortion from guerrillas and felt betrayed by the central government's favorable view of political competition, agrarian reforms and peace talks \citep{Romero2005, GutierrezBaron2005, Lopez2010cap1}. In 1997, paramilitary factions formed a national coalition called the United Self-Defense Groups of Colombia (AUC). Its creation considerably increased the effectiveness of the paramilitaries. 
 
 In 2002, with the arrival of a new president who eventually offered de-facto amnesty to paramilitaries, the level of violence began to decline, and by the end of the 2000s, the severity of the conflict decreased significantly. In July 2016, the  Colombian government and the FARC signed a historic peace deal,  which earned then-Colombian President Juan Manuel Santos the 2016 Nobel Peace Prize.

Colombia's armed conflict came with extraordinary levels of human rights abuses, with civilians by far the principal victims.\footnote{The best estimates show that the conflict claimed at least 220,000 lives, with civilians accounting for about 81\% of this number \cite[see][pp. 31-32]{GMH2013}.}  Civilians have routinely been the victims of kidnappings, forced displacement and targeted assassinations. Civilians died in two primary ways: intentional, targeted killings, in which an armed group enters a village and executes one or several pre-selected inhabitants, and unintended deaths resulting from another action.


\subsection{Colombian marketplace for religion}

In the last few decades, while suffering from an intense armed conflict, Colombia also experienced a dramatic change to its religious landscape. The Catholic monopoly that prevailed for four centuries --- which was strongly protected by state-enforced barriers to entry --- started to decline, as Protestant groups made significant inroads. From 1950 to 1970, the percentage of Colombians who identified as Catholic oscillated between  91\% and 95\%. It had fallen to 79\% by 2014 \cite[see][p. 27]{PewResearchCenter2014}. The fall in Catholic church memberships was accompanied by a stark rise in Protestant ones: in 2014, when adherence to Catholicism fell by a net of 13  percentage points,\footnote{Net gains and losses are defined as the difference between the percentage of the population in each country that was raised in a religious group and the percentage currently in that group \citep[see][p. 33]{PewResearchCenter2014}.} Protestantism rose by 8 percentage points. With some small differences, particularly regarding timing, this pattern has been seen throughout all of Latin America \citep[see][]{PewResearchCenter2014, SommaBargstedValenzuela2017}.

Importantly, the religious shift in Colombia coincided with the opening of the first non-Catholic churches in many municipalities. Figure \ref{percmpiosiglesiasnctimeseries20032013} in Web Appendix \ref{appspatialtemporalncc} plots the evolution of the proportion of municipalities with at least one non-Catholic church from 1996 to 2017. The figure shows a stark rise, which is consistent with the changes in religious adherence. Figure \ref{mapdumiglesiasnc2017_map} shows that this phenomenon occurred broadly across Colombia and was not confined to a particular region.

Pentecostals are the fastest-growing Protestant group in Colombia. In 2014, at least 56\% of Colombian Protestants either said that they belong to a church that is part of a Pentecostal denomination or that they personally identify as a Pentecostal Christian \citep[see][p. 62]{PewResearchCenter2014}.  Scholars have identified several factors that may explain Pentecostalism's success in Colombia, including i) urbanization, ii) the new needs of the population for hope that results from the difficulties associated with the urban transition, iii) the poor response to this phenomenon by the Catholic Church, and iv) the greater efficiency of Pentecostal organizations in this  regard \cite[see][]{Bastian2005, Beltran2013}.\footnote{In our empirical analysis, we find evidence that supports this explanation, but only in municipalities with less political competition (see footnote \ref{ftdetfirshistorical} in Section \ref{Robustness} and Table \ref{dfirstiglesiasnchist_tab} in the Appendix (\ref{appdeterminants})).} The greater efficiency of Pentecostal groups in responding appropriately to the demands and needs of the population has been explained by their charismatic authority, their flexibility, and the strategic use of marketing and mass media \cite[see][]{Bastian2005, Beltran2013}. In addition, Colombian Pentecostal groups have been characterized as ideologically heterogeneous, fragmented and economically weak \citep[see][p. 140]{Beltran2013}, with the majority of them having emerged from the initiative of religious leaders who decided to found their own church \citep[see][p. 19]{Tejeiro2010}.

\subsection{Religious competition and armed conflict}\label{backgroundrelcomp}

To the extent that religious competition increased across Colombia, it is reasonable to consider its consequences in regions where armed groups operated. Scholars have noted the efforts of religious leaders --- Catholic and non-Catholic --- to identify as ``apolitical." In practice, that has meant avoiding expressions that could be interpreted as sympathy or support for any of the armed actors \citep[see][p. 195]{Beltran2013}.\footnote{According to the US Commission on International Religious Freedom (USCIRF), ``most religious groups reported that due to threats from guerrillas and other illegal armed groups, many religious authorities were forced to refrain from publicly discussing the internal conflict" \cite{USCIRF2005}.} However, and importantly, most religious bodies  unconditionally accepted the authority of the official government --- including that of local elites --- which led those bodies to explicitly discourage the extralegal and violent activities promoted by armed groups. Two examples are the recruitment of young people as soldiers and the participation in coca cultivation: significant anecdotal evidence suggests that religious groups campaigned repeatedly and explicitly against these activities (see \citealt[p. 196]{Beltran2013} and \citealt{USCIRF2009, USCIRF2010, USCIRF2011}).

Because religious movements discouraged the crimes promoted by armed groups, it appears these groups --- particularly the FARC --- viewed religious groups as not only an obstacle to their insurgency, but as competition and, importantly, as a military target (see \citealt[p. 196]{Beltran2013} and \citealt{SemanaMARCH202005}).  In this respect, \cite{Beltran2013} reports that insurgent groups usually saw the evangelical ministers newly established in a municipality as i) exploiters of the population, enriching themselves personally from members' contributions to the church; and ii) an obstacle to the work of insurgency, insofar as those ministers opposed socialism \citep[see][p. 196]{Beltran2013}. The US Commission on International Religious Freedom (USCIRF) reports that the armed groups ``generally targeted religious leaders and practitioners for political or financial rather than religious reasons,'' and that ``nearly all killings of priests by terrorist groups could be attributed to leftist guerrillas, particularly the FARC" and that ``the FARC is responsible for 90 percent of the murders of Protestant religious leaders" \citep{USCIRF2005}.


\section{Theory}\label{Theory}

In this section, we formally analyze the effects of religious competition on armed conflict. Based on the anecdotal evidence mentioned in the previous section, we propose that religious competition affects armed conflict in two different but related ways: motivating an armed group's members and collaborators to abandon the armed group or reduce their support, and providing the armed group with incentives to use greater violence to prevent a potential fall in its membership and support.\footnote{In Section \ref{Othermechanismsandadditionalevidence}, we examine other ways that religious competition may affect armed conflict. We argue that although alternative explanations may be possible, the evidence appears to be more consistent with the model proposed in this section.} Our analysis uses the framework developed in \cite{BarrosGaroupa2002}, where individuals form  communities around churches, each of which chooses a level of religious strictness in a linear space,\footnote{That the strictness space is linear can be easily generalized to a circle model, with individuals and churches located along the circumference of a circle.} and provides a local public good. We focus on how religious competition affects participation in non-religious groups, and add the possibility that there is an armed group that uses violence to prevent potential losses if it expects to be negatively affected by religious competition.

Consider a society in which each individual, $x$, has a preferred level of religious strictness that takes a value from zero to one.\footnote{As will be discussed later, since what matters in our model is contributions from individuals to the organization they belong to, an individual's preferred level of religious strictness can also be interpreted as that individual's willingness to actively collaborate with their chosen church.}  Let $x$ denote individual $x$'s preferred level of strictness. A value of zero (minimal strictness) is interpreted as the individual aligning with a non-religious group, whereas a value of one represents a very religious individual (in our model, this person would align with the Catholic Church). Since members (or collaborators) of armed groups do not actively participate in the religious activities of a church (nor do they provide financial contributions), these individuals are plausibly among those who we classify as belonging to a non-religious group; to simplify the exposition, we assume that all of them belong to the most powerful armed group, which we denote using $F$.\footnote{As previously mentioned, in Colombia, this group is plausibly the FARC. This approach abstracts from a scenario in which there are multiple armed groups, with competition between these groups for recruitment.}

Individuals' preferences for religious strictness are uniformly distributed over $[0,1]$. Individuals can join a church, $C$, or the armed group $F$. 
An individual with an ideal strictness of $x$, affiliated with church $C$ that has a strictness of $c$, has a utility function of $u_x=1-|x-c|$, where $1$ is the utility that the individual gets from consuming a ``religious'' local public good. Each church chooses a strictness to maximize its objective function, which is the sum of the contributions from its members. We assume that each contribution is a linear function of each member's satisfaction, so each church's objective function can be understood as the sum of the welfare of its members.

We consider two scenarios. First, one in which the religious market is served by a single church, $A$,  with strictness $a$.  This could represent a church that is protected from competition by the state. As previously mentioned, this is consistent with the virtual monopoly of the Catholic Church in Colombia before the '90s. In the second scenario, there are two churches, $A$ and $B$, with a strictness of $a$ and $b$, respectively. $B$ could represent an evangelical Protestant church, which describes the great majority of non-Catholic churches in Colombia in the period we examine.  Without loss of generality, we assume that $a<b$, so only $A$ and $F$ will compete for members --- given that $F$'s strictness is fixed at zero.

Importantly, we assume that recruiting new members or contributors from $F$ is costly to $A$. Specifically, we assume that through violence, the armed group reduces by a fraction $\beta\in (0,1)$ the revenue that $A$ would otherwise get by persuading some members of $F$ to reduce their collaboration to the armed group and collaborate more with $A$.\footnote{That only church $A$ can recruit contributors from $F$ (and bear the costs of this action) is also without loss of generality; as previously mentioned, if instead we had assumed a circular space, this assumption would not be necessary.}  $F$ uses violence to deter the leaders of a church from trying to recruit the armed groups's members and collaborators.\footnote{Even though $F$'s actions are not explicitly modeled, and are assumed to be exogenous, this is consistent with a scenario in which in an earlier stage, $F$ decides whether to use violence, with the decision only depending on whether $F$ expects that there are people willing to abandon $F$ to join $A$.}  Note that conflict-related violence occurs only if $F$ expects that there are people willing to reduce their collaboration with $F$ and collaborate more with $A$ as a consequence of the introduction of religious competition. We examine whether, in equilibrium, this happens.

Let $f^m$ denote the equilibrium proportion of people collaborating with the armed group when the religious market is served by a monopoly $A$. Let $f^{c}$ be the (Nash) equilibrium proportion of people collaborating with the armed group when there are two churches that compete in the religious marketplace.  And let $\Delta f \equiv f^m-f^{c}$ denote the proportion of people that, in equilibrium, are willing to reduce their collaboration with the armed group and collaborate more with the church.  In the Appendix (\ref{appmodel}) we show that  $f^m=1/5$ and $ f^{c}=7\beta/(9+15\beta)$, with which
\begin{equation}
\label{propdes0}
\Delta f=\frac{9-20\beta}{5(9+15\beta)}
\end{equation}
Note from (\ref{propdes0}) that there is a unique level of conflict-related violence equal to $9/20$, such that $\Delta f\geq 0$ for all $\beta\leq 9/20$. This means that if the cost to $A$ of violence done to $A$ is sufficiently small, religious competition decreases collaboration with $F$ and increases collaboration with $A$, and this means that conflict-related violence occurs. Note that for $\beta=9/20$, there is a positive level of conflict violence such that $\Delta f=0$, i.e. $F$ successfully prevents a decrease to its support. Figure \ref{hotelling1} illustrates this scenario.

\footnotesize
\begin{figure}[h!]
\begin{center}
\caption{}\label{hotelling1}
\begin{tikzpicture}
\draw (0,0) -- (10,0);
\foreach \x in {0,1.8, 3.4,10}
\draw (\x cm,3pt) -- (\x cm,-3pt);
\foreach \x in {4,7.5}
\draw (\x cm,15pt) -- (\x cm,-3pt);
\foreach \x in {6.5}
\draw [dashed] (\x cm,40pt) -- (\x cm,-3pt);
\foreach \x in {3.4}
\draw [dashed] (\x cm,40pt) -- (\x cm,-3pt);
\foreach \x in {1.8}
\draw (\x cm,15pt) -- (\x cm,-3pt);
\foreach \x in {1.8,4,7.5}
\draw (\x cm,-17pt) -- (\x cm,-30pt);
\draw (0,0) node[below=1pt] {$ 0 $} node[above=3pt] {$ $};
\draw (6.5,0) node[below=1pt] {$ a^m $} node[above=3pt] {$ $};
\draw (2.7,0) node[below=1pt] {$ $} node[above=7pt] {$ \Delta f $};
\draw (1.1,0) node[below=1pt] {$ $} node[above=8pt] {$ F^{c} $};
\draw (3.4,0) node[below=1pt] {$ f^m  $} node[above=3pt] {$ $};
\draw (1.8,0) node[below=1pt] {$ f^{c}  $} node[above=3pt] {$ $};
\draw (4.1,0) node[below=1pt] {$a^{c}$} node[above=3pt] {$ $};
\draw (4.6,0) node[below=1pt] {$  $} node[above=3pt] {$ $};
\draw (7.5,0) node[below=1pt] {$ b^{c}$} node[above=3pt] {$ $};
\draw (10,0) node[below=1pt] {$ 1 $} node[above=3pt] {$ $};
\draw (1.9,0) node[below=1pt] {$ $} node[above=30pt] {$F^m $};
\draw[decorate, decoration = {brace, amplitude = 5pt}, xshift = 0pt, yshift = 5pt] (1.8, 0) -- (3.4, 0);
\draw[decorate, decoration = {brace, amplitude = 5pt}, xshift = 0pt, yshift = 5pt] (0, 0) -- (1.8, 0);
\draw[decorate, decoration = {brace, amplitude = 5pt}, xshift = 0pt, yshift = 23pt] (0, 0) -- (3.4, 0);
\draw[<-] (2.2,-0.4) -- (2.9,-0.4) node(xline)[right] {$$};
\draw[<-] (4.7,-0.4) -- (6,-0.4) node(xline)[right] {$$};
\end{tikzpicture}
\end{center}
\end{figure}
\normalsize
\vspace{-0.6cm}

The intuition for this result is as follows. Religious competition, insofar as it forces churches to compete with each other for adherents, motivates churches to compensate for possible losses by trying to recruit new adherents from the armed group's collaborators. However, this can be costly for the churches, because an expected higher proportion of people willing to reduce their collaboration with the armed group and adhere to the church implies an increase in the chance that the armed group uses violence to reduce its expected loss of support. In equilibrium, religious competition is expected to increase conflict-related violence if the cost that violence imposes on the churches is sufficiently low.

This result constitutes the main empirical prediction of the model: the introduction of religious competition will coincide with a higher probability of conflict-related violence when the armed group expects a drop in support to arise from the religious fracture.


\section{Data and Empirical Strategy}\label{DataandEmpiricalStrategy}


\subsection{Data}\label{subdata}

Our primary data observes the establishment of the first non-Catholic church in a municipality and conflict-related violence. The church establishment data comes from the \emph{Public Registry of Religious Organizations},\footnote{In Spanish, it is called the \emph{Registro P\'ublico de Entidades Religiosas}. This data is publicly available from \url{https://asuntosreligiosos.mininterior.gov.co/mision/asuntos-religiosos/registro-publico-de-entidades-religiosas}.} provided by the  Colombian Ministry of the Interior. It records the time and place at which a non-Catholic church gains legal ``personhood.'' In Colombia, non-Catholic churches must apply for legal status (the Catholic Church already has legal status), and it takes at least 60 working days to receive a response from the government. The request is usually granted. Legal personhood allows a church to sign any contract (e.g. open bank accounts, hire employees, pay the leaders of the organization, buy properties and qualify for tax incentives) and to collect contributions from its members (see Law 133 of 1994). This data is available for each year from 1995 to 2017, and includes the date that the church gained legal status and the primary municipality in which this church is active.\footnote{Once a church becomes a legal entity, its activity may be legally extended to the entire country. Note that this could bias our measure of religious competition, since a church that is the first to obtain legal personhood in a municipality may not be the first church in operation, as another church in the same municipality may have obtained its legal status elsewhere. Thus, we should be cautious when interpreting our estimates, as they may be biased toward zero. However,  in Section \ref{Othermechanismsandadditionalevidence} we show that our results are robust to focusing on non-Catholic churches that are not affiliated with others that already have legal status, which makes up approximately 85\% of our sample. This proportion is consistent with anecdotal evidence showing that Colombian non-Catholic churches are very fragmented (see \citealt[][p. 140]{Beltran2013}  and \citealt[][p. 19]{Tejeiro2010}).}

The data on civil war violence comes from the Conflict Analysis Resource Center (CERAC), which includes information about violent episodes in almost all Colombian municipalities from 1988 onwards. CERAC is a private research organization specializing in data-intensive studies of conflict and criminal violence. The CERAC data use media reports from major newspapers, and cross-check events with other official sources, including data from the National Police and reports by Human Rights Watch and Amnesty International.\footnote{For more information about the collection procedure, see \cite{RestrepoSpagatVargas2004}; see also \cite{DubeVargas2013}, who extensively use this data.}  The CERAC data focuses on attacks and clashes between groups, including information about unilateral actions such as incursions into villages where civilians were intentionally killed, but without distinguishing them from events such as the bombing of pipelines, bridges and other infrastructure, the destruction of police stations or military bases, and ambushes of military convoys.\footnote{Unfortunately, data that is publicly available only includes dummy variables for the occurrence of an event, and ends in 2009. Due to these limitations, in our main results we focus on the extensive margin.} 

We also use of data from  Colombia's National Centre for Historical Memory (\emph{Centro Nacional de Memoria Hist\'orica}, NCHM),  a national and public entity  created to produce a historical account of the armed conflict.\footnote{This data can be found at \scriptsize{\url{http://centrodememoriahistorica.gov.co/observatorio/ bases-de-datos/}}.} The NCHM data we use focuses on civilian victims,  and includes the number of killings, with which we can estimate effects on rates. However, and importantly, this information is very imprecise (e.g. for more than 40\% of the observations, the most likely perpetrator is unknown), and their observations have not been systematically cross-checked against other sources. Given these limitations, we prefer the CERAC database, and use the NCHM database mainly to check robustness.\label{fnGMHdata}

Other controls include the number of Catholic churches in each municipality (which only exists for 1995), municipal population (rural and urban), the proportion of people with unsatisfied basic needs (used as a proxy for poverty), coca crops, internally displaced people, ethnic minority population, and homicide and robbery rates. Sources for these controls are listed in the note attached to Table \ref{precharacteristics_tab}.


\subsection{Empirical Strategy}

To estimate the impact of religious competition on conflict-related violence, we use a linear two-way (unit and time) fixed effects (FE) estimator.  Our main specification models a given outcome $y_{i,t}$ (e.g. probability of a conflict-related event) in municipality $i$ and year $t$ as a function of the arrival of the first non-Catholic church:

\begin{equation}
\label{baseline}
y_{i,t}=\alpha_i+\beta_t+ \sum_{\tau=-K}^{-2}\gamma_\tau D_{i,t}^\tau+\sum_{\tau=0}^{L}\gamma_\tau D_{i,t}^\tau+\epsilon_{i,t}
\end{equation}
where $D_{i,t}^\tau$ is an indicator for $\tau$ years for municipality $i$'s treatment (which in our case is the establishment of the first non-Catholic church in municipality $i$, which occurs at $\tau=0$),\footnote{Note that negative values of $\tau$ indicate the arrival of the first non-Catholic church $|\tau|$ years in the future (so it represents the ``lead'' effect, which is represented by $K$), and positive values indicate its arrival $\tau$ years in the past (i.e, the ``lag,'', which is represented by $L$). Thus,  in this dynamic specification we allow the treatment to have different effects in the adoption year than in subsequent years. Also note that we do not include all possible relative time indicators in Eq. (\ref{baseline}); this is due to multicolinearities, as discussed by \cite{laporte2005estimation}, \cite{BorusyakJaravel2017} and \cite{AbrahamSun2018}. As for which relative time to exclude, we follow the common practice and normalize relative to the period prior to treatment. In addition, in the main specifications we exclude 1995 to allow for heterogenous effects using variables that only exist for that year. In addition, excluding 1995 helps avoid the potential problem of the first non-Catholic church being established in a municipality before 1995.} $\alpha_i$ and $\beta_t$ are municipality and year fixed effects, and $\epsilon_{i,t}$ is the error term. The coefficients of interest are $\gamma_\tau$. The main specifications also include department-year fixed effects, municipality-specific linear trends, and municipality-year controls.\footnote{In all specifications, we cluster the standard errors at the municipal level to control for potential serial correlation over time. However, our results are robust to clustering at the department level. This is a fairly strict test since the cross-sectional variation in our key variables is at the municipal level, and 405 municipalities in our final sample are grouped into 32 departments.}

The primary identifying assumption of the dynamic two-way fixed-effects model in (\ref{didbaseline}) is that changes in ``control'' municipalities (i.e. those that have not yet  experienced the arrival of a non-Catholic church, and those that already experienced it $L$ periods ago) provide a valid counterfactual for the changes that would have occurred in adopting municipalities if they had never experienced the arrival of a non-Catholic church (i.e. there are no time-varying unobserved effects). This assumption, which is fundamental to interpreting our estimates as causal effects, may be invalid if, for example, the arrival of a non-Catholic church was preceded (and explained) by an unobserved conflict-related event. 

We also assume no reverse causality between conflict-related violence and the establishment of a municipality's first non-Catholic church. This implies that it is non-Catholic church's establishment that leads to conflict-related violence, not the other way around. 

The main specification also assumes that the treatment effects are homogenous. This implies that the change in the probability of conflict-related violence associated with the establishment of a non-Catholic church is the same for all municipalities (particularly for those treated at different times). If the homogeneous treatment effects assumption is not satisfied, the estimates obtained using a dynamic two-way fixed effects model may not be causally interpretable. Indeed, in some recent works on identification of treatment effects using multiple time periods and staggered adoption (\citealt{AbrahamSun2018}; \citealt{CallawaySantAnna2018}; \citealt{deChaisemartinDHaultfoeuille2018}; \citealt{GoodmanBacon2018}), in settings where treatment timing varies (as in our case) or  the treatment effect is expected to be non-homogenous, the usual fixed-effects estimator might not recover the average treatment effect. These works show that when treatment effects are heterogeneous across units, estimation of parameters such as $\gamma_\tau$ in Eq.  (\ref{baseline}) recovers a weighted average of all possible pairs of underlying difference-in-differences estimators, so the estimated parameters may not be causally interpretable (see for instance Prop. 2 in \citealt{AbrahamSun2018}).

We examine the empirical validity of these assumptions in a variety of ways.  For example, we explore the robustness of our findings by conditioning on a variety of possibly confounding municipality-year variables. To verify our hypothesis that there is no reverse causality, we show that the presence of conflict-related violence (and related factors) prior to the arrival of a non-Catholic church in a municipality is not correlated with the church's arrival. 

In addition, we explore robustness of our estimations to heterogeneity across units  by using \cite{AbrahamSun2018}'s estimator, which is based on the interaction-weighted estimator of \cite{GoodmanBacon2018} and relies on the absence of anticipation. Finally and importantly, insofar as \cite{AbrahamSun2018}'s estimator assumes a quasi-experimental setting (i.e. a difference-in-differences setting), which may not correspond to our case, we follow \cite{Tchuente_Wind2019} and use an alternative estimator which, besides being robust to cross-unit heterogeneity, allows us to recover the average treatment effects.

In the robustness check section, we consider alternative specifications to Eq. (\ref{baseline}).\footnote{For example, we consider specifications in which different lag and lead effects are included, the treatment effects are heterogenous across municipalities, and $\gamma_\tau$ does not necessarily equal 0 for all $\tau< 0$.} A particularly important alternative specification is the following: 
\begin{equation}
\label{didbaseline}
y_{i,t}=\alpha_i+\beta_t+\gamma D_{i,t}+\epsilon_{i,t}
\end{equation}
where $D_{i,t}$  is a treatment indicator that is equal to one if municipality $i$ is treated at time $t$ or before  and zero otherwise.  We refer to this specification as \emph{static}, and we interpret $\gamma$ as capturing longer-run effects. Compared  to Eq. (\ref{baseline}), the specification in Eq. (\ref{didbaseline}) imposes constant treatment effects for all $\tau\geq 0$, which may not hold if treatment effects grow or decay over time. Given the prevalence of specification (\ref{didbaseline}), its potential usefulness in capturing long-term effects, and that it allows our heterogeneous effects analysis to be presented more clearly,  we always report results obtained using this specification (in addition to those obtained using Eq. (\ref{baseline})). 


\section{Main Results}\label{MainResults}

\subsection{Baseline results}

\begin{figure}[H]
             \caption{Effect of first non-Catholic church on the prob. of a conflict-related event}
        \label{PANYPGUEPPARannual_fig}
\begin{subfigure}{0.5\textwidth}
\caption{Attack by any armed group} \label{PANYannual_fig}
\includegraphics[width=\linewidth]{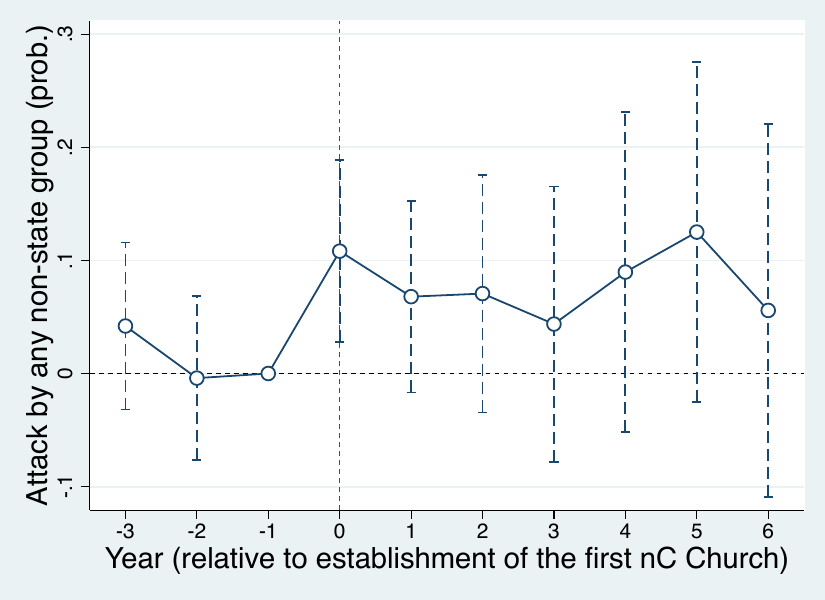}
\end{subfigure}\hspace*{\fill}
\begin{subfigure}{0.5\textwidth}
\caption{Attack by a guerrilla group} \label{PGUEannual_fig}
\includegraphics[width=\linewidth]{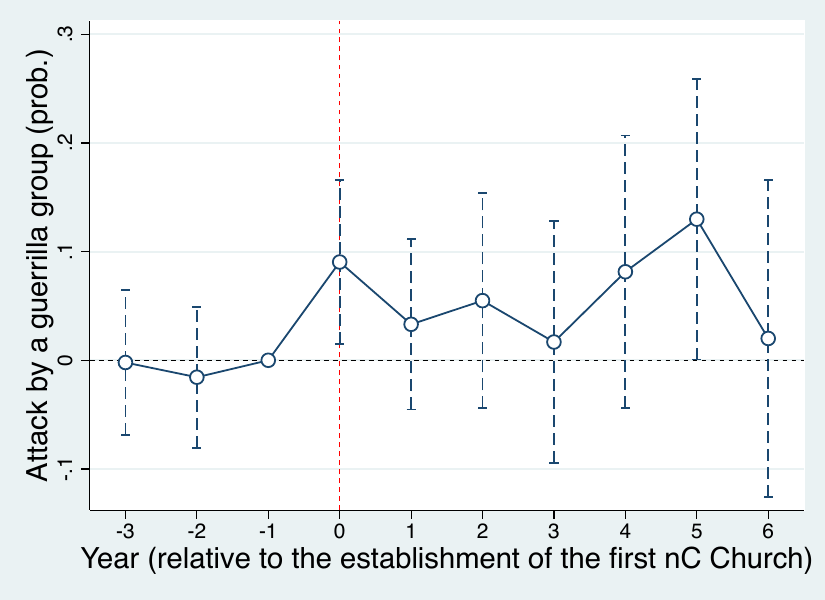}
\end{subfigure}

\medskip
\begin{subfigure}{0.5\textwidth}
\caption{Attack by a paramilitary group} \label{PPARannual_fig}
\includegraphics[width=\linewidth]{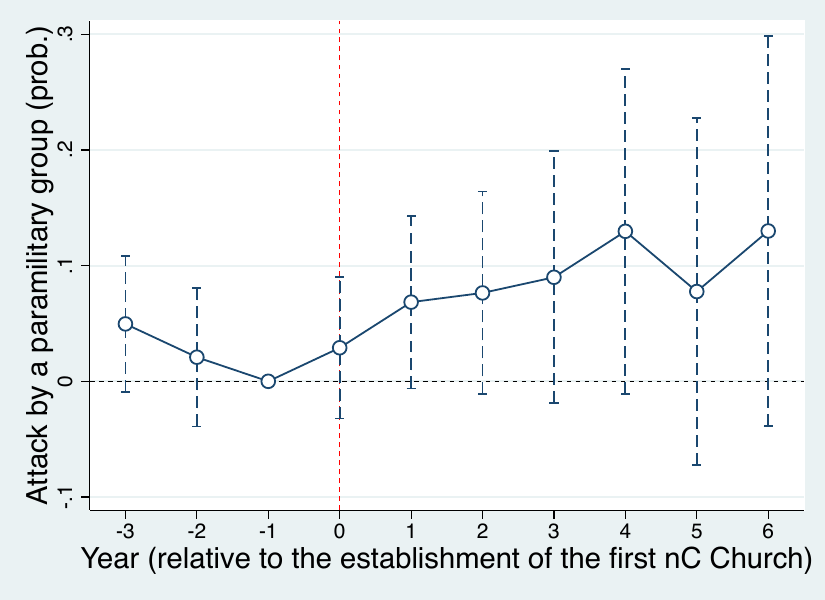}
\end{subfigure}\hspace*{\fill}
\begin{subfigure}{0.5\textwidth}
\caption{Assassination by any armed group} \label{dkillanyannual19962017_fig}
\includegraphics[width=\linewidth]{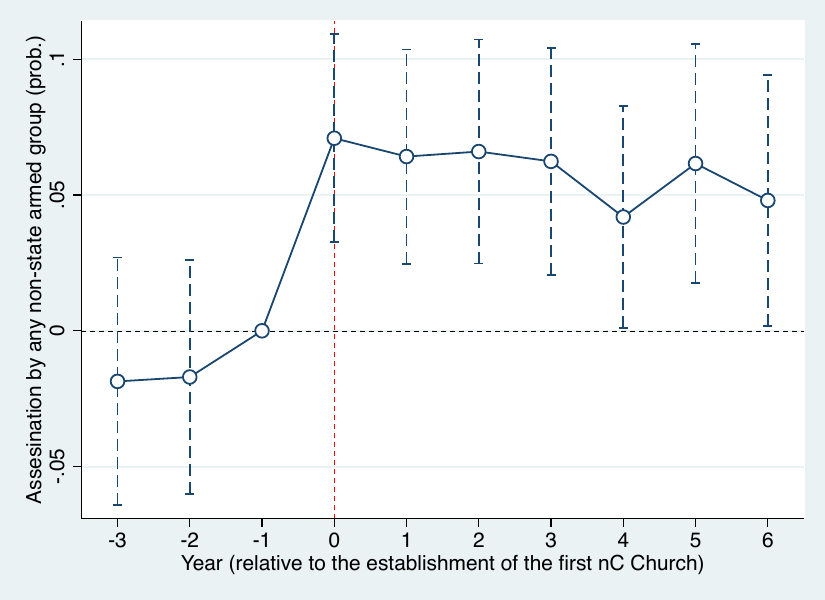}
\end{subfigure}
     \begin{minipage}{16cm} \footnotesizes These figures show the two-way fixed effects estimates from Eq. (\ref{baseline}), in a specification that includes municipality and year  fixed effects, department $\times$ year fixed effects, municipality-specific linear trends and the following (lagged) covariates:  log of total population, proportion of the population living in rural areas, proportion with unsatisfied basic needs (used as a proxy for poverty), proportion of ethnic minority population and homicide rate. Vertical lines indicate 90\% confidence intervals.
\end{minipage}
\end{figure}

Figure \ref{PANYannual_fig} represents the effect of the establishment of the first non-Catholic church in a municipality on the probability of an attack by a non-state armed group. It shows a statistically significant increase in the probability of an attack. Columns (1) and (2) in Panel A of Table \ref{PGUEPANYPGUEPPAR_tab} show estimates for the same effect using Eq. (\ref{baseline}). The specification in column (2) corresponds to that used in Figure \ref{PANYannual_fig}, and includes  municipality and year fixed effects, department-year fixed effects, municipality-specific linear trends and a series of municipality-year controls. The estimates confirm those from Figure \ref{PANYannual_fig}: the probability of an attack by a non-state armed group is 10 percentage points higher in the year that the first non-Catholic church is established. This increase corresponds to 0.2 of a standard deviation, a large effect. The estimates for the following years are also positive, but a little smaller and noisier.\footnote{To simplify the exposition, Table \ref{PGUEPANYPGUEPPAR_tab} only includes estimates for $\tau=-3,-2, 0, 1, 2, 3, 4$ and $5$. Table \ref{PGUErobust1_tab}  in Web Appendix \ref{approbustnessmain} shows estimates for all lags and leads. Table \ref{PGUErobust1redlagsleads_tab} shows that the estimates in Table \ref{PGUEPANYPGUEPPAR_tab} are robust to using fewer lags and leads.} 

Figure \ref{PGUEannual_fig} and \ref{PPARannual_fig}, and columns (3) to (6) in Panel A of Table \ref{PGUEPANYPGUEPPAR_tab}, disaggregate the previous results by perpetrator: a guerrilla group or a paramilitary group.\footnote{In Table \ref{PEST_tab} we report results for attacks by the Colombian national army. The results show no effect.} The results for a guerrilla attack (Figure \ref{PGUEannual_fig} and columns (3) and (4) in Table \ref{PGUEPANYPGUEPPAR_tab}) show a positive, relatively large and statistically significant effect. The results for a paramilitary attack (Figure \ref{PGUEannual_fig} and columns (5) and (6) in Table \ref{PGUEPANYPGUEPPAR_tab})  exhibit a positive but significantly smaller and statistically insignificant effect.\footnote{In Table \ref{PANYdgue_tab} we show  that the estimates in columns (3) and (4) are statistically different from those in columns (5) and (6). We do this by adding interaction terms between the treatment indicators and a dummy variable for a guerrilla attack to the specifications in columns (1) and (2) in Table \ref{PGUEPANYPGUEPPAR_tab}, and showing that all estimates for the coefficient on the interaction terms are positive and statistically significant.} 

Panel B of Table \ref{PGUEPANYPGUEPPAR_tab} reports results  for the same effects from the \emph{static} specification described in Eq. (\ref{didbaseline}). While columns (1) to (4) show  positive, relatively large and statistically significant effects, columns (5) and (6) show no effects. Importantly, all the estimates in Panel B have the same sign and roughly the same magnitude as in Panel A, which may indicate that treatment effects do not change much over time.

Figure \ref{dkillanyannual19962017_fig}  and Table \ref{NCHMyearly_tab} explore robustness to using the alternative dataset from the NCHM, which focuses on the assassination of civilians by a non-state armed group.  As previously mentioned, this data includes information on the intensity of the killings and of the perpetrators (which is however very imprecise). Figure \ref{dkillanyannual19962017_fig} and columns (1) and (2) in Panels A and B of Table \ref{NCHMyearly_tab} present estimates for the probability of a killing by any non-state armed group that are consistent with the results in columns (1) and (2) in Table \ref{PGUEPANYPGUEPPAR_tab}: we find positive, relatively large and statistically significant effects in both the dynamic and static specifications.  In addition, the effects are of about the same magnitude.  This indicates that our earlier results are not spuriously generated by sample selection bias.\footnote{The estimates in Panel A of Table \ref{NCHMyearly_tab}, for the years that follow the establishment of the first non-Catholic church, are less noisy than those in Panel A of Table \ref{PGUEPANYPGUEPPAR_tab}. A possible explanation is that Table \ref{NCHMyearly_tab} uses a larger sample.} 

Columns (3) to (6) in Table \ref{NCHMyearly_tab} focus on  killings for which the most likely perpetrator is a guerrilla group (columns (3) and (4)) or a paramilitary group (columns (5) and (6)). The estimates in Panels A and B are still consistent with those in Table \ref{PGUEPANYPGUEPPAR_tab}: a positive and statistically significant effect for a guerrilla killing, and no effect for a paramilitary killing. However, the magnitudes of these estimates are smaller than those found in  Table \ref{PGUEPANYPGUEPPAR_tab}: for instance, while the probability of an attack by the guerrillas is approximately 8 percentage points higher after the first non-Catholic church is established in a municipality, the probability of a killing by the guerrillas is approximately 3 percentage points higher after the same event. A possible explanation for this difference is the high degree of imprecision in the data about perpetrators: as mentioned in Section \ref{subdata},  for more than 40\% of the killings, the most likely perpetrator is unknown, so guerrilla groups may be behind an important percentage of these killings. This explanation is consistent with the estimates in columns (7) and (8) in Table \ref{NCHMyearly_tab}, which focus on  killings for which the most likely perpetrator is an unknown armed group. The estimates in these columns show positive,  relatively large, and statistically significant effects. Finally, Panels C and D in Table \ref{NCHMyearly_tab} present estimates for the killing rates (per 10,000 inhabitants). No estimate is statistically significant, which indicates that the effect found in Table \ref{PGUEPANYPGUEPPAR_tab}  and Panels A and B of Table \ref{NCHMyearly_tab} is specific to the extensive margin.\footnote{Table \ref{PGUEPANYPGUEPPARrobutscontrols_tab}  explores the robustness of our main results by considering the following alternative specifications: including controls for past electoral outcomes (columns (1), (4) and (7)), including controls for past attacks  (columns (2), (5) and (8)), and excluding municipality-specific linear trends (columns (3), (6) and (9)).  Table \ref{PGUEPANYPGUEPPARclusterdepto_tab}  repeats the regressions in Table \ref{PGUEPANYPGUEPPAR_tab} but clustering the standard errors at the department level (instead of at the municipal level). All the results in Tables \ref{PGUEPANYPGUEPPARrobutscontrols_tab} and \ref{PGUEPANYPGUEPPARclusterdepto_tab} are virtually the same as those in Table \ref{PGUEPANYPGUEPPAR_tab}. Section \ref{Robustness} contains additional robustness checks.} 

The results in this section show that establishing the first non-Catholic church in a municipality substantially increases the probability of an attack by a non-state armed group. In addition, the results show that this effect seems to be specific to attacks by guerrilla groups, and to the extensive margin. These results are consistent with left-wing guerrilla groups being more negatively affected by the presence of non-Catholic churches, thereby having a greater incentive to react violently. As previously argued, this is consistent with anecdotal evidence that identifies left-wing guerrilla groups, and in particular the FARC, as being primarily responsible for the murders of religious leaders.


\subsection{Evidence on our proposed mechanism}

We argued in Section \ref{Theory} that the introduction of a competing church primarily affects conflict-related violence because competition for church adherents affects those would-be adherents' collaboration with non-state armed groups, particularly guerrilla groups. 

Empirically assessing this theory is complex. We examine the plausibility of our hypothesis by looking at one of its most immediate implications: the effect of religious competition on guerrilla violence should be larger in places where guerrilla groups expect to be more affected by a decrease in their recruitment and member retention capacities.
 
To identify these places, we use historical data on coca crops and the existence of cases of forced recruitment. We hypothesize that in municipalities with coca crops or where armed groups routinely recruit combatants, armed groups see the establishment of the first non-Catholic church as more threatening and should therefore react more violently to prevent any drop in collaboration. This hypothesis is consistent with anecdotal evidence mentioned in Section \ref{backgroundrelcomp} that shows that during our period of study, religious organizations repeatedly campaigned against the recruitment of soldiers and against participation in coca cultivation, and this was one of the reasons why armed groups saw religious organizations as an obstacle to their goals.

\begin{figure}[H]
             \caption{Effect of first non-Catholic church on the prob. of a guerrilla attack: heterogeneous effects by existence of cases of forced recruitment and by presence of coca crops}
        \label{PANYPGUEPPARhetero_fig}
\begin{subfigure}{0.5\textwidth}
\caption{Forced recruitment before 1996} \label{PANYPGUEPPARheteroA_fig}
\includegraphics[width=\linewidth]{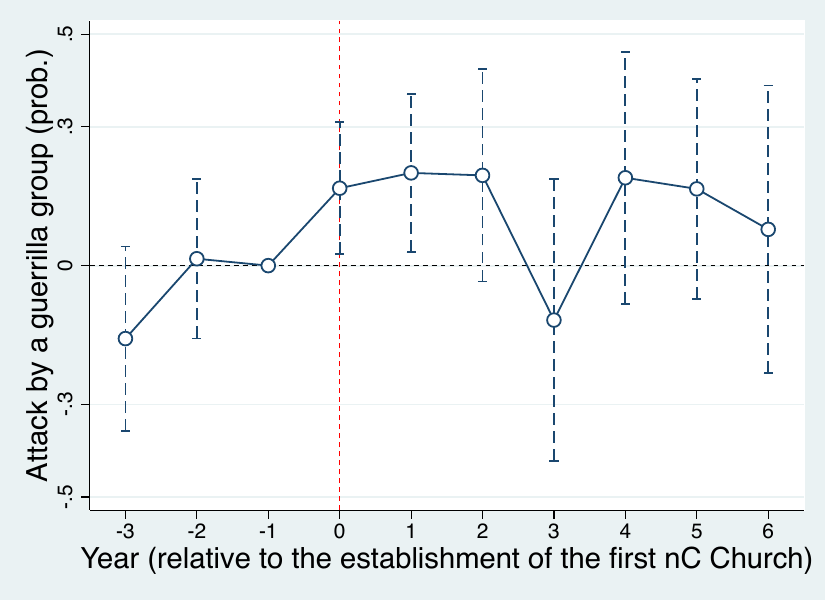}
\end{subfigure}\hspace*{\fill}
\begin{subfigure}{0.5\textwidth}
\caption{No forced recruitment before 1996} \label{PANYPGUEPPARheteroB_fig}
\includegraphics[width=\linewidth]{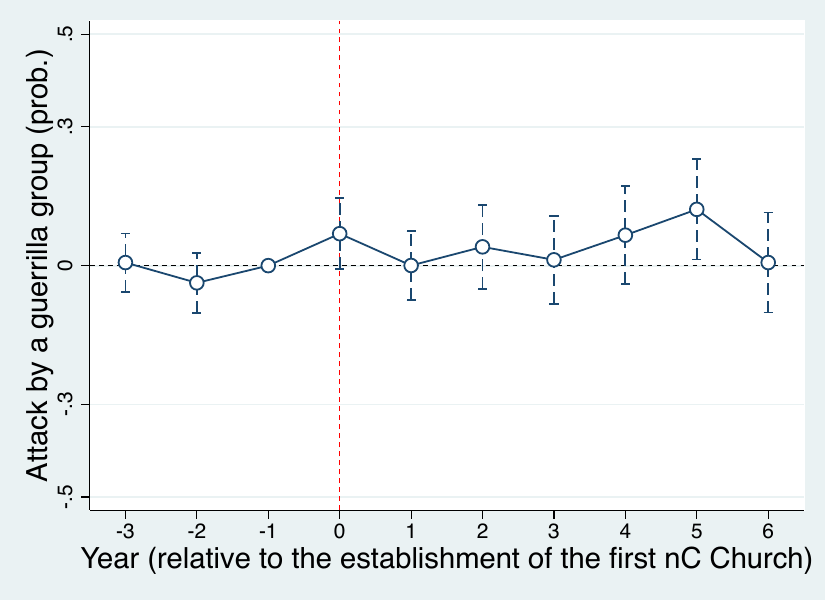}
\end{subfigure}

\medskip
\begin{subfigure}{0.5\textwidth}
\caption{Presence of coca crops in 2000} \label{PANYPGUEPPARheteroC_fig}
\includegraphics[width=\linewidth]{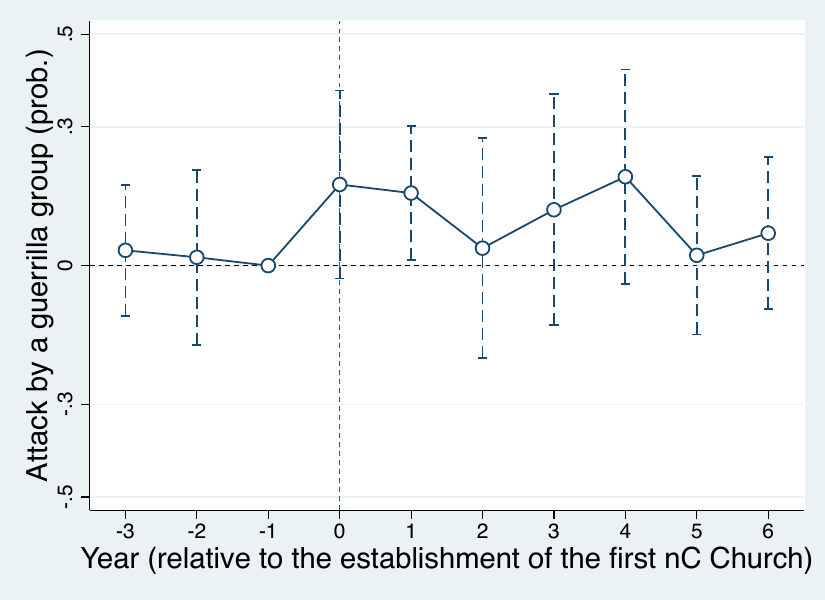}
\end{subfigure}\hspace*{\fill}
\begin{subfigure}{0.5\textwidth}
\caption{No presence of coca crops in 2000} \label{PANYPGUEPPARheteroD_fig}
\includegraphics[width=\linewidth]{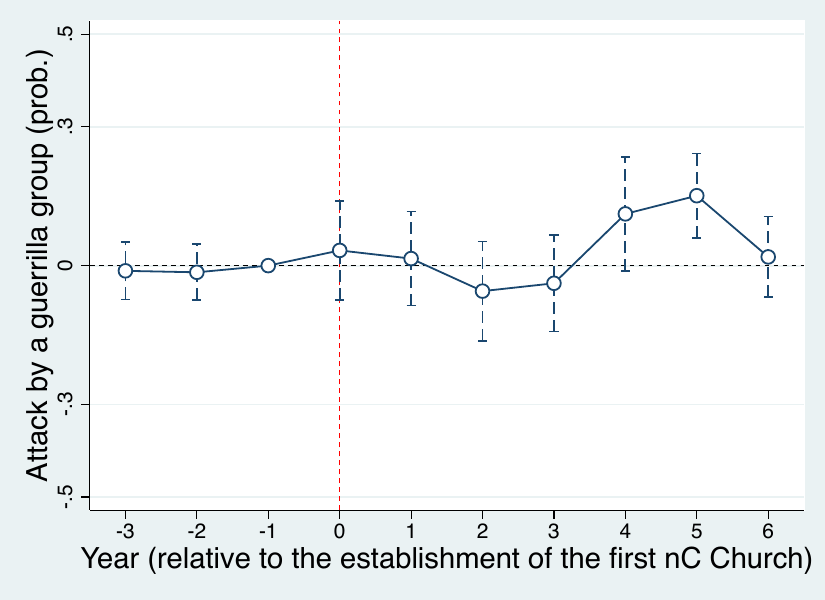}
\end{subfigure}
     \begin{minipage}{16cm} \footnotesizes These figures show the two-way fixed effects estimates from Eq. (\ref{baseline}), in a specification that includes municipality and year  fixed effects, department $\times$ year fixed effects,  and the following (lagged) covariates:  log of total population, proportion of the population living in rural areas, proportion with unsatisfied basic needs (used as a proxy for poverty), proportion of ethnic minority population and homicide rate. Vertical lines indicate 90\% confidence intervals.
\end{minipage}
\end{figure}

Figures \ref{PANYPGUEPPARheteroA_fig} and \ref{PANYPGUEPPARheteroB_fig} (and Table \ref{PANYPGUEPPARrecruit_tab} in Web Appendix \ref{approbustnessmech})  show estimates of the same specifications used in column (4) in Panel A of Table \ref{PGUEPANYPGUEPPAR_tab}, but distinguish between those municipalities with and without past cases of forced recruitment (i.e. before 1996). We find that the effect of religious competition on the probability of an attack by a guerrilla group is substantially larger in those municipalities with at least one case of forced recruitment. Figures \ref{PANYPGUEPPARheteroC_fig} and \ref{PANYPGUEPPARheteroD_fig} (and Table \ref{DIDPANYPGUEPPARcocain2000_tab}  in Web Appendix \ref{approbustnessmech})  repeat the same exercise but distinguish between those municipalities with and without coca crops in 2000 (this is the first year for which data on coca crops is available for the whole country). We find that the effect of religious competition on the probability of an attack by a guerrilla group is larger in those municipalities with a past presence of coca crops.  

Table \ref{PANYPGUEPPARheterostatic_tab} explores robustness to using the \emph{static} specification described in Eq. (\ref{didbaseline}). This table shows estimates of the same specifications used in Panel B of Table \ref{PGUEPANYPGUEPPAR_tab}, but  interacting  the treatment indicator with dummies for municipalities with  past cases of forced recruitment (Panel A) and past presence of coca crops (Panel B). The estimates are consistent with those in Figure \ref{PANYPGUEPPARhetero_fig}: for guerrilla attacks (column (2)), the coefficients of the interaction terms are positive and large.\footnote{Figure \ref{dvsvictkillanyyearhetero_fig} and Tables \ref{DIDNCHMrecruit_tab} and \ref{DIDNCHMcocain2000_tab} in Web Appendix \ref{approbustnessmech} explore robustness to using data from the NCHM, which focuses on conflict-related killings. Figure \ref{dvsvictkillanyyearhetero_fig} and columns (1) and (5) in Panel A of Tables  \ref{DIDNCHMrecruit_tab} and \ref{DIDNCHMcocain2000_tab} show estimates from the dynamic specification. The results are generally consistent with those in Figure \ref{PANYPGUEPPARheteroA_fig} and Tables \ref{PANYPGUEPPARrecruit_tab} and \ref{DIDPANYPGUEPPARcocain2000_tab}: although they are noisier, the estimates for the effect of religious competition on the probability of assassination by a non-state armed group are substantially larger in those municipalities with at least one case of forced recruitment and in municipalities with a past presence of coca crops. Columns (1) and (2) in Panel B of Tables \ref{DIDNCHMrecruit_tab} and \ref{DIDNCHMcocain2000_tab} show results for the static specification. These results are noisier but generally consistent with those in Table \ref{PANYPGUEPPARheterostatic_tab}.}  

The results in Figure \ref{PANYPGUEPPARhetero_fig} and  Table \ref{PANYPGUEPPARheterostatic_tab}  support the theory that our main results are due to guerrilla groups' expectations that more religious competition leads to less collaboration. An immediate question that follows from our hypothesis is how religious competition might affect the capacity of guerrillas to recruit new members and to deter existing members from deserting. The theory in Section \ref{Theory}, as well as the argument in the previous paragraphs, predicts an ambiguous effect. The establishment of the first non-Catholic church in a municipality could imply a decrease in recruitment and an increase in desertion, because greater competition between the churches for adherents includes convincing potential guerrilla recruits and existing guerrilla members to abandon the group for the church. However, if the violent responses by guerrilla groups are effective, the effect should be very small, non-existent or may even be reversed.

Panel A of Table \ref{forcedrecruitdesertion_tab}  shows estimates of Eq. (\ref{baseline}) where the dependent variable is a dummy equal to one if there is at least one case of forced recruitment or a desertion from a guerrilla group, in a municipality-year (the desertion data is only available for 2001 and later, so we restrict the sample from 2001 to 2009 for this outcome). They show no effect in the year that the first non-Catholic church is established, and an increase in subsequent years. Panel B of Table \ref{forcedrecruitdesertion_tab} confirms these results by showing estimates from the \emph{static} specification described in Eq. (\ref{didbaseline}): it shows positive and statistically significant effects in the long run. 

The results in Table \ref{forcedrecruitdesertion_tab}  can be explained as follows. First, in the year that the first non-Catholic church is established (when we found that guerrilla groups reacted more violently), we do not observe any statistically significant effect on the probability of forced recruitment or desertion. This result is consistent with guerrilla groups being effective in maintaining their recruiting capacity and in discouraging desertion. Second, in the years that follow the increase in religious competition, we observe an increase in the chances of forced recruitment and desertion. The increase in the chance of desertion is consistent with the presence of an additional (and perhaps unexpected) cost to guerrilla groups associated with their violent reaction to the increase in religious competition: it may encourage desertion because it may weaken the guerrilla groups internally.\footnote{An important factor that may favor desertion is the ideological deterioration associated with guerrilla groups' involvement in drug trafficking \cite[see][]{verdadabiertaSEPT202010}. Since the guerrillas' violent reaction to the increase in religious competition is greater in municipalities with coca crops, violent actions seeking to protect this  income source may encourage desertion in these municipalities.} And the increased likelihood of forced recruitment is consistent with guerrilla groups increasing forced recruitment to replace deserters \cite[for anecdotal evidence supporting this claim, see][]{eltiempoJUL082019}.


\section{Other mechanisms and additional evidence}
\label{Othermechanismsandadditionalevidence}

The most intuitive alternative explanation for the results in the last section is that the increase in violence by non-state armed groups is only due to ideological differences between the armed groups, particularly guerrilla groups, and non-Catholic churches. According to this hypothesis, the ideology of a typical non-Catholic church --- which, as argued in Section \ref{Background}, includes a defense of the economic \emph{status quo} --- is incompatible with guerrillas' Marxist ideology, and this incompatibility is sufficient to explain the increase in conflict-related violence.

\begin{figure}[H]
             \caption{Effect of first nC church on the prob. of a guerrilla attack: heterogeneous effects by number of Catholic churches in 1995 and Conservative vote share before 1996}
        \label{PGUEhowconscatho_fig}

\begin{subfigure}{0.5\textwidth}
\caption{Above-median \# of Catholic churches} \label{PGUEhowconscathoA_fig}
\includegraphics[width=\linewidth]{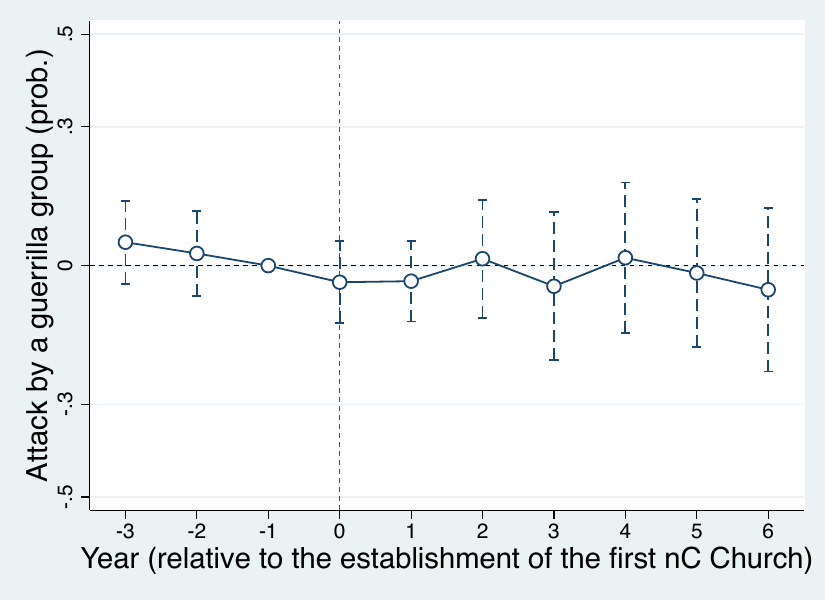}
\end{subfigure}\hspace*{\fill}
\begin{subfigure}{0.5\textwidth}
\caption{Below-median \# of Catholic churches} \label{PGUEhowconscathoB_fig}
\includegraphics[width=\linewidth]{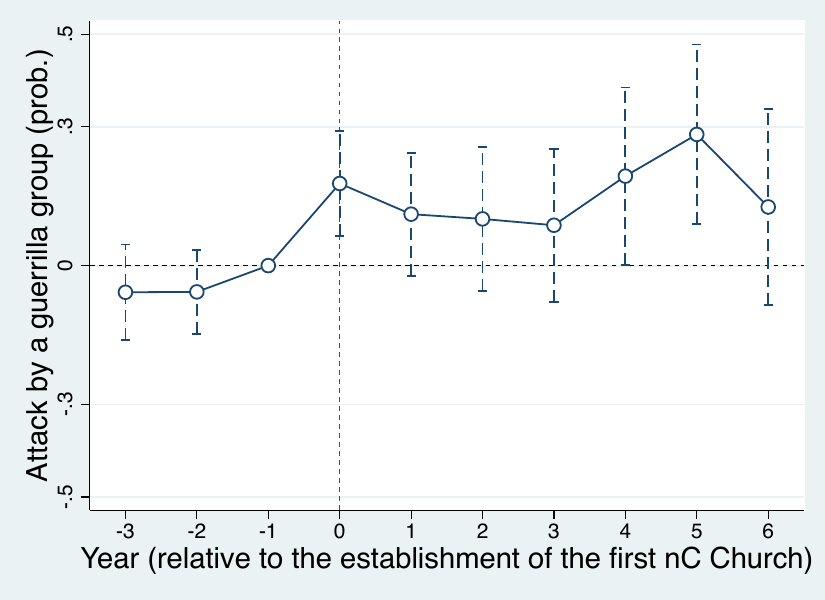}
\end{subfigure}

\medskip
\begin{subfigure}{0.5\textwidth}
\caption{Above-median conservative vote share} \label{PGUEhowconscathoC_fig}
\includegraphics[width=\linewidth]{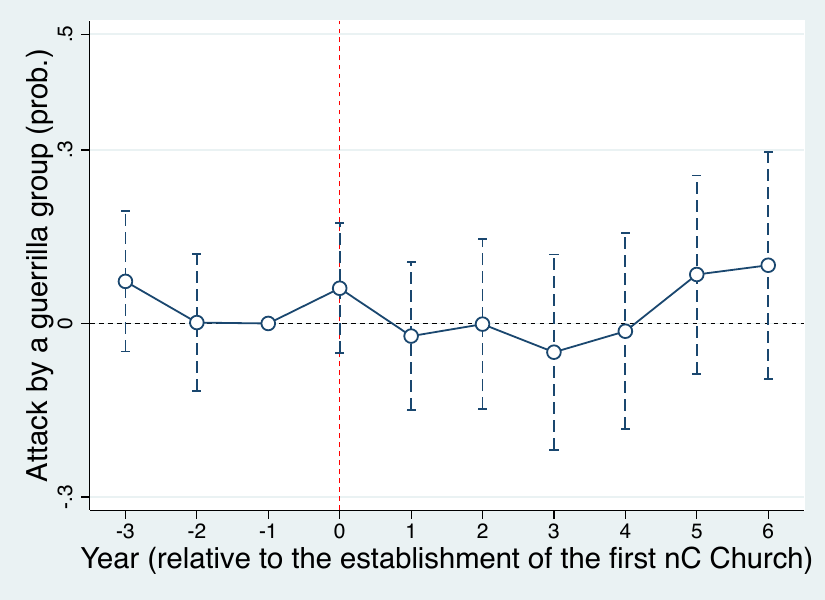}
\end{subfigure}\hspace*{\fill}
\begin{subfigure}{0.5\textwidth}
\caption{Below-median conservative vote share} \label{PGUEhowconscathoD_fig}
\includegraphics[width=\linewidth]{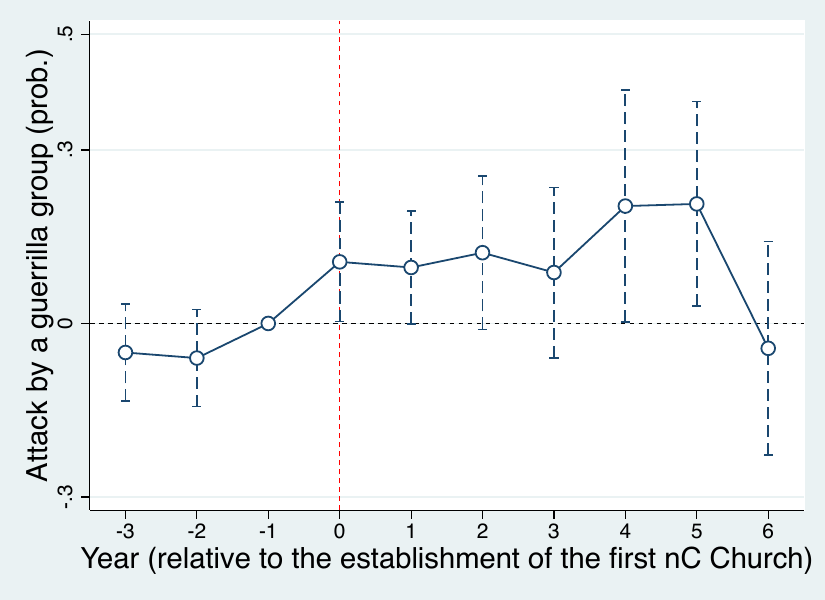}
\end{subfigure}
     \begin{minipage}{16cm} \footnotesizes These figures show the two-way fixed effects estimates from Eq. (\ref{baseline}), in a specification that includes municipality and year  fixed effects, department $\times$ year fixed effects, and the following (lagged) covariates:  log of total population, proportion of the population living in rural areas, proportion with unsatisfied basic needs (used as a proxy for poverty), proportion of ethnic minority population and homicide rate. Vertical lines indicate 90\% confidence intervals.
\end{minipage}
\end{figure}

Empirically assessing the extent to which the ideology of guerrilla groups is incompatible with that of a non-Catholic church is difficult. A first argument against this alternative explanation was provided in the last section: even though the effect of the first non-Catholic church on the probability of an attack by a right-wing paramilitary group is small and noisy, it still appears to be positive (e.g. see Figure \ref{PPARannual_fig} and columns (5) and (6) in Table \ref{PGUEPANYPGUEPPAR_tab}).  A second argument is based on the idea that if ideological differences alone mattered, the intensity of religious competition should not be particularly relevant to the results in Table \ref{PGUEPANYPGUEPPAR_tab}. In Figure \ref{PGUEhowconscatho_fig} and Table \ref{PGUEhowconscatho_tab}, we examine whether this is the case by presenting estimates for municipalities where the past presence of the Catholic church (as measured by the number of Catholic churches in 1995 per 100,000 inhabitants) is above (or below) the median, and where the historical level of conservatism (as measured by the Conservative party's vote share in mayoral elections before 1996) is above (or below) the median. We hypothesize that in municipalities that had relatively few Catholic churches or that have not been very conservative, the first non-Catholic church has a greater chance of attracting new adherents, so religious competition there should be more intense. Consistent with our hypothesis, we find that the effect of religious competition on the probability of an attack by a guerrilla group is substantially greater in municipalities that had nor been strongly Catholic (Figures \ref{PGUEhowconscathoA_fig} and \ref{PGUEhowconscathoB_fig} and Panel A of Table \ref{PGUEhowconscatho_tab}) nor very conservative (Figures \ref{PGUEhowconscathoC_fig} and \ref{PGUEhowconscathoD_fig} and Panel B of Table \ref{PGUEhowconscatho_tab}). In addition, for strongly Catholic or conservative municipalities, the effect is not only smaller but statistically insignificant. We interpret these results as providing additional evidence that intense religious competition --- rather than only ideological differences --- is crucial to the story.

A second alternative explanation is that the increase in guerrilla violence is due to guerrilla groups' interests in preventing newly established non-Catholic groups from gaining political representation. Although this explanation is compatible with that proposed in Section \ref{Theory}, and we see it as complementing our preferred mechanism, anecdotal evidence does not seem to support it: between the mid-1990s and mid-2000s, guerrilla groups (particularly the FARC) typically tried to sabotage local elections instead of sponsoring left-wing candidates \cite[see][pp. 257-267]{GMHFARC2013}.\footnote{In Tables \ref{DIDPANYPGUEPPARdpropleftvote_tab} and \ref{nextelectoral_tab}  we present indirect evidence that is consistent with the hypothesis that political influence is not crucial for our main results. In Table \ref{DIDPANYPGUEPPARdpropleftvote_tab}, we compare the effect of religious competition on the probability of an attack by a guerrilla group in municipalities where the historical support for left-wing parties (as measured by vote share for left-wing parties in mayoral elections before 1996) is above versus below the median. Although very noisy given the small number of municipalities with historical support for left-wing parties, the results show no differences between municipalities with above-median historical support for left-wing parties and municipalities with below-median support. In Table \ref{nextelectoral_tab}, we look at the effect of the establishment of the first non-Catholic church on the  the vote share obtained by Liberals, Conservatives or left-wing parties in the next election. We do not find any statistically significant effect.}

\begin{figure}[H]
             \caption{Effect of first nC church on the prob. of a guerrilla attack: heterogeneous effects by ethnic minority population in 1993}
        \label{PGUEindigblacktotprop_fig}

\begin{subfigure}{0.5\textwidth}
\caption{Above-median ethnic minority population} \label{PGUEindigblacktotpropA_fig}
\includegraphics[width=\linewidth]{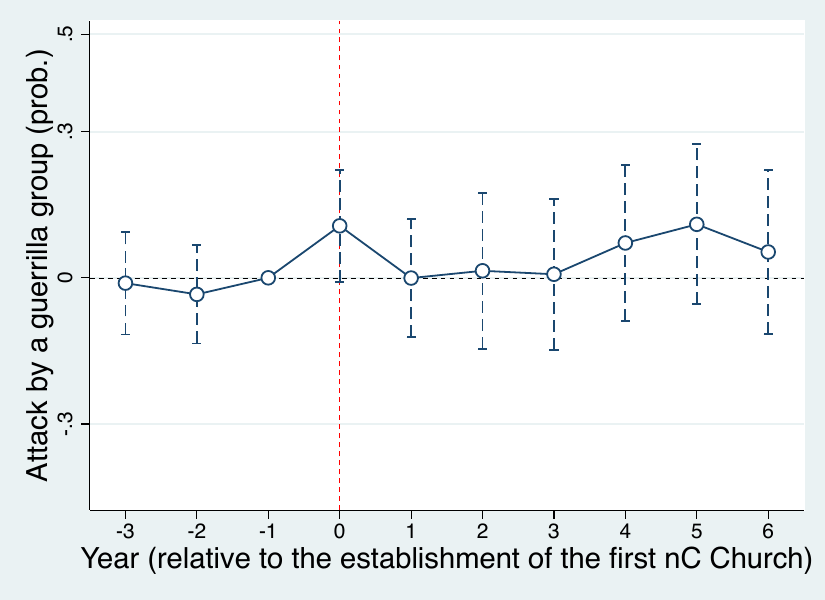}
\end{subfigure}\hspace*{\fill}
\begin{subfigure}{0.5\textwidth}
\caption{Below-median ethnic minority population} \label{PGUEindigblacktotpropB_fig}
\includegraphics[width=\linewidth]{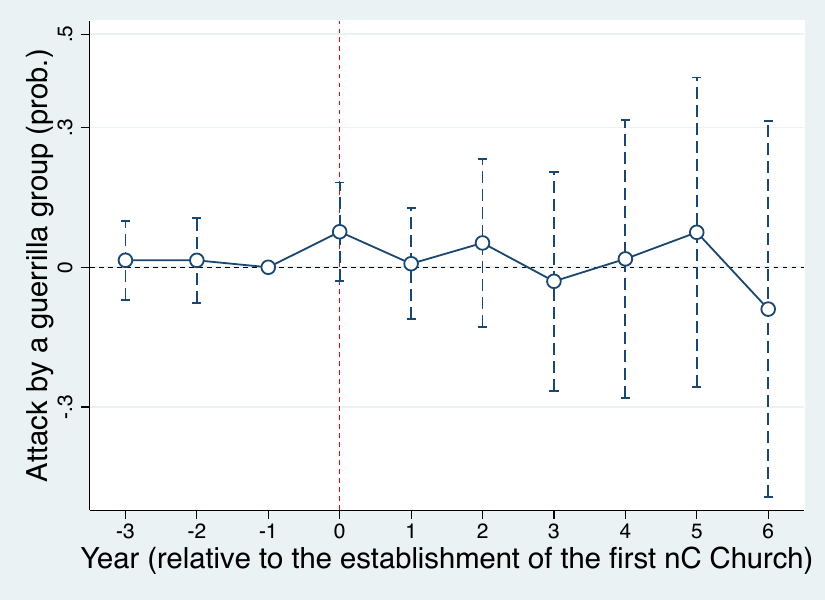}
\end{subfigure}
     \begin{minipage}{16cm} \footnotesizes These figures show the two-way fixed effects estimates from Eq. (\ref{baseline}), in a specification that includes municipality and year  fixed effects, department $\times$ year fixed effects, and the following (lagged) covariates:  log of total population, proportion of the population living in rural areas, proportion with unsatisfied basic needs (used as a proxy for poverty),  and homicide rate. Vertical lines indicate 90\% confidence intervals.
\end{minipage}
\end{figure}

A  third alternative explanation relates to the role of ethnic identity in violence. Although in the next section we will argue that it is unlikely that the establishment of a first non-Catholic church in a municipality is due to a previous change in its ethnic diversity,  the ethnic tensions associated with this diversity may nevertheless be exacerbated by increased religious competition (e.g. through the ``sacralization'' of group identity, as \citealt{Seul1999} suggests). This may result in more violence if armed groups are instrumental to the goals of some of the conflicting communities.\footnote{We thank a referee for suggesting this alternative explanation.}  In Figure \ref{PGUEindigblacktotprop_fig} and Panel A of Table \ref{indigblacktotprop_tab}, we empirically assess the plausibility of this explanation by comparing the effect of religious competition on the probability of an attack by a guerrilla group in municipalities where the past share of ethnic minority population (as measured by the proportion of Afro-Colombian and indigenous population in 1993) is above and below the median.\footnote{In the period we focus on, information on ethnic identity  is only available for two years, 1993 and 2005, and is limited to two ethnic minority groups: Afro-Colombian and indigenous (together, these two groups account for no more than 15\% of the population).} The results show no difference.  In Panel B of Table \ref{indigblacktotprop_tab}, we examine whether the establishment of a first non-Catholic church increases ethnic tensions by altering the ethnic diversity of the municipality (instead of just exacerbating existing tensions). The results show no effect. Although some anecdotal evidence documents the success of non-Catholic churches in recruiting people from ethnic minority groups (in particular from Afro-Colombian communities, as documented by \citealt[][p. 192]{Beltran2013}),  Figure \ref{PGUEindigblacktotprop_fig} and Table \ref{indigblacktotprop_tab} indicates that this phenomenon is either too small to be observable in our data, or it is not relevant in explaining our main results.\footnote{These results are consistent with the standard narrative on the role of ethnic minority groups in the Colombian conflict, which is that the vast majority of these groups consistently refused to become directly involved in the conflict as violent actors or by aligning with existing violent actors \citep{Penaranda2015}.}

A fourth alternative explanation is related to the possible effect that religious competition might have on the ``moral'' behavior of individuals, which while not directly related to their collaboration with guerrilla groups, may nonetheless affects guerrillas' incentives to use violence. Specifically, consider a scenario in which the introduction of religious competition makes individuals more predisposed to committing crimes --- such as murders or robberies --- because their spiritual standards are diluted (as in \citealt{Eswaran2011}). An increase in crime could lower the cost of violence for guerrillas, by exhausting police resources and weakening local institutions, leading them to an increase in guerrilla violence. Table \ref{homicidespcyear} examines the plausibility of this hypothesis by looking at the effect of religious competition on the rates of homicides and robberies. Importantly, no estimate in Table \ref{homicidespcyear} is statistically different from zero, which provides evidence against this explanation.

\smallskip

\begin{figure}[H]
             \caption{Effect of first nC church on the prob. of a guerrilla attack: heterogeneous effects by type of non-Catholic church}
        \label{PGUEspeext_fig}
\begin{subfigure}{0.5\textwidth}
\caption{Non-affiliated non-Catholic churches} \label{PGUEspe_fig}
\includegraphics[width=\linewidth]{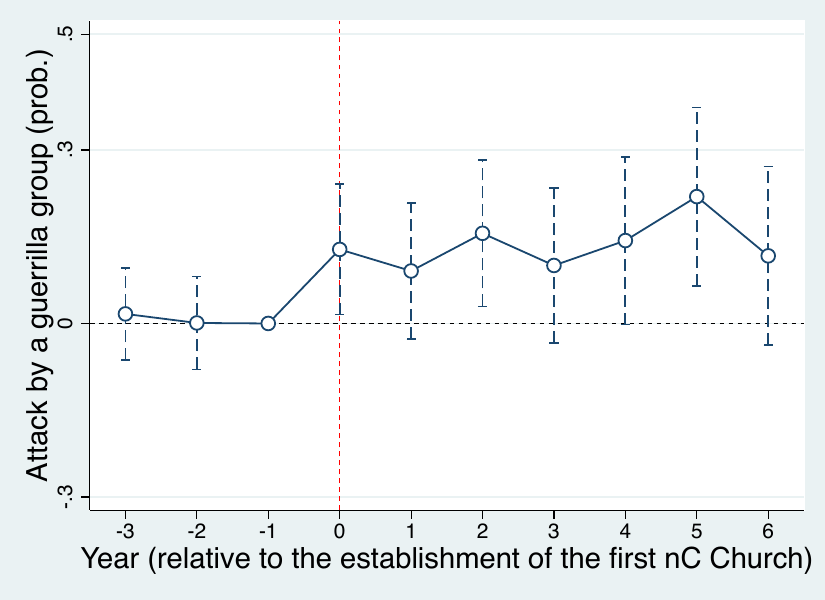}
\end{subfigure}\hspace*{\fill}
\begin{subfigure}{0.5\textwidth}
\caption{Affiliated non-Catholic churches} \label{PGUEext_fig}
\includegraphics[width=\linewidth]{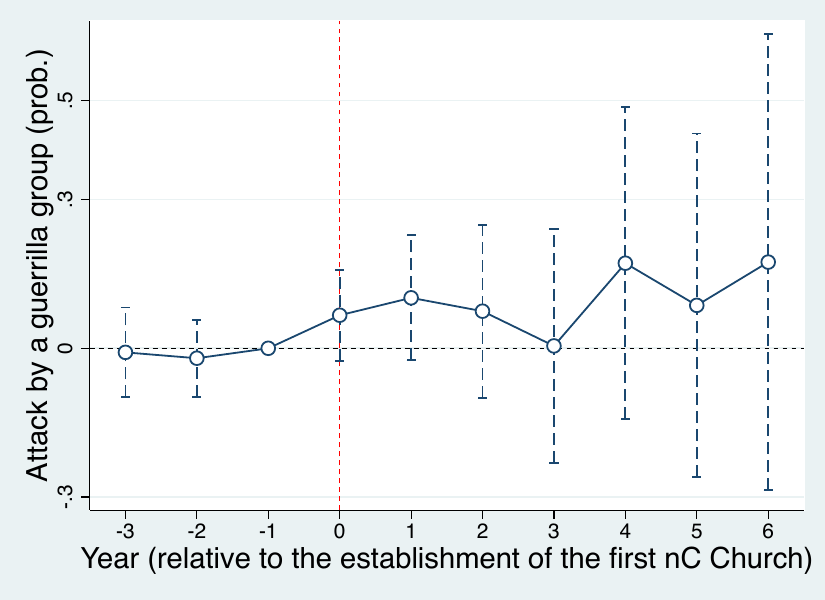}
\end{subfigure}
     \begin{minipage}{16cm} \footnotesizes These figures show the two-way fixed effects estimates from Eq. (\ref{baseline}), in a specification that includes municipality and year  fixed effects, department $\times$ year fixed effects, and the following (lagged) covariates:  log of total population, proportion of the population living in rural areas, proportion with unsatisfied basic needs (used as a proxy for poverty), proportion of ethnic minority population and homicide rate. Vertical lines indicate 90\% confidence intervals.
\end{minipage}
\end{figure}

Finally, we empirically examine the relevance of a key differentiator among non-Catholic churches for which data is available: whether or not these churches are affiliated with others that already have legal status.\footnote{According to Law 133 of 1994, affiliated religious organizations must apply for legal status (i.e. ``extended legal personhood''). Our data allows us to identify this type of application.} This distinction may be relevant in assessing the plausibility of our preferred mechanism if, for instance, our main results are concentrated on non-Catholic churches affiliated with an small number of ``mega-churches''; such evidence would be more consistent with an explanation based on competition for political representation (between the armed groups and the mega-churches).  First, we find that  approximately 85\% of the first non-Catholic churches in our sample are not affiliated with another church. This provides empirical support for our claim that Colombian non-Catholic churches are very fragmented. Second, we repeat the regressions in  Table \ref{PGUEPANYPGUEPPAR_tab} but distinguish between affiliated and non-affiliated churches. Figure \ref{PGUEspeext_fig}  (and Tables \ref{PGUEPANYPGUEPPARspe_tab} and \ref{PGUEPANYPGUEPPARext_tab} in Web Appendix \ref{approbustnessmech}) show the results. Importantly, we find that our main results are robust and even stronger for non-affiliated churches (Figure \ref{PGUEspe_fig}), and that the effect still exists (but with a somewhat smaller magnitude) for affiliated churches. We interpret this result as being consistent with our preferred mechanism.


 \section{Additional Robustness Checks}
 \label{Robustness}

As mentioned in Section \ref{DataandEmpiricalStrategy}, our empirical strategy relies on four main identifying assumptions: the absence of time-varying selection on unobservables, the absence of reverse causality, treatment effect homogeneity and no anticipatory behavior. In this section, we discuss these assumptions and alternative estimation procedures that relax them.

Regarding the absence of selection on unobservables, we do the following. First, we show that our results are robust to the inclusion of a variety of possibly key confounding municipality-year variables. They include the log of total population; the proportion of the population living in rural areas; the proportion of the population with unsatisfied basic needs (used as a proxy for poverty); the proportion of the population that is an ethnic minority; the homicide rate; the share of the vote for the Liberal Party, the Conservative party and left-wing parties; and the prior presence of a non-state armed group.  The results in columns (2), (3) and (5) of Table \ref{PGUEPANYPGUEPPAR_tab}, which come from a specification that includes the first five of these controls,  show that the estimates are virtually the same as those obtained without controlling for these variables. Table \ref{PGUEPANYPGUEPPARrobutscontrols_tab} in Web Appendix \ref{approbustnessmain}  expands this robustness check by presenting results from specifications that include the other controls previously listed. The results remain the same. In addition, the specifications in Tables \ref{PGUEPANYPGUEPPAR_tab} and \ref{PGUEPANYPGUEPPARrobutscontrols_tab} not only include municipality and year fixed effects, but also department-year fixed effects; importantly,  some specifications include municipality-specific linear trends (which allow for the possibility that conflict-related violence was already on different trajectories in different municipalities prior to the arrival of a non-Catholic church). 

Second, we show that the presence of conflict-related violence (and related factors) in the year prior to the establishment of the first non-Catholic church in a municipality is not correlated with that church's arrival. Column (1) in Table \ref{dfirstiglesiasnc_tab} presents the correlation between the establishment of the first non-Catholic church in a municipality and an attack by a non-state armed group in the previous year, using a specification that includes municipality, year, department-year fixed effects and municipality-specific linear trends.  Column (2) regresses the same outcome on lags of total population, the proportion of the population living in rural areas, unsatisfied basic needs,  the proportion of the population that is an ethnic minority, and the homicide rate. Column (3) combines regressors from columns (1) and (2) and adds electoral outcomes for the last mayoral election (which reduces the sample). Column (4) includes lags of the rate of internally displaced persons (outflows and inflows), of the occurrence of at least one desertion from a guerrilla group and an event of forced recruitment. Column (5) combines the regressors from columns (1) and (4). Column (6) adds the interaction of the internal coffee price with coffee intensity (which, following  \citealp{DubeVargas2013}, we interpret as a proxy for income shocks).\footnote{\cite{DubeVargas2013} use this interaction to estimate the effect of commodity price shocks on armed conflict in Colombian municipalities. They find that coffee price shocks are negatively related to conflict. By noting that agricultural commodities are labor intensive, and by showing that coffee price shocks also affect labor market outcomes, \citeauthor{DubeVargas2013} argue in favor of an opportunity cost mechanism: commodity prices affect conflict because they alter the opportunity cost of armed recruitment. Since coffee price shocks can similarly affect religious adherence to newly established non-Catholic churches, the inclusion of this variable is important for identification.} Finally,  column (7) combines the regressors from all the previous columns. Importantly, we find that in all specifications, factors closely related to present attacks (e.g. violence in the previous year, internally displaced persons, guerrilla desertion,  income shocks) are statistically insignificant. These results are robust to using the NCHM data (see Table \ref{dfirstiglesiasncNCHMyearly_tab} in Web Appendix \ref{approbustnessmain}).

In Table \ref{dfirstiglesiasnc_tab}, we focus on specifications where the sample coincides most closely with our main results. The estimates show that the only variable that is correlated (slightly negatively) with the establishment of the first non-Catholic church is total population. We interpret these results as evidence that conditional on the controls (which includes an important group of fixed effects), the arrival of a non-Catholic church in a municipality is unlikely to be related to any unobserved time-varying factors specific to a municipality that are also closely correlated to guerrilla violence.\footnote{\label{ftdetfirshistorical}The results in Table  \ref{dfirstiglesiasnc_tab} do not shed any light on why a non-Catholic church establishes in a municipality. In addition, they do not seem consistent with what other scholars  have mentioned as the main reason for the success of non-Catholic churches in Colombia: the urban transition and the need for hope that results from this phenomenon (see Section \ref{Background}).  Since it is reasonable to expect that the establishment of the first non-Catholic church in a municipality is not random, in Web Appendix \ref{appdeterminants} we propose an explanation that is consistent with the results in Table  \ref{dfirstiglesiasnc_tab}, and provide evidence in its favor.  Specifically, we hypothesize that the urban transition is still relevant, but only in places where, for historical reasons, other channels for social (or political) action are not available (see Table \ref{dfirstiglesiasnchist_tab}).  These additional results allow us to argue that in our main specification, we account for the key factors that plausibly explain the success of non-Catholic churches in Colombia: a historical characteristic, captured by municipality fixed effects and whose evolution may be captured by the municipality-specific trends, and the proportion of the population living in rural areas, the level of poverty and attacks by guerrillas, which we include as controls.}

Third, an important assumption of the two-way flexible model proposed for our analysis is the absence of reverse causality.  To verify this assumption, we provide three kinds of evidence. First, we extend the analysis presented in Tables \ref{dfirstiglesiasnc_tab} and \ref{dfirstiglesiasncNCHMyearly_tab}, which, as previously discussed, indicates that conflict-related violence in the year preceding the establishment  of a first non-Catholic church has no effect on that church opening. Even though this evidence is consistent with the absence of reverse causality, it may be still the case that conflict-related attacks preceded the non-Catholic church's opening if they occurred in the same year that the church opened.  Table \ref{dfirstiglesiasncNCHMyearly_tab} in Web Appendix \ref{approbustnessmain}  implements the same specification as in Table \ref{dfirstiglesiasnc_tab}, but using \emph{monthly} data (which is available only from the NCHM dataset). Crucially, the results are consistent with those in Tables \ref{dfirstiglesiasnc_tab} and \ref{dfirstiglesiasncNCHMyearly_tab}.  Second, we examine the relationship between conflict-related attacks and non-Catholic church establishment occurring in the same calendar year. Specifically, we add to Eq. (\ref{baseline}) the interaction between  each lag and lead and a set of dummies for the month of the establishment of each church.\footnote{We thank a referee for suggesting this robustness check.}   Our hypothesis is that in absence of reverse causality,  churches established early in the year should have a relatively larger effect on attacks than churches established late in the year. Table \ref{PGUEPANYPGUEPPARmonthofest_tab} in Web Appendix \ref{approbustnessmain} shows results that are consistent with our hypothesis: the effect in the year of installation is larger for churches established in January. Third, we examine the effect of being $\tau$ \emph{months} prior to the establishment of the first non-Catholic church on conflict-related assassinations. Table \ref{NCHMmonhtly_tab} in Web Appendix \ref{approbustnessmain}  presents the results. Although the estimates are much noisier, we find no effect, which is consistent with our hypothesis. These results provide crucial support for the absence of reverse causality.

Regarding the assumption of treatment effect homogeneity, we implement an interaction-weighted estimator whose identification relies on assumptions specific to difference-in-differences frameworks. As previously mentioned, we use the specification proposed by \cite{GoodmanBacon2018} and implemented by \cite{AbrahamSun2018}.\footnote{The \cite{AbrahamSun2018} estimator essentially uses an interacted specification saturated with relative time and cohort indicators, and takes appropriate weighted averages of the resulting estimates. Specifically, we first estimate
\vspace{-0.3cm}
\begin{equation}
\label{iwestimator}
y_{i,t}=\alpha_i+\beta_t+ \sum_{e=1996}^{2009} \sum_{\tau\neq -1}\delta_{e, \tau}(1\{E_i=e\} \cdot D_{i,t}^\tau)+\nu_{i,t}
\end{equation}
\vspace{-0.3cm}

where $e$ denotes the year of the initial treatment, $1\{E_i=e\}$ is a dummy variable equal to one for the set of units treated at $e$, $D_{i,t}^\tau$ is an indicator for being $\tau$ years relative to $i$'s treatment, and $\alpha_i$ and $\beta_t$ are municipality and year fixed effects. Second, we estimate a set of appropriate weights that are the sample share of each cohort $e$ across cohorts that are observed $\tau$ periods after the establishment of the first non-Catholic church. Third and finally, we take weighted averages of the estimates from (\ref{iwestimator}), $\hat{\delta}_{e,\tau}$, to form average treatment effect estimates with the weight estimates previously obtained. Like \cite{AbrahamSun2018}, we exclude the first cohort from the sample (i.e. municipalities treated in 1996) because we do not observe this cohort when untreated. Unlike \cite{AbrahamSun2018}, we do not need to drop the last period because there are municipalities that are not treated in the last period (but that will be treated in a future period). In addition, our sample contains a year in which no municipality is treated (2000), which we need to exclude. As in (\ref{baseline}), we include department-year fixed effects, but do not include municipality-specific linear trends nor any other control variable.}   Panel A in Figure \ref{PGUEiwaands_fig} and columns (3) and (4) in  Table \ref{PANYPGUEPPARiwaands_tab} in Web Appendix \ref{approbustnessmain} show the estimates (which we call ``AS''). The AS estimates are consistent with those in Table \ref{PGUEPANYPGUEPPAR_tab}. 

Additionally, we relax both the homogeneity across  municipalities assumption and the ``no anticipatory behavior'' assumption by proposing an alternative specification from \cite{Tchuente_Wind2019}.  In Appendix  \ref{appidentificationTW}, we summarize this specification, which focuses on the delayed and anticipated average treatment effects (as in \citealt{laporte2005estimation}) rather than on the ``cohort-specific average treatment"  (as in \citealt{AbrahamSun2018}).\footnote{The \cite{Tchuente_Wind2019} estimator essentially splits the sample in two groups: one for influenced periods (denoted by $L$), and other for which $D_{i,t}^{\tau}=0$. For the first group, we estimate
\vspace{-0.0cm}
\begin{equation}
\label{ }
y_{i,t}-\bar{y}_{i}^L= \sum_{\tau \in \{-K,...,L\}}\theta^{\tau,L}D_{i,t}^{\tau}+\upsilon_{i,t}
\end{equation}
\vspace{-0.2cm}

where $\bar{y}_{i}^L=1/(T-\bar{L}_i)\sum_{t=1}^{T}y_{i,t}(1-L_{i,t})$, $L_{i,t}= \mathbf{1}\{ t \in \{  -K,... ,e_i,...L\}\}$ is an indicator of the fact that period $t$ is in the neighborhood of the event, and $\bar{L}_{i}=\sum_{t=1}^{T}L_{i,t}$.  For the second group, we estimate
\begin{equation}
\label{ }
y_{i,t}-\bar{y}_{i}^L= \theta^{\tau,D}L_{i,t}+\vartheta_{i,t}
\end{equation}
Finally, we compute $\hat{\gamma}_{FEL}^{\tau}= \hat{\theta}^{\tau,L}+ \hat{\theta}^{\tau,D}$.} Panel B in Figure \ref{PGUEiwaands_fig} and columns (5) and (6) in Table \ref{PANYPGUEPPARiwaands_tab} in Web Appendix \ref{approbustnessmain} show the estimates (which we call ``TW'').  The TW estimates are also consistent with those in Table \ref{PGUEPANYPGUEPPAR_tab}.   These results, as well as those obtained using the AS estimator,  provide a crucial robustness check for the estimates presented in Section \ref{MainResults}.

Finally, concerning the no anticipatory behavior assumption, in addition to the results in Table \ref{dfirstiglesiasnc_tab} (which are consistent with non-anticipation), we show that there is no effect of being $\tau$ years prior to the establishment of  the first non-Catholic church. In our specification, this means that the estimates for $\gamma_\tau$ with $\tau<0$ are equal to zero.  Table \ref{PGUEPANYPGUEPPAR_tab} shows the estimates for $\gamma_\tau$ with $\tau=-3$ and $\tau=-2$ and Table \ref{PGUErobust1_tab} in Web Appendix \ref{approbustnessmain} includes the estimates for all leads. None of the relevant estimates for $\tau<0$ are statistically significant. 


\section{Conclusion}\label{Conclusion}

This study examines how religious competition affects armed conflict.  It focuses on Colombia, a deeply religious country that has suffered one of the world's longest-running domestic conflicts, and that in the last few decades also experienced a significant increase in religious competition. Two-way fixed effects estimates show that religious competition substantially increases the probability of an attack by a non-state armed group, particularly a guerrilla group. Further analysis suggests that the increase in attacks by guerrilla groups is associated with guerrillas' expectation that their membership will drop because of more competition for religious adherents.

Several opportunities exist for future research. One could examine the effect of religious competition on political participation.  It would also be interesting to examine whether other organizations and local institutions are affected. Finally, there are the questions about how religious groups react to the increase in violence, and whether armed conflict influences the evolution of individuals' religious identities.


\hbox {} \newpage
\section*{Figures and Tables}


 \begin{table}[H]
 \renewcommand{\arraystretch}{0.8}
\setlength{\tabcolsep}{2pt}
\hspace{-1cm}\begin{center}
\caption {Descriptive Statistics}  \label{precharacteristics_tab}
\small
 \begin{tabular}{lcccccc}
\hline\hline \addlinespace[0.15cm]
& &&& \multicolumn{3}{c}{Municipalities where}\\
& &&& \multicolumn{3}{c}{first non-Catholic}\\
& &&& \multicolumn{3}{c}{church was established}\\
& \multicolumn{3}{c}{All municipalities} & \multicolumn{3}{c}{ in 1996-2017}\\\cmidrule[0.2pt](l){2-4}\cmidrule[0.2pt](l){5-7}
    & \multicolumn{1}{c}{Obs.} & \multicolumn{1}{c}{Mean}  & \multicolumn{1}{c}{St. dev.} & \multicolumn{1}{c}{Obs.} & \multicolumn{1}{c}{Mean}  & \multicolumn{1}{c}{St. dev.} \\\cmidrule[0.2pt](l){2-2}\cmidrule[0.2pt](l){3-3} \cmidrule[0.2pt](l){4-4} \cmidrule[0.2pt](l){5-5} \cmidrule[0.2pt](l){6-6}  \cmidrule[0.2pt](l){7-7}
            &         (1)&         (2)&         (3)&         (4)&         (5)&         (6)\\
\midrule
\emph{\textbf{Armed conflict}}&            &            &            &            &            &            \\
Attack by non-state group (prob.)&       15417&       0.368&       0.482&        5600&       0.448&       0.497\\
Attack by guerrillas (prob.)&       15417&       0.327&       0.469&        5600&       0.386&       0.487\\
Attack by paramilitaries (prob.)&       15417&       0.131&       0.338&        5600&       0.201&       0.401\\
Assassination by non-state group (prob.)&       24382&       0.403&       0.491&        8839&       0.508&       0.500\\
Assassination by guerrillas (prob.)&       24382&       0.185&       0.388&        8839&       0.225&       0.418\\
Assassination by paramilitaries (prob.)&       24382&       0.238&       0.426&        8839&       0.313&       0.464\\
Assassination by unknown group (prob.)&       24382&       0.283&       0.450&        8839&       0.401&       0.490\\
Assassination by non-state group (rate)&       24284&      18.995&      48.518&        8819&      17.486&      37.042\\
Assassination by guerrillas (rate)&       24284&       3.487&      12.686&        8819&       2.174&       7.205\\
Assassination by paramilitaries (rate)&       24284&       9.291&      30.990&        8819&       9.299&      24.709\\
Assassination by unknown group (rate)&       24284&       5.256&      14.338&        8819&       5.323&      12.068\\
Desertion from guerrillas (prob.)&       13999&       0.125&       0.330&        5126&       0.170&       0.375\\
Forced recruitment (prob.)&       24382&       0.102&       0.302&        8839&       0.131&       0.337\\
Internally displaced outflows (rate)&       24037&    1441.335&    4355.259&        8757&    1152.142&    2729.684\\
Internally displaced inflows (rate)&       24037&     767.639&    2410.893&        8757&     839.672&    1701.072\\
Coca crops (prob.)&       20917&       0.161&       0.368&        7592&       0.140&       0.347\\
\emph{\textbf{Crime}}&            &            &            &            &            &            \\
Homicides (rate)&       23794&      42.483&      61.156&        8702&      44.428&      48.697\\
Robberies (rate)&       16606&      87.141&     129.647&        6020&     139.809&     167.418\\
\emph{\textbf{Sociodemographics and economy}}&            &            &            &            &            &            \\
Catholic churches (in 1995)&       24382&       2.007&      10.479&        8839&       3.719&      17.250\\
Total population&       24284&   39494.805&  240480.158&        8819&   89015.736&  394049.686\\
Proportion of rural population&       24284&       4.258&      12.785&        8819&       1.419&       2.196\\
Proportion of ethnic minority population&       24284&       0.089&       0.200&        8819&       0.104&       0.211\\
Unsatisfied basic needs&       24283&      48.671&      21.012&        8821&      42.873&      21.311\\
\emph{\textbf{Election outcomes}}&            &            &            &            &            &            \\
Vote for Liberals (share, last election)&       23705&       0.224&       0.273&        8703&       0.260&       0.275\\
Vote for Conservatives (share, last election)&       23705&       0.177&       0.252&        8703&       0.121&       0.204\\
Vote for the Left (share, last election)&       23705&       0.019&       0.084&        8703&       0.021&       0.079\\

 \addlinespace[0.15cm]
\hline\hline
\multicolumn{7}{p{16.5cm}}{\footnotesizes{\textbf{Notes:} The sample in all columns is restricted to data from 1996 onwards.  The sample in columns (1)-(3) includes all municipalities. The sample in columns (4)-(6) is limited to municipalities where a non-Catholic church obtained legal status for the first time between 1996 and 2017. Data on attacks by guerrillas and paramilitaries comes from the Conflict Analysis Resource Center (CERAC). Data on conflict-related assassination and forced recruitment is from Colombia's National Centre for Historical Memory (NCHM). Data on desertion from a guerrilla group, internally displaced people, coca crops and crime (homicides and robberies) is from the Center for Studies of Economic Development (CEDE). Data on population, ethnic minority population, people with unsatisfied basic needs (used as a proxy for poverty) are from the National Administrative Department of Statistics (DANE). Electoral data is from the from the Colombian Electoral Agency.} }
\end{tabular}
\end{center}
\end{table}


\newpage


\begin{table}[H]
\begin{center}
{
\renewcommand{\arraystretch}{0.8}
\setlength{\tabcolsep}{7pt}
\caption {Impact of first non-Catholic church on the prob. of a conflict-related attack}  \label{PGUEPANYPGUEPPAR_tab}
\small
\centering  \begin{tabular}{lcccccc}
\hline\hline \addlinespace[0.15cm]
    & \multicolumn{6}{c}{Dep. variable = 1 if an attack by} \\\addlinespace[0.15cm]
    & \multicolumn{2}{c}{Any group} & \multicolumn{2}{c}{Guerrilla} & \multicolumn{2}{c}{Paramilitaries} \\\cmidrule[0.2pt](l){2-3}\cmidrule[0.2pt](l){4-5}\cmidrule[0.2pt](l){6-7}
& (1)& (2) & (3) & (4) & (5) & (6)  \\   \addlinespace[0.15cm] \hline \addlinespace[0.15cm]
            \multicolumn{1}{l}{\emph{\underline{Panel A}:}}            & \multicolumn{6}{c}{\emph{Dynamic} specification}\\\addlinespace[0.15cm]
Effect at t=-3      &       0.041         &       0.042         &       0.004         &      -0.002         &       0.049         &       0.050         \\
                    &     (0.044)         &     (0.045)         &     (0.040)         &     (0.040)         &     (0.035)         &     (0.036)         \\
Effect at t=-2      &      -0.011         &      -0.004         &      -0.018         &      -0.016         &       0.021         &       0.021         \\
                    &     (0.043)         &     (0.044)         &     (0.038)         &     (0.039)         &     (0.035)         &     (0.036)         \\
Effect at t=-1      &       0.000         &       0.000         &       0.000         &       0.000         &       0.000         &       0.000         \\
                    &         (.)         &         (.)         &         (.)         &         (.)         &         (.)         &         (.)         \\
Effect at t=0       &       0.108\sym{**} &       0.108\sym{**} &       0.083\sym{*}  &       0.090\sym{**} &       0.040         &       0.029         \\
                    &     (0.047)         &     (0.049)         &     (0.044)         &     (0.046)         &     (0.037)         &     (0.037)         \\
Effect at t=+1      &       0.071         &       0.068         &       0.037         &       0.033         &       0.069         &       0.068         \\
                    &     (0.049)         &     (0.051)         &     (0.046)         &     (0.047)         &     (0.044)         &     (0.045)         \\
Effect at t=+2      &       0.089         &       0.071         &       0.073         &       0.055         &       0.075         &       0.076         \\
                    &     (0.062)         &     (0.064)         &     (0.057)         &     (0.060)         &     (0.052)         &     (0.053)         \\
Effect at t=+3      &       0.064         &       0.044         &       0.036         &       0.017         &       0.087         &       0.090         \\
                    &     (0.072)         &     (0.074)         &     (0.064)         &     (0.068)         &     (0.064)         &     (0.066)         \\
Effect at t=+4      &       0.095         &       0.090         &       0.086         &       0.081         &       0.119         &       0.130         \\
                    &     (0.084)         &     (0.086)         &     (0.073)         &     (0.076)         &     (0.083)         &     (0.085)         \\
Effect at t=+5      &       0.125         &       0.125         &       0.128\sym{*}  &       0.130\sym{*}  &       0.073         &       0.078         \\
                    &     (0.090)         &     (0.091)         &     (0.075)         &     (0.078)         &     (0.090)         &     (0.091)         \\
R-sq                &       0.574         &       0.579         &       0.596         &       0.603         &       0.458         &       0.463         \\
Observations        &        5530         &        5357         &        5530         &        5357         &        5530         &        5357         \\

\addlinespace[0.15cm]\hline\addlinespace[0.15cm]
            \multicolumn{1}{l}{\emph{\underline{Panel B}:}}            & \multicolumn{6}{c}{\emph{Static} specification (long-term effect)}\\\addlinespace[0.15cm]
Effect at t$\geq 0$ &       0.080\sym{*}  &       0.080\sym{*}  &       0.078\sym{**} &       0.082\sym{**} &       0.040         &       0.036         \\
                    &     (0.042)         &     (0.043)         &     (0.037)         &     (0.038)         &     (0.032)         &     (0.032)         \\
R-sq                &       0.572         &       0.576         &       0.594         &       0.600         &       0.454         &       0.460         \\
Observations        &        5530         &        5357         &        5530         &        5357         &        5530         &        5357         \\

 \addlinespace[0.15cm]\hline \addlinespace[0.15cm]
Baseline controls & No & Yes& No& Yes& No& Yes \\
\addlinespace[0.15cm]\hline\hline
\multicolumn{7}{p{13cm}}{\footnotesizes{\textbf{Notes:} All columns in Panel A report the estimates from Eq. (\ref{baseline}) and all columns in Panel B report the estimates from Eq. (\ref{didbaseline}). All models include municipality and year fixed effects, department $\times$ year fixed effects and municipality-specific trends.  The models in Panel A include 13 lags and 13 leads, normalized to the period prior to treatment. The models with baseline controls (columns (2), (4) and (6)) include the following (lagged) covariates:  log of total population, proportion of the population living in a rural area, proportion of the population with unsatisfied basic needs, proportion of ethnic minority population and the homicide rate. Samples for regression models use data from 1996 to 2009. Robust standard errors (in parentheses) are clustered by municipality. * denotes statistically significant estimates at 10\%, ** denotes significant at 5\% and *** denotes significant at 1\%.} }
\end{tabular}
}
\end{center}
\end{table}


\begin{table}[H]
\renewcommand{\arraystretch}{0.6}
\setlength{\tabcolsep}{7.5pt}
\caption {Impact of first non-Catholic church on conflict-related assassination}  \label{NCHMyearly_tab}
\vspace{-0.35cm}
\smalls
\centering  \begin{tabular}{lcccccccc}
\hline\hline \addlinespace[0.1cm]
    & \multicolumn{2}{c}{Any group} & \multicolumn{2}{c}{Guerrilla} & \multicolumn{2}{c}{Paramilitaries} & \multicolumn{2}{c}{Unknown}  \\\cmidrule[0.2pt](l){2-3}\cmidrule[0.2pt](l){4-5}\cmidrule[0.2pt](l){6-7}\cmidrule[0.2pt](l){8-9}
& (1)& (2) & (3) & (4) & (5) & (6)  & (7) & (8)  \\   \addlinespace[0.10cm] \hline\hline \addlinespace[0.05cm]
            \multicolumn{1}{l}{\emph{\underline{Panel A}:}}            & \multicolumn{8}{c}{Assassination probability (\emph{dynamic} specification)}\\\addlinespace[0.1cm]
Effect at t=-3      &      -0.025         &      -0.019         &      -0.022         &      -0.018         &       0.005         &       0.001         &      -0.041         &      -0.037         \\
                    &     (0.027)         &     (0.028)         &     (0.023)         &     (0.024)         &     (0.017)         &     (0.017)         &     (0.028)         &     (0.028)         \\
Effect at t=-2      &      -0.018         &      -0.018         &      -0.017         &      -0.016         &       0.022         &       0.020         &      -0.033         &      -0.033         \\
                    &     (0.026)         &     (0.026)         &     (0.021)         &     (0.021)         &     (0.014)         &     (0.014)         &     (0.025)         &     (0.025)         \\
Effect at t=-1      &       0.000         &       0.000         &       0.000         &       0.000         &       0.000         &       0.000         &       0.000         &       0.000         \\
                    &         (.)         &         (.)         &         (.)         &         (.)         &         (.)         &         (.)         &         (.)         &         (.)         \\
Effect at t=0       &       0.066\sym{***}&       0.071\sym{***}&       0.020         &       0.023         &       0.014         &       0.011         &       0.036         &       0.043\sym{*}  \\
                    &     (0.023)         &     (0.023)         &     (0.019)         &     (0.019)         &     (0.012)         &     (0.012)         &     (0.025)         &     (0.025)         \\
Effect at t=+1      &       0.059\sym{**} &       0.064\sym{***}&       0.039\sym{*}  &       0.040\sym{*}  &       0.016         &       0.015         &       0.010         &       0.015         \\
                    &     (0.023)         &     (0.023)         &     (0.021)         &     (0.022)         &     (0.014)         &     (0.014)         &     (0.024)         &     (0.024)         \\
Effect at t=+2      &       0.067\sym{***}&       0.064\sym{***}&       0.044\sym{**} &       0.043\sym{**} &       0.013         &       0.007         &       0.028         &       0.028         \\
                    &     (0.024)         &     (0.024)         &     (0.021)         &     (0.021)         &     (0.015)         &     (0.015)         &     (0.024)         &     (0.024)         \\
Effect at t=+3      &       0.062\sym{***}&       0.061\sym{**} &       0.032         &       0.029         &      -0.013         &      -0.021         &       0.053\sym{**} &       0.058\sym{**} \\
                    &     (0.024)         &     (0.024)         &     (0.021)         &     (0.021)         &     (0.016)         &     (0.016)         &     (0.024)         &     (0.024)         \\
Effect at t=+4      &       0.050\sym{**} &       0.044\sym{*}  &       0.042\sym{*}  &       0.035         &       0.009         &       0.001         &       0.018         &       0.024         \\
                    &     (0.024)         &     (0.024)         &     (0.023)         &     (0.024)         &     (0.016)         &     (0.016)         &     (0.025)         &     (0.025)         \\
Effect at t=+5      &       0.062\sym{**} &       0.065\sym{**} &       0.051\sym{**} &       0.050\sym{**} &      -0.004         &      -0.006         &       0.031         &       0.037         \\
                    &     (0.026)         &     (0.025)         &     (0.024)         &     (0.025)         &     (0.018)         &     (0.018)         &     (0.026)         &     (0.026)         \\
R-sq                &       0.720         &       0.724         &       0.565         &       0.571         &       0.771         &       0.776         &       0.650         &       0.654         \\
Observations        &        9071         &        8878         &        9071         &        8878         &        9071         &        8878         &        9071         &        8878         \\

\addlinespace[0.05cm]\hline\addlinespace[0.05cm]
 \multicolumn{1}{l}{\emph{\underline{Panel B}:}}            & \multicolumn{8}{c}{Assassination probability (\emph{static} specification)}\\\addlinespace[0.1cm]
Effect at t$\geq 0$ &       0.085\sym{***}&       0.082\sym{***}&       0.036\sym{**} &       0.031\sym{**} &       0.005         &       0.002         &       0.063\sym{***}&       0.065\sym{***}\\
                    &     (0.020)         &     (0.020)         &     (0.015)         &     (0.015)         &     (0.011)         &     (0.011)         &     (0.019)         &     (0.019)         \\
R-sq                &       0.709         &       0.713         &       0.560         &       0.566         &       0.767         &       0.772         &       0.640         &       0.644         \\
Observations        &        8729         &        8538         &        8729         &        8538         &        8729         &        8538         &        8729         &        8538         \\

 \addlinespace[0.05cm]
 \hline \addlinespace[0.05cm]
             \multicolumn{1}{l}{\emph{\underline{Panel C}:}}            & \multicolumn{8}{c}{Assassination rate (\emph{dynamic} specification)}\\\addlinespace[0.05cm]
Effect at t=-3      &       0.871         &       1.280         &      -0.505         &      -0.466         &       0.757         &       0.919         &       0.133         &       0.249         \\
                    &     (1.639)         &     (1.672)         &     (0.315)         &     (0.326)         &     (0.910)         &     (0.924)         &     (0.745)         &     (0.756)         \\
Effect at t=-2      &       2.459         &       2.293         &      -0.213         &      -0.220         &       1.904\sym{*}  &       1.818\sym{*}  &       0.657         &       0.578         \\
                    &     (1.677)         &     (1.634)         &     (0.252)         &     (0.259)         &     (1.101)         &     (1.080)         &     (0.652)         &     (0.637)         \\
Effect at t=-1      &       0.000         &       0.000         &       0.000         &       0.000         &       0.000         &       0.000         &       0.000         &       0.000         \\
                    &         (.)         &         (.)         &         (.)         &         (.)         &         (.)         &         (.)         &         (.)         &         (.)         \\
Effect at t=0       &       0.370         &       0.366         &       0.010         &       0.017         &       0.692         &       0.570         &      -0.212         &      -0.140         \\
                    &     (0.842)         &     (0.849)         &     (0.238)         &     (0.243)         &     (0.550)         &     (0.532)         &     (0.453)         &     (0.461)         \\
Effect at t=+1      &       0.184         &       0.047         &       0.113         &       0.134         &       0.449         &       0.240         &      -0.431         &      -0.396         \\
                    &     (0.976)         &     (0.955)         &     (0.269)         &     (0.271)         &     (0.637)         &     (0.610)         &     (0.507)         &     (0.515)         \\
Effect at t=+2      &       0.220         &      -0.146         &       0.345         &       0.341         &       0.015         &      -0.235         &      -0.206         &      -0.290         \\
                    &     (1.115)         &     (1.048)         &     (0.240)         &     (0.238)         &     (0.775)         &     (0.741)         &     (0.489)         &     (0.477)         \\
Effect at t=+3      &      -0.294         &      -0.421         &       0.053         &       0.055         &       0.304         &       0.143         &      -0.677         &      -0.660         \\
                    &     (1.472)         &     (1.377)         &     (0.244)         &     (0.247)         &     (1.142)         &     (1.106)         &     (0.532)         &     (0.508)         \\
Effect at t=+4      &      -1.751         &      -1.930         &      -0.044         &      -0.073         &      -0.655         &      -0.833         &      -1.097\sym{*}  &      -1.065\sym{*}  \\
                    &     (1.684)         &     (1.585)         &     (0.259)         &     (0.266)         &     (1.258)         &     (1.215)         &     (0.600)         &     (0.581)         \\
Effect at t=+5      &      -2.577         &      -2.549         &      -0.125         &      -0.068         &      -1.325         &      -1.509         &      -1.128\sym{*}  &      -0.994\sym{*}  \\
                    &     (1.826)         &     (1.712)         &     (0.292)         &     (0.290)         &     (1.351)         &     (1.302)         &     (0.591)         &     (0.565)         \\
R-sq                &       0.730         &       0.744         &       0.561         &       0.564         &       0.691         &       0.702         &       0.623         &       0.639         \\
Observations        &        9051         &        8878         &        9051         &        8878         &        9051         &        8878         &        9051         &        8878         \\

\addlinespace[0.15cm]\hline\addlinespace[0.05cm]
 \multicolumn{1}{l}{\emph{\underline{Panel D}:}}            & \multicolumn{8}{c}{Assassination rate (\emph{static} specification)}\\\addlinespace[0.1cm]
Effect at t$\geq 0$ &      -0.499         &      -0.693         &       0.264         &       0.231         &       0.192         &       0.036         &      -0.470         &      -0.459         \\
                    &     (1.652)         &     (1.510)         &     (0.245)         &     (0.246)         &     (1.049)         &     (0.984)         &     (0.646)         &     (0.617)         \\
R-sq                &       0.729         &       0.744         &       0.561         &       0.564         &       0.690         &       0.702         &       0.622         &       0.638         \\
Observations        &        8709         &        8538         &        8709         &        8538         &        8709         &        8538         &        8709         &        8538         \\

 \addlinespace[0.05cm]\hline \addlinespace[0.05cm]
 Baseline controls & No & Yes& No& Yes& No& Yes & No& Yes \\
\addlinespace[0.1cm]\hline\hline
\multicolumn{9}{p{16.5cm}}{\footnotesizes{\textbf{Notes:} Panels A and C report the estimates from Eq. (\ref{baseline}) and Panels B and D report the estimates from Eq. (\ref{didbaseline}).  All models include municipality and year fixed effects, department $\times$ year fixed effects and municipality-specific trends.  The models in Panels A and C include 17 lags and 17 leads. The models with baseline controls  include the following (lagged) covariates:  log of total pop., proportion of the pop. living in a rural area, proportion of the pop. with unsatisfied basic needs, proportion of ethnic minority pop. and the homicide rate. Samples for regression models use data from 1996 to 2017. Robust standard errors (in parentheses) are clustered by municipality. * denotes statistically significant estimates at 10\%, **  significant at 5\% and ***  significant at 1\%.} }
\end{tabular}
\end{table}


\begin{table}[H]
\begin{center}
{
\renewcommand{\arraystretch}{0.7}
\setlength{\tabcolsep}{11pt}
\caption {Effect of first non-Catholic church on the prob. of a conflict-related attack: heterogeneous effects  by presence of forced recruitment and coca crops (\emph{static} specification)}  \label{PANYPGUEPPARheterostatic_tab}
\small
\vspace{-0.3cm}\centering  \begin{tabular}{lccc}
\hline\hline \addlinespace[0.15cm]
    & \multicolumn{3}{c}{Dep. variable = 1 if an attack by} \\ \addlinespace[0.15cm]
     & \multicolumn{1}{c}{any group}& \multicolumn{1}{c}{guerrilla} & \multicolumn{1}{c}{paramilitaries}  \\\cmidrule[0.2pt](l){2-2} \cmidrule[0.2pt](l){3-3} \cmidrule[0.2pt](l){4-4}
    & (1)& (2) & (3)\\   \addlinespace[0.15cm]\hline\addlinespace[0.15cm]
    \multicolumn{1}{l}{\emph{\underline{Panel A}:}}  \\\addlinespace[0.15cm]
Non-Catholic church &       0.062\sym{*}  &       0.071\sym{**} &       0.027         \\
                    &     (0.036)         &     (0.031)         &     (0.024)         \\
Non-Catholic church &      -0.011         &       0.134\sym{**} &      -0.126         \\
\hspace{0.2cm} $\times$ Forced recruitment bf 1996&     (0.082)         &     (0.068)         &     (0.104)         \\
R-sq                &       0.535         &       0.563         &       0.422         \\
Observations        &        5250         &        5250         &        5250         \\

    \addlinespace[0.15cm]\hline\addlinespace[0.15cm]
           \multicolumn{1}{l}{\emph{\underline{Panel B}:}}    &\\\addlinespace[0.15cm]
Non-Catholic church &       0.019         &       0.027         &       0.014         \\
                    &     (0.049)         &     (0.042)         &     (0.031)         \\
Non-Catholic church &       0.114         &       0.117         &       0.014         \\
\hspace{0.2cm} $\times$ Coca crops in 2000&     (0.100)         &     (0.077)         &     (0.090)         \\
R-sq                &       0.566         &       0.600         &       0.464         \\
Observations        &        3413         &        3413         &        3413         \\

\addlinespace[0.15cm]\hline\hline
\multicolumn{4}{p{13.2cm}}{\footnotesizes{\textbf{Notes:} All columns in report the estimates from Eq. (\ref{didbaseline}).  All models include municipality and year fixed effects, department $\times$ year fixed effects, and  the following (lagged) covariates:  log of total population, proportion of the population living in a rural area, proportion of the population with unsatisfied basic needs, proportion of ethnic minority population and the homicide rate. Samples for regression models in Panel A use data from 1996 to 2009. Samples for regression models in Panel B use data from 2001 to 2009. Robust standard errors (in parentheses) are clustered by municipality.  * denotes statistically significant estimates at 10\%, ** denotes significant at 5\% and *** denotes significant at 1\%.} }
\end{tabular}
}
\end{center}
\end{table}


\begin{table}[H]
\begin{center}
{
\renewcommand{\arraystretch}{0.7}
\setlength{\tabcolsep}{13pt}
\caption {Impact of first nC church on the prob. of a forced recruitment and a desertion}  \label{forcedrecruitdesertion_tab}
\small
\centering  \begin{tabular}{lcccc}
\hline\hline \addlinespace[0.15cm]
    \multicolumn{5}{c}{Dep. variable = 1 if there is at least one case of} \\\addlinespace[0.15cm]
    & \multicolumn{2}{c}{forced recruitment} & \multicolumn{2}{c}{desertion}  \\\cmidrule[0.2pt](l){2-3}\cmidrule[0.2pt](l){4-5}
& (1)& (2) & (3) & (4)   \\   \addlinespace[0.15cm] \hline \addlinespace[0.15cm]
            \multicolumn{1}{l}{\emph{\underline{Panel A}:}}            & \multicolumn{4}{c}{\emph{Dynamic} specification}\\\addlinespace[0.15cm]
Effect at t=-3      &      -0.016         &      -0.016         &      -0.026         &      -0.022         \\
                    &     (0.020)         &     (0.020)         &     (0.065)         &     (0.066)         \\
Effect at t=-2      &       0.021         &       0.020         &       0.053         &       0.057         \\
                    &     (0.019)         &     (0.019)         &     (0.066)         &     (0.068)         \\
Effect at t=-1      &       0.000         &       0.000         &       0.000         &       0.000         \\
                    &         (.)         &         (.)         &         (.)         &         (.)         \\
Effect at t=0       &      -0.001         &      -0.001         &       0.079         &       0.086         \\
                    &     (0.016)         &     (0.016)         &     (0.069)         &     (0.071)         \\
Effect at t=+1      &       0.029         &       0.029         &       0.171\sym{**} &       0.176\sym{**} \\
                    &     (0.018)         &     (0.019)         &     (0.080)         &     (0.080)         \\
Effect at t=+2      &       0.017         &       0.017         &       0.101         &       0.095         \\
                    &     (0.018)         &     (0.019)         &     (0.096)         &     (0.099)         \\
Effect at t=+3      &       0.043\sym{**} &       0.041\sym{**} &       0.162         &       0.156         \\
                    &     (0.019)         &     (0.020)         &     (0.130)         &     (0.135)         \\
Effect at t=+4      &       0.037\sym{*}  &       0.036         &       0.058         &       0.053         \\
                    &     (0.022)         &     (0.023)         &     (0.112)         &     (0.117)         \\
Effect at t=+5      &       0.063\sym{**} &       0.062\sym{**} &       0.088         &       0.082         \\
                    &     (0.026)         &     (0.026)         &     (0.098)         &     (0.102)         \\
R-sq                &       0.477         &       0.480         &       0.618         &       0.620         \\
Observations        &        8729         &        8538         &        3135         &        3083         \\

 \addlinespace[0.15cm]\hline \addlinespace[0.15cm]
            \multicolumn{1}{l}{\emph{\underline{Panel B}:}}            & \multicolumn{4}{c}{\emph{Static} specification}\\\addlinespace[0.15cm]
Effect at t$\geq 0$ &       0.029\sym{*}  &       0.028\sym{*}  &       0.119\sym{*}  &       0.121\sym{**} \\
                    &     (0.015)         &     (0.016)         &     (0.061)         &     (0.061)         \\
R-sq                &       0.472         &       0.476         &       0.614         &       0.615         \\
Observations        &        8729         &        8538         &        3135         &        3083         \\

\addlinespace[0.15cm]\hline\addlinespace[0.15cm]\
Controls & No & Yes& No& Yes  \\
\addlinespace[0.15cm]\hline\hline
\multicolumn{5}{p{11.5cm}}{\footnotesizes{\textbf{Notes:} All columns in Panel A report the estimates from Eq. (\ref{baseline}) and all columns in Panel B report the estimates from Eq. (\ref{didbaseline}).  All models include municipality and year fixed effects, department $\times$ year fixed effects and municipality-specific trends.  The models in columns (1) and (2) use data from 1996 to 2017, and those in Panel A include 13 lags and 13 leads normalized to the period prior to treatment. The models in columns (3) and (4) use data from 2001 to 2008, and those in Panel A include 8 lags and 8 leads normalized to the period prior to treatment. The models with controls (columns (2) and (4)) include the following (lagged) covariates:  log of total population, proportion of the population living in a rural area, proportion of the population with unsatisfied basic needs, proportion of ethnic minority population, the homicide rate and the share of vote for liberals, conservatives and the left. Robust standard errors (in parentheses) are clustered by municipality. * denotes statistically significant estimates at 10\%, ** denotes significant at 5\% and *** denotes significant at 1\%.} }
\end{tabular}
}
\end{center}
\end{table}


\begin{table}[H]
\begin{center}
{
\renewcommand{\arraystretch}{0.7}
\setlength{\tabcolsep}{11pt}
\caption {Effect of first nC church on the prob. of a guerrilla attack: heterogeneous effects by number of Catholic churches in 1995 and Conservative vote share before 1996 (\emph{static} specification)}  \label{PGUEhowconscatho_tab}
\small
\vspace{-0.3cm}\centering  \begin{tabular}{lccc}
\hline\hline \addlinespace[0.15cm]
    & \multicolumn{3}{c}{Dep. variable = 1 if an attack by} \\ \addlinespace[0.15cm]
     & \multicolumn{1}{c}{any group}& \multicolumn{1}{c}{guerrilla} & \multicolumn{1}{c}{paramilitaries}  \\\cmidrule[0.2pt](l){2-2} \cmidrule[0.2pt](l){3-3} \cmidrule[0.2pt](l){4-4}
    & (1)& (2) & (3)\\   \addlinespace[0.15cm]\hline\addlinespace[0.15cm]
    \multicolumn{1}{l}{\emph{\underline{Panel A}:}}  \\\addlinespace[0.15cm]
    Non-Catholic church &       0.113\sym{**} &       0.123\sym{***}&       0.026         \\
                    &     (0.049)         &     (0.041)         &     (0.036)         \\
Non-Catholic church &      -0.109         &      -0.121\sym{**} &       0.009         \\
\hspace{0.2cm} $\times$ High number of Catholic churches&     (0.067)         &     (0.057)         &     (0.047)         \\
R-sq                &       0.557         &       0.581         &       0.447         \\
Observations        &        5281         &        5281         &        5281         \\

    \addlinespace[0.15cm]\hline\addlinespace[0.15cm]
           \multicolumn{1}{l}{\emph{\underline{Panel B}:}}    &\\\addlinespace[0.15cm]
      Non-Catholic church &       0.091\sym{**} &       0.098\sym{***}&       0.031         \\
                    &     (0.040)         &     (0.035)         &     (0.032)         \\
Non-Catholic church &      -0.055         &      -0.040         &      -0.010         \\
\hspace{0.2cm} $\times$ Large conservative vote share&     (0.070)         &     (0.062)         &     (0.045)         \\
R-sq                &       0.558         &       0.583         &       0.439         \\
Observations        &        5321         &        5321         &        5321         \\

\addlinespace[0.15cm]\hline\hline
\multicolumn{4}{p{14cm}}{\footnotesizes{\textbf{Notes:} All columns in report the estimates from Eq. (\ref{didbaseline}).  All models include municipality and year fixed effects, department $\times$ year fixed effects, and  the following (lagged) covariates:  log of total population, proportion of the population living in a rural area, proportion of the population with unsatisfied basic needs, proportion of ethnic minority population and the homicide rate. Samples for regression models in Panel A use data from 1996 to 2009. Samples for regression models in Panel B use data from 2001 to 2009. Robust standard errors (in parentheses) are clustered by municipality.  * denotes statistically significant estimates at 10\%, ** denotes significant at 5\% and *** denotes significant at 1\%.} }
\end{tabular}
}
\end{center}
\end{table}


\begin{table}[H]
\begin{center}
{
\renewcommand{\arraystretch}{0.7}
\setlength{\tabcolsep}{16pt}
\caption {Effect of first non-Catholic church on homicide and robbery rates}  \label{homicidespcyear}
\small
\centering  \begin{tabular}{lcccc}
\hline\hline \addlinespace[0.15cm]
    & \multicolumn{4}{c}{Dep. variable:} \\\cmidrule[0.2pt](l){2-5}
        & \multicolumn{2}{c}{Homicide rate} & \multicolumn{2}{c}{Robbery rate}  \\
    \cmidrule[0.2pt](l){2-3}\cmidrule[0.2pt](l){4-5}
& (1)& (2) & (3) & (4)  \\\addlinespace[0.15cm] \hline \addlinespace[0.15cm]
           \multicolumn{1}{l}{\emph{\underline{Panel A}:}}    & \multicolumn{4}{c}{\emph{Dynamic} specification}\\\addlinespace[0.15cm]
Effect at t=-3      &      -1.963         &      -0.499         &      -6.887         &      -6.554         \\
                    &     (4.661)         &     (4.383)         &     (9.011)         &     (9.263)         \\
Effect at t=-2      &      -0.084         &       0.393         &      -9.140         &      -9.099         \\
                    &     (3.832)         &     (3.671)         &     (8.014)         &     (8.025)         \\
Effect at t=-1      &       0.000         &       0.000         &       0.000         &       0.000         \\
                    &         (.)         &         (.)         &         (.)         &         (.)         \\
Effect at t=0       &       1.426         &       1.575         &     -10.327         &      -9.497         \\
                    &     (3.744)         &     (3.659)         &    (14.892)         &    (15.017)         \\
Effect at t=+1      &       1.071         &       1.143         &       0.767         &       2.331         \\
                    &     (4.357)         &     (3.895)         &    (20.917)         &    (21.429)         \\
Effect at t=+2      &      -0.816         &      -0.773         &      13.510         &      14.688         \\
                    &     (5.054)         &     (4.658)         &    (29.852)         &    (30.430)         \\
Effect at t=+3      &      -0.803         &      -0.480         &      18.506         &      20.697         \\
                    &     (5.414)         &     (4.932)         &    (43.415)         &    (44.259)         \\
Effect at t=+4      &      -2.666         &      -2.314         &      11.254         &      15.476         \\
                    &     (6.131)         &     (5.664)         &    (58.902)         &    (60.031)         \\
Effect at t=+5      &      -7.586         &      -6.711         &     132.375         &     137.615         \\
                    &     (7.189)         &     (6.669)         &    (88.075)         &    (89.374)         \\
R-sq                &       0.705         &       0.717         &       0.809         &       0.810         \\
Observations        &        5383         &        5348         &        2163         &        2147         \\

\addlinespace[0.15cm]
\hline\addlinespace[0.15cm]
               \multicolumn{1}{l}{\emph{\underline{Panel B}:}}    & \multicolumn{4}{c}{\emph{Static} specification}\\\addlinespace[0.15cm]
 Effect at t$\geq 0$ &       3.104         &       2.963         &     -11.419         &     -10.896         \\
                    &     (4.582)         &     (4.541)         &    (13.878)         &    (13.958)         \\
R-sq                &       0.704         &       0.705         &       0.806         &       0.807         \\
Observations        &        5383         &        5377         &        2163         &        2162         \\
              
    \addlinespace[0.15cm]\hline\addlinespace[0.15cm]
Baseline controls & No & Yes& No& Yes \\
\addlinespace[0.15cm]\hline\hline
\multicolumn{5}{p{12.7cm}}{\footnotesizes{\textbf{Notes:} All columns in Panel A report the estimates from Eq. (\ref{baseline}) and all columns in Panel B report the estimates from Eq. (\ref{didbaseline}). All models are normalized relative to the period prior to treatment, and include municipality and year fixed effects, department $\times$ year fixed effects and municipality-specific linear trends.  The models in Panel A, columns (1) and (2) include 13 lags and 13 leads, and contain data from 1996 to 2009. The models in Panel B, columns (3) and (4) include 6 lags and 6 leads, and data from 2003 to 2009. The models with controls (columns (2) and (4)) include the following (lagged) covariates:  log of total population, proportion of the population living in a rural area, proportion with unsatisfied basic needs (used as a proxy for poverty) and  proportion of ethnic minority population.  Robust standard errors (in parentheses) are clustered by municipality. * denotes statistically significant estimates at 10\%, ** denotes significant at 5\% and *** denotes significant at 1\%.} }
\end{tabular}
}
\end{center}
\end{table}


\begin{table}[H]
\begin{center}
{
\renewcommand{\arraystretch}{0.7}
\setlength{\tabcolsep}{10pt}
\caption {Rol of ethnic minority population}  \label{indigblacktotprop_tab}
\small
\centering  \begin{tabular}{lcccccc}
\hline\hline \addlinespace[0.15cm]
                \multicolumn{1}{l}{\emph{\underline{Panel A}:}}    & \multicolumn{6}{c}{Dep. variable = 1 if an conflict-related attack  (\emph{static} specification)}\\\cmidrule[0.2pt](l){2-7}   \addlinespace[0.15cm]
 & \multicolumn{2}{c}{any group}& \multicolumn{2}{c}{guerrilla} & \multicolumn{2}{c}{paramilitaries}  \\\cmidrule[0.2pt](l){2-3} \cmidrule[0.2pt](l){4-5} \cmidrule[0.2pt](l){6-7} 
Non-Catholic church &       0.047         &       0.048         &       0.057         &       0.058         &       0.029         &       0.030         \\
                    &     (0.041)         &     (0.043)         &     (0.038)         &     (0.040)         &     (0.024)         &     (0.025)         \\
Non-Catholic church &       0.047         &       0.053         &       0.034         &       0.036         &      -0.015         &      -0.022         \\
\hspace{0.2cm} $\times$ Ethnic minority pop.&     (0.063)         &     (0.066)         &     (0.057)         &     (0.060)         &     (0.051)         &     (0.051)         \\
R-sq                &       0.548         &       0.552         &       0.569         &       0.575         &       0.422         &       0.429         \\
Observations        &        5460         &        5288         &        5460         &        5288         &        5460         &        5288         \\

  \addlinespace[0.15cm]\hline\hline\addlinespace[0.15cm]     
  \multicolumn{1}{l}{\emph{\underline{Panel B}:}}    & \multicolumn{6}{c}{Dep. variable: share of population} \\\cmidrule[0.2pt](l){2-7}
        & \multicolumn{2}{c}{Afro-Colombian} & \\
        & \multicolumn{2}{c}{and Indigenous} & \multicolumn{2}{c}{Afro-Colombian} & \multicolumn{2}{c}{Indigenous}  \\
    \cmidrule[0.2pt](l){2-3}\cmidrule[0.2pt](l){4-5}\cmidrule[0.2pt](l){6-7}
& (1)& (2) & (3) & (4)  & (5) & (6)  \\\addlinespace[0.15cm] \hline \addlinespace[0.15cm]
 Non-Catholic church &      -0.000         &      -0.005         &       0.002         &      -0.000         &      -0.004         &      -0.006         \\
$\times$ dummy for year 2005&     (0.018)         &     (0.019)         &     (0.017)         &     (0.018)         &     (0.005)         &     (0.006)         \\
R-sq                &       0.499         &       0.552         &       0.389         &       0.425         &       0.366         &       0.572         \\
Observations        &         780         &         748         &         780         &         748         &         780         &         748         \\
  
    \addlinespace[0.15cm]\hline\hline\addlinespace[0.15cm]
Baseline controls & No & Yes& No& Yes & No& Yes \\
\addlinespace[0.15cm]\hline\hline
\multicolumn{7}{p{16.2cm}}{\footnotesizes{\textbf{Notes:}  All columns in Panel A report the estimates from Eq. (\ref{didbaseline}),  include municipality and year fixed effects, and department $\times$ year fixed effects, and use data from 1996 to 2009.   All models in Panel B contain data from years 1993 and 2005, and all the columns in this panel present estimates for $\lambda$ from $y_{i}=\alpha+\beta T+\gamma D_i+\lambda (D_{i}\times T)+\epsilon_{i}$, where $T$ is a dummy variable that is equal to one when  $t=2005$ and to zero when $t=1993$, and where $D_{i}$  is a treatment indicator that is equal to one if municipality $i$ is treated before 2005, and other otherwise. The models  with baseline controls include the following (lagged) covariates:  log of total population, proportion of the population living in a rural area, proportion with unsatisfied basic needs and homicide rate, and for those in Panel B, we add area, elevation, and distance to Bogota and to the capital of the department, and (lagged) attacks by guerrillas and paramilitaries.  Robust standard errors (in parentheses) are clustered by municipality. * denotes statistically significant estimates at 10\%, ** denotes significant at 5\% and *** denotes significant at 1\%.} }
\end{tabular}
}
\end{center}
\end{table}


\begin{table}[H]
\begin{center}
{
\renewcommand{\arraystretch}{0.8}
\setlength{\tabcolsep}{2pt}
\caption {Determinants of first non-Catholic church}  \label{dfirstiglesiasnc_tab}
\small
\centering  \begin{tabular}{lccccccc}
\hline\hline \addlinespace[0.15cm]
    & \multicolumn{7}{c}{Dep. variable = 1 when first} \\
    & \multicolumn{7}{c}{ non-Catholic church is established} \\\cmidrule[0.2pt](l){2-8}
& (1)& (2) & (3)  & (4) & (5)& (6)& (7) \\    \addlinespace[0.15cm] \hline \addlinespace[0.15cm]
Attack by guerrilla (prob.) at t-1&      -0.001         &                     &      -0.001         &                     &      -0.001         &      -0.004         &      -0.004         \\
                    &     (0.007)         &                     &     (0.008)         &                     &     (0.008)         &     (0.008)         &     (0.008)         \\
Attack by paramilitaries (prob.) at t-1&      -0.001         &                     &      -0.003         &                     &      -0.003         &      -0.006         &      -0.006         \\
                    &     (0.006)         &                     &     (0.007)         &                     &     (0.007)         &     (0.007)         &     (0.007)         \\
Internally displaced pop. (outflow, rate) at t-1&                     &                     &                     &       0.000         &       0.000         &      -0.000         &      -0.000         \\
                    &                     &                     &                     &     (0.000)         &     (0.000)         &     (0.000)         &     (0.000)         \\
Internally displaced pop. (inflow, rate) at t-1&                     &                     &                     &       0.000         &       0.000         &       0.000         &       0.000         \\
                    &                     &                     &                     &     (0.000)         &     (0.000)         &     (0.000)         &     (0.000)         \\
Desertion from the guerrilla (prob.) at t-1&                     &                     &                     &      -0.005         &      -0.005         &      -0.005         &      -0.006         \\
                    &                     &                     &                     &     (0.008)         &     (0.008)         &     (0.009)         &     (0.009)         \\
Forced recruitment (prob.) at t-1&                     &                     &                     &       0.003         &       0.003         &       0.004         &       0.003         \\
                    &                     &                     &                     &     (0.009)         &     (0.009)         &     (0.010)         &     (0.010)         \\
Log of total population at t-1&                     &      -0.211\sym{*}  &      -0.203\sym{*}  &      -0.204\sym{*}  &      -0.203\sym{*}  &      -0.173         &      -0.162         \\
                    &                     &     (0.110)         &     (0.114)         &     (0.118)         &     (0.118)         &     (0.119)         &     (0.121)         \\
Prop. of rural population at t-1&                     &       0.012         &       0.010         &       0.012         &       0.012         &       0.014         &       0.013         \\
                    &                     &     (0.024)         &     (0.025)         &     (0.026)         &     (0.026)         &     (0.026)         &     (0.027)         \\
Unsatisfied Basic Needs at t-1&                     &       0.000         &       0.000         &       0.000         &       0.000         &      -0.000         &      -0.000         \\
                    &                     &     (0.001)         &     (0.001)         &     (0.001)         &     (0.001)         &     (0.001)         &     (0.001)         \\
Ethinic minority population at t-1&                     &      -0.018         &      -0.022         &      -0.019         &      -0.020         &      -0.004         &      -0.007         \\
                    &                     &     (0.034)         &     (0.035)         &     (0.034)         &     (0.034)         &     (0.036)         &     (0.036)         \\
Homicide (rate) at t-1&                     &       0.000         &       0.000         &       0.000         &       0.000         &       0.000         &       0.000         \\
                    &                     &     (0.000)         &     (0.000)         &     (0.000)         &     (0.000)         &     (0.000)         &     (0.000)         \\
Vote share for the Liberal party at t-1&                     &                     &       0.023         &                     &                     &                     &       0.009         \\
                    &                     &                     &     (0.016)         &                     &                     &                     &     (0.016)         \\
Vote share for the Conservative party at t-1&                     &                     &       0.007         &                     &                     &                     &       0.013         \\
                    &                     &                     &     (0.020)         &                     &                     &                     &     (0.020)         \\
Vote share for leftwing parties at t-1&                     &                     &       0.061         &                     &                     &                     &       0.055         \\
                    &                     &                     &     (0.047)         &                     &                     &                     &     (0.049)         \\
Coffee int. x log coffee price at t-1&                     &                     &                     &                     &                     &       0.010         &       0.008         \\
                    &                     &                     &                     &                     &                     &     (0.011)         &     (0.010)         \\
R-sq                &       0.255         &       0.255         &       0.258         &       0.258         &       0.258         &       0.238         &       0.238         \\
Observations        &        5526         &        5357         &        5276         &        5313         &        5313         &        4755         &        4702         \\

\addlinespace[0.15cm]\hline\hline
\multicolumn{8}{p{17cm}}{\footnotesizes{\textbf{Notes:} All models include municipality, year and department $\times$ year fixed effects, as well as  municipality-specific linear trends.  Samples for regression models include data from 1996 to 2009.   Robust standard errors (in parentheses) are clustered by municipality. * denotes statistically significant estimates at 10\%, ** denotes significant at 5\% and *** denotes significant at 1\%.} }
\end{tabular}
}
\end{center}
\end{table}


  \begin{figure}[H]
\caption{Effect of first non-Catholic church on prob. of guerrilla attack: alternative estimators}\label{PGUEiwaands_fig}
\begin{subfigure}[b]{.5\textwidth}
\centering
\caption*{Panel A: AS estimator}
\includegraphics[width=1\linewidth]{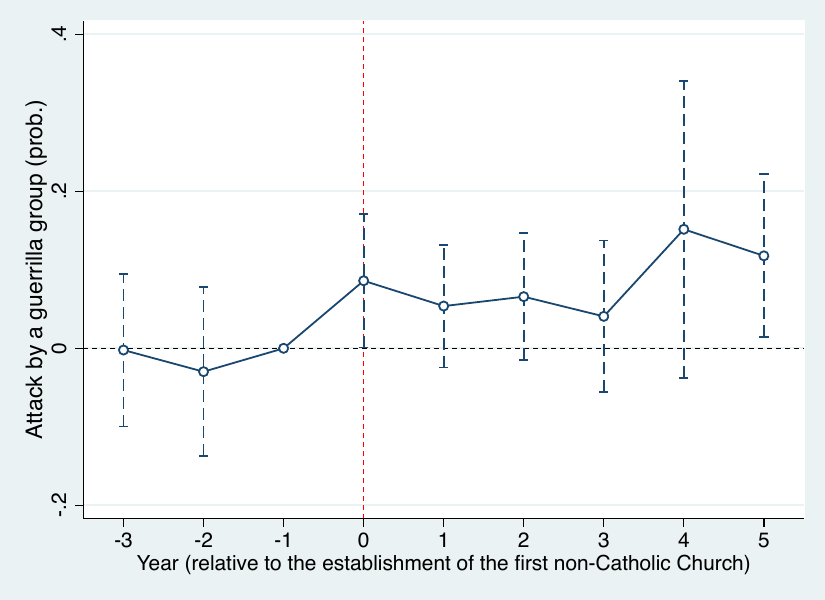}
\end{subfigure}
\begin{subfigure}[b]{.5\textwidth}
\centering
\caption*{Panel B: TW estimator}
\includegraphics[width=1\linewidth]{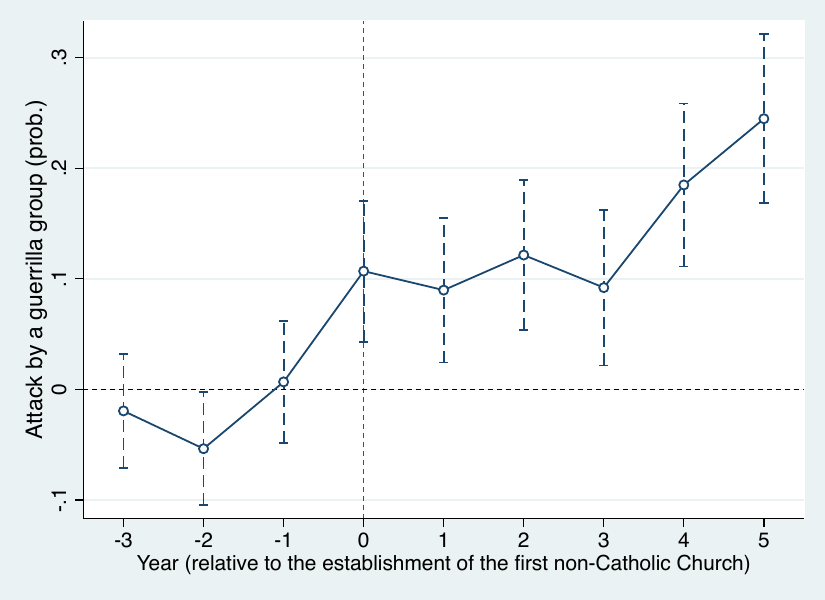}
\end{subfigure}
\begin{minipage}{15.5cm} \footnotesizes The figure in panel A shows \cite{AbrahamSun2018}'s interaction-weighted estimator (AS).  The figure in panel B shows \cite{Tchuente_Wind2019}'s estimator (TW).  Both AS and TW estimators are implemented following the procedures described in the Appendix (\ref{appidentificationTW}).
\end{minipage}
\end{figure}


\hbox {}
\hbox {} \newpage
\appendix
\section{Appendix}

\subsection{Model}\label{appmodel}
\small
In this section, we develop the theoretical model for which the results are described in Section \ref{Theory}.  As previously mentioned,  we use a Hotelling-like framework. First, we consider the scenario in which the religious market is served by a monopoly $A$ with strictness $a$. Note that an individual located to the right of $a$ always prefers (and chooses) the church. Also note that there is an indifferent individual $i$ such that any individual located to the left of $i$ prefers  (and chooses)  the armed group, $F$, to $A$, and any individual located to the right of $i$ prefers  (and chooses) $A$ to $F$. For an individual located at $j\in(0,a)$ who is indifferent between joining $F$ and $A$, we have that  $1-(a-i)=1-(i-0)$,  or, equivalently, $i=a/2$.

The church chooses its level of strictness, $a$, to maximize the sum of the contributions of its members, i.e., it solves
\begin{equation}
\label{ }
\max_{a} \int_{a/2}^{a}[1-(a-x)]dx+\int_{a}^{1}[1-(x-a)]dx
\end{equation}
Differentiating the last  expression with respect to $a$, we have the first-order condition
\small
\begin{equation}
\label{ }
-\int_{a/2}^{a}dx+[1-(a-a)]-[1-(a-a/2)](1/2)+\int_{a}^{1}dx+[1-(1-a)]0-[1-(a-a)]=0
\end{equation}
\normalsize
which, rearranging, and solving for $a$, is equivalent to $a^m=2/5$, where $m$ represents the monopoly scenario.\footnote{It is easy to see that $a^m$ corresponds to a maximum, since differentiating with respect to $a$ against the first-order condition, we have that $(1/2)(2-1/2)-1<0$.} Importantly, note that the proportion of people belonging to $F$ in this scenario, which we denote by $f^m$, is
\small
\begin{equation}
\label{am}
f^m=\frac{1}{5}
\end{equation}
\normalsize

Now we consider the scenario in which there are two churches, $A$ and $B$, with strictnesses of $a$ and $b$ respectively, and where, without loss of generality, $b\geq a$.  First, note that for an individual $j$ located at $j\in(a,b)$ who is indifferent between joining a church located at $a$ and a church located at $b$, we have that $1-(j-a)=1-(b-j)$ or, equivalently, $j=(a+b)/2$.

As for  an individual $j$ located at $j\in (0,a)$, who is indifferent between joining $F$ and $A$, we know that $i=a/2$. Importantly, note that if $ a/2\leq 1/5$, those individuals in $(a/2,1/5)$ decide to abandon $F$ to join $A$. We assume that $A$ pays a cost associated with its recruitment efforts. Specifically, we assume that $A$ loses a fraction $\beta$ of the contributions by the individuals who decide to abandon $F$ to join $A$. We examine the conditions under which $a/2\leq 1/5$. In this case, $A$ must solve
\small
\begin{equation}
\label{ }
\max_{a}V^A=(1-\beta)\int_{a/2}^{1/5}[1-(a-x)]dx+\int_{1/5}^{a}[1-(a-x)]dx+\int_{a}^{(a+b)/2}[1-(x-a)]dx
\end{equation}
\normalsize
the first-order condition of which is
\small
\begin{equation}
\label{ }
\begin{split}
-(1-\beta)\int_{a/2}^{1/5}dx-(1-\beta)[1-(a-a/2)](\frac{1}{2})\\
-\int_{1/5}^{a}dx+1+\int_{a}^{(a+b)/2}dx+[1-((a+b)/2-a)](\frac{1}{2})-1=0
\end{split}
\end{equation}
\normalsize

Rearranging, and solving for $a$, we have that $A$'s best response function is
\small
\begin{equation}
\label{brA}
a(b)=\frac{b}{(5-3(1-\beta))}+\frac{12-14(1-\beta)}{5(5-3(1-\beta))}
\end{equation}
\normalsize
As for church $B$, it solves the problem
\small
\begin{equation}
\label{ }
\max_{b}V^B=\int_{(a+b)/2}^{b}[1-(b-x)]dx+\int_{b}^{1}[1-(x-b)]dx
\end{equation}
\normalsize
where the first-order condition is
\small
\begin{equation}
\label{ }
-b/2+a/2-(1/2)+b/4-a/4+1-b=0
\end{equation}
\normalsize
which rearranging and solving for $b$ yields $B$'s best response function
\small
\begin{equation}
\label{brB}
b(a)=\frac{(a+2)}{5}
\end{equation}
\normalsize
Combining (\ref{brA}) and (\ref{brB}), and solving for $a$, we have  that $a^{c}=14\beta/(9+15\beta)$, where $c$ represents the Nash solution competition scenario. Importantly, note that in this case the proportion of people belonging to $F$, which we denote by $f^{c}$, is $a^{c}/2$, or, equivalently,
\small
\begin{equation}
\label{ane}
f^{c}=\frac{7\beta}{9+15\beta}
\end{equation}
\normalsize
Finally, we compare $f^m$ in (\ref{am}) and $f^{c}$ in (\ref{ane}). Note that if $f^{c}<f^m$, a marketplace for religion decreases the proportion of the population joining the armed group $F$.  From (\ref{am}) and (\ref{ane}), we have that $f^{c}\leq f^m$ implies that
\small
\begin{equation}
\label{ }
\frac{7\beta}{9+15\beta}\leq \frac{1}{5}
\end{equation}
\normalsize
which, solving for $\beta$, is equivalent to\footnote{Note that when (\ref{betafin}) is satisfied, $a^{c}/2\leq 1/5\leq a^{c}$.}
\small
\begin{equation}
\label{betafin}
\beta\leq  \frac{9}{20}
\end{equation}
\normalsize
 Let $\Delta f(\beta)=f^m-f^{c}$ be the proportion of individuals abandoning $F$ to join $A$. We know that $\Delta f(\beta)\geq 0$ when (\ref{betafin}) is satisfied. From  (\ref{am}) and (\ref{ane}), we have that
\small
\begin{equation}
\label{propdes}
\Delta f(\beta)=\frac{9-20\beta}{5(9+15\beta)}
\end{equation}
\normalsize
where it is easy to see that  $d \Delta f(\beta)/d \beta<0$. Thus, an increase in $\beta$ decreases the proportion of people willing to abandon $F$ to join $A$.

\subsection{Identification} \label{appidentificationTW}
\small
In this section, we propose an alternative specification (to \cite{GoodmanBacon2018}'s) from \cite{Tchuente_Wind2019}, the estimates for which are described in Section \ref{Robustness}.

Consider an outcome $y_{i,t}$ (e.g. the probability of a conflict-related event) in municipality $i$ and year $t$, and model it as a function of $D_{i,t}$, where $D_{i,t}$ is an indicator of the state of municipality $i$ and year $t$. There are $L+K+2$ states.
Assume that for any time period,  a municipality can be in only one of the  $L+K+2$ states. In this scenario, the potential outcomes are  $y_{i,t}(\tau)$ with $ \tau\in {-K,...,-1,e,1,...,L}$, and $ y_{i,t}(0)$, where $y_{i,t}(0)$ corresponds to the outcome out of the influence area of the treatment. Assume that the treatment status is an absorption state; in the context of this paper,  this assumption is plausibly satisfied as we focus on the establishment of the first non-Catholic church in a municipality.

Let us define now the following quantities.

\medskip

   \textbf{Treatment effect at time of event}:
\begin{equation}
\label{ }
\gamma_{ie}=y_{i,t}(e)-y_{i,t}(0)
\end{equation}

 \textbf{$\tau$ period anticipated treatment effect}:
\begin{equation}
\label{ }
\gamma_{i\tau^-}=y_{i,t}(-\tau)-y_{i,t}(0)
\end{equation}

\textbf{$\tau$ period delayed treatment effect}:
\begin{equation}
\label{ }
\gamma_{i\tau^+}=y_{i,t}(\tau)-y_{i,t}(0)
\end{equation}
Note that the observed outcome can be written as
\begin{equation}
\label{ }
y_{i,t}=\sum_{s \in \{-K,...,L,e\}}\mathbf 1\{s=S_{i,t}\}y_{i,t}(s)
\end{equation}
Assume that the outcome out of the influence area of the treatment can be written as
\begin{equation}\label{modelpanelFlex}
  y_{i,t}(0)=\alpha_i + \alpha_t + \alpha X_{i,t} + \varepsilon_{i,t}
\end{equation}
and that
\begin{equation}\label{modelpanelFlexesti}
  y_{i,t}=\alpha_i + \alpha_t +\alpha X_{i,t} +\sum_{\tau \in \{-K,...,L\}}\gamma_{i\tau}D_{i,t}^{\tau}+ \varepsilon_{i,t}
\end{equation}
where $D_{i,t}^{\tau}$ is a dummy variable indicating if individual $i$ is in state $\tau$ at period $t$. Defining $\gamma_i=(\gamma_{i-K},...,\gamma_{iL})$, $S_{i,t}=(D_{i,t}^{-K},...,D_{i,t}^L)$ and $\gamma= E(\gamma_i)$, it is easy to see that the model in (\ref{modelpanelFlexesti}) is equivalent to
\begin{equation}\label{modelpanelFlexestiM}
  y_{i,t}=\alpha_i + \alpha_t + \sum_{\tau \in \{-K,...,L\}}\gamma_{\tau}D_{i,t}^{\tau} + v_{i,t}
\end{equation}
with $v_{i,t}=(\gamma_i-\gamma) S_{i,t}'+ \varepsilon_{i,t}$ and where, to simplify the exposition,  we have removed $X_{i,t}$.

We are interested in the estimation of $\gamma$.  As recent works on identification of dynamic treatment effects have shown (e.g. \citealt{abraham2019estimating}; \citealt{deChaisemartinDHaultfoeuille2018}, and \citealt{GoodmanBacon2018}), in the presence of heterogenous treatment effects, the estimation of (\ref{modelpanelFlexestiM}) does not give the average treatment effect.

In a recent work, \cite{Tchuente_Wind2019} show that the variable $S_{i,t}$ is a real random vector with special properties. For instance, if $S_{i,t}=(1,0...0)$, then $S_{it+1}=(0,1,0...0)$. Motivated by this observation, \cite{Tchuente_Wind2019} propose an estimator that can solve the problem of the failure of the identification in estimating (\ref{modelpanelFlexestiM}). In the rest of this section, we will briefly describe this estimator, which as previously mentioned, we call TW.

\medskip

Let $L_{i,t}= \mathbf{1}\{ t \in \{  -K,... ,e_i,...L\}, \hspace{0.2cm}for \hspace{0.2cm} individual \hspace{0.2cm}i\}$ be an indicator of the fact that period $t$ is in the neighborhood of the event, and define $\bar{L}_{i}=\sum_{t=1}^{T}L_{i,t}$ as the number of influenced periods. Note that the average outcome in a period far from the event, which we will use as a benchmark, can be written as
\begin{equation}
\label{ }
\bar{y}_{i}^L=\frac{1}{T-\bar{L}_i}\sum_{t=1}^{T}y_{i,t}(1-L_{i,t})
\end{equation}
Note that the following quantity identifies the average treatment effect
\begin{equation}
\label{ateTW}
\gamma_{FEL}^{\tau} = E[y_{i,t}-\bar{y}_{i}^L| D_{i,t}^{\tau}=1, L_{i,t}=1]-E[y_{i,t}-\bar{y}_{i}^L| D_{i,t}^{\tau}=0, L_{i,t}=0]
\end{equation}
Since $\gamma_{FEL}^{\tau} $ in Eq. (\ref{ateTW})  cannot be estimated using a linear panel model, we propose a transformation that uses a split sample strategy. First, note that
\begin{eqnarray*}
  \gamma_{FEL}^{\tau} &=& E[y_{i,t}-\bar{y}_{i}^L| D_{i,t}^{\tau}=1, L_{i,t}=1]-E[y_{i,t}-\bar{y}_{i}^L| D_{i,t}^{\tau}=0, L_{i,t}=0] \\
   &=& E[y_{i,t}-\bar{y}_{i}^L| D_{i,t}^{\tau}=1, L_{i,t}=1]-E[y_{i,t}-\bar{y}_{i}^L| D_{i,t}^{\tau}=0, L_{i,t}=1]\\
   &+& E[y_{i,t}-\bar{y}_{i}^L| D_{i,t}^{\tau}=0, L_{i,t}=1]-E[y_{i,t}-\bar{y}_{i}^L| D_{i,t}^{\tau}=0, L_{i,t}=0]
\end{eqnarray*}
From the last expression, define $\theta^{\tau,L}=E[y_{i,t}-\bar{y}_{i}^L| D_{i,t}^{\tau}=1, L_{i,t}=1]-E[y_{i,t}-\bar{y}_{i}^L| D_{i,t}^{\tau}=0, L_{i,t}=1]$ and $\theta^{\tau,D}=E[y_{i,t}-\bar{y}_{i}^L| D_{i,t}^{\tau}=0, L_{i,t}=1]-E[y_{i,t}-\bar{y}_{i}^L| D_{i,t}^{\tau}=0, L_{i,t}=0]$, with which
\begin{equation}
\label{ }
\gamma^\tau_{FEL}= \theta^{\tau,L}+ \theta^{\tau,D}
\end{equation}
We can split the sample in two groups. First, we create a sub-sample for which $L_{i,t}=1$, and estimate the following linear panel data model:
\begin{equation}
\label{thetal1}
y_{i,t}-\bar{y}_{i}^L= \sum_{\tau \in \{-K,...,L\}}\theta^{\tau,L}D_{i,t}^{\tau}+\upsilon_{i,t}
\end{equation}
Second, we consider the sub-sample for which $D_{i,t}^{\tau}=0$, and estimate the model
\begin{equation}
\label{thetad0}
y_{i,t}-\bar{y}_{i}^L= \theta^{\tau,D}L_{i,t}+\vartheta_{i,t}
\end{equation}
Finally, we compute $\hat{\gamma}_{FEL}^{\tau}= \hat{\theta}^{\tau,L}+ \hat{\theta}^{\tau,D}$, where $\hat{\theta}^{\tau,D}$ and $\hat{\theta}^{\tau,L}$ are obtained from (\ref{thetal1}) and (\ref{thetad0}), respectively.


\hbox {} \newpage
\setcounter{table}{0}
\setcounter{figure}{0}
\setcounter{subsection}{0}
\renewcommand{\thefigure}{\Alph{section}\arabic{figure}}
\renewcommand{\thetable}{\Alph{section}\arabic{table}}

\section{Web Appendix: Additional Figures and Tables}


\subsection{Spatial and temporal distribution of non-Catholic churches}\label{appspatialtemporalncc}

%
  \begin{figure}[H]
\caption{Evolution of the proportion of municipalities with at least one non-Catholic church (nCc) and Geographical distribution of non-Catholic churches in 2017}\label{PGUEiwaands_fig}
\begin{subfigure}[b]{0.9\textwidth}
\centering
\includegraphics[width=0.6\linewidth]{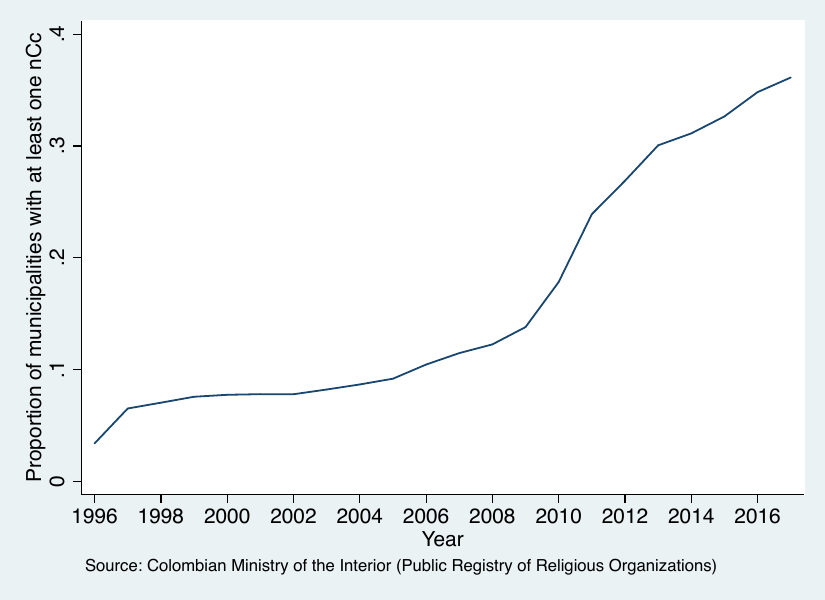}
\caption{}\label{percmpiosiglesiasnctimeseries20032013}
\end{subfigure}
\begin{subfigure}[b]{0.9\textwidth}
\centering
\includegraphics[width=0.6\linewidth]{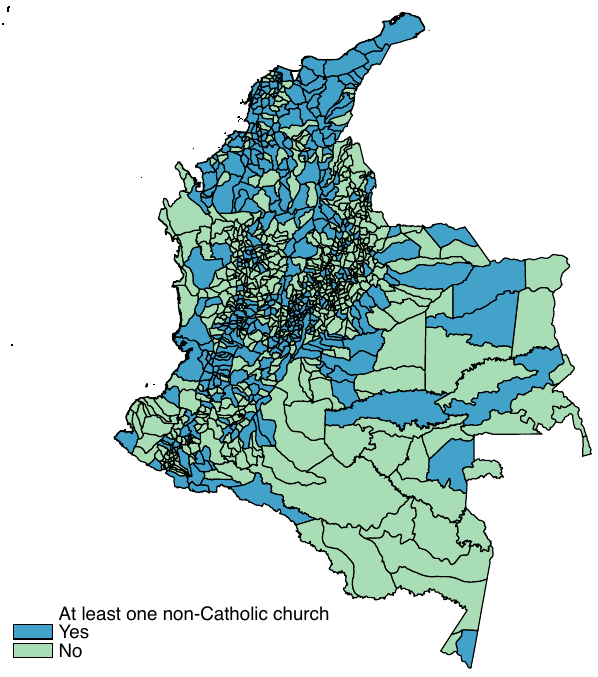}
\caption{}\label{mapdumiglesiasnc2017_map}
\end{subfigure}
\end{figure}


\subsection{Additional robustness checks for the main results}\label{approbustnessmain}

\enlargethispage{5\baselineskip}
\vspace{-0.4cm}
\begin{table}[H]
\begin{center}
{
\renewcommand{\arraystretch}{0.1}
\setlength{\tabcolsep}{15pt}
\caption {Effect of first nC church on the prob. of a guerrilla attack: all lags and leads}  \label{PGUErobust1_tab}
\footnotesizes
\vspace{-0.3cm}\centering  
\begin{tabular}{lccccc}
\hline\hline
    & \multicolumn{5}{c}{Dep. variable = 1 if an attack by a guerrilla group} \\\cmidrule[0.2pt](l){2-6}
& (1)& (2) & (3) & (4) & (5)  \\     \hline \addlinespace[0.15cm]
Effect at t=-13     &      -0.050\sym{*}  &      -0.054\sym{*}  &      -0.054\sym{*}  &      -0.058\sym{*}  &      -0.056\sym{*}  \\
                    &     (0.029)         &     (0.030)         &     (0.033)         &     (0.033)         &     (0.033)         \\
Effect at t=-12     &      -0.022         &      -0.015         &      -0.027         &      -0.022         &      -0.024         \\
                    &     (0.030)         &     (0.030)         &     (0.035)         &     (0.035)         &     (0.036)         \\
Effect at t=-11     &      -0.010         &      -0.001         &      -0.015         &      -0.007         &      -0.005         \\
                    &     (0.029)         &     (0.029)         &     (0.037)         &     (0.036)         &     (0.037)         \\
\end{tabular}
}
\end{center}
\end{table}


\begin{table}[H]
\begin{center}
{
\renewcommand{\arraystretch}{0.4}
\setlength{\tabcolsep}{10pt}
\caption {Effect of first nC church on the prob. of a guerrilla attack: fewer lags and leads}  \label{PGUErobust1redlagsleads_tab}
\footnotesize
\vspace{-0.3cm}\centering  \begin{tabular}{lccccc}
\hline\hline
    & \multicolumn{5}{c}{Dep. variable = 1 if an attack by a guerrilla group} \\\cmidrule[0.2pt](l){2-6}
& (1)& (2) & (3) & (4) & (5)   \\     \hline \addlinespace[0.15cm]
        & \multicolumn{1}{c}{$L=K=12$} & \multicolumn{1}{c}{$L=K=11$}  & \multicolumn{1}{c}{$L=K=10$}  & \multicolumn{1}{c}{$L=K=9$}  & \multicolumn{1}{c}{$L=K=8$}  \\
    \cmidrule[0.2pt](l){2-2}\cmidrule[0.2pt](l){3-3}\cmidrule[0.2pt](l){4-4}\cmidrule[0.2pt](l){5-5}\cmidrule[0.2pt](l){6-6}
Effect at t=-12     &       0.003         &                     &                     &                     &                     \\
                    &     (0.030)         &                     &                     &                     &                     \\
Effect at t=-11     &       0.019         &       0.018         &                     &                     &                     \\
                    &     (0.033)         &     (0.030)         &                     &                     &                     \\
Effect at t=-10     &       0.019         &       0.018         &       0.011         &                     &                     \\
\end{tabular}
}
\end{center}
\end{table}


\begin{table}[H]
\begin{center}
{
\renewcommand{\arraystretch}{0.8}
\setlength{\tabcolsep}{8pt}
\caption {Impact of first nC church on the prob. of an attack by the Colombian Army}  \label{PEST_tab}
\small
\centering  \begin{tabular}{lcccccc}
\hline\hline \addlinespace[0.15cm]
    & \multicolumn{5}{c}{Dep. variable = 1 if an attack by the} \\
    & \multicolumn{5}{c}{Colombian national army} \\
    \cmidrule[0.2pt](l){2-6}\addlinespace[0.15cm]
& (1)& (2) & (3) & (4) & (5)  \\   \addlinespace[0.15cm] \hline \addlinespace[0.15cm]
            \multicolumn{1}{l}{\emph{\underline{Panel A}:}}            & \multicolumn{6}{c}{\emph{Dynamic} specification}\\\addlinespace[0.15cm]
   Effect at t=-3      &      -0.028         &      -0.037         &      -0.039         &      -0.039         &      -0.025         \\
                    &     (0.044)         &     (0.046)         &     (0.046)         &     (0.046)         &     (0.039)         \\
Effect at t=-2      &      -0.002         &      -0.006         &      -0.005         &      -0.007         &       0.003         \\
                    &     (0.042)         &     (0.044)         &     (0.044)         &     (0.044)         &     (0.039)         \\
\end{tabular}
}
\end{center}
\end{table}


\begin{table}[H]
\begin{center}
{
\renewcommand{\arraystretch}{0.9}
\setlength{\tabcolsep}{12pt}
\caption {Impact of first non-Catholic church on the prob. of a conflict related attack: comparison between the effects for the guerrillas and the paramilitaries}  \label{PANYdgue_tab}
\small
\centering  \begin{tabular}{lcccccc}
\hline\hline \addlinespace[0.15cm]
    & \multicolumn{4}{c}{Dep. variable = 1 if an attack by any group} \\\addlinespace[0.15cm]
    & \multicolumn{2}{c}{\emph{Static} specification}  & \multicolumn{2}{c}{\emph{Dynamic} specification}\\\cmidrule[0.2pt](l){2-3}\cmidrule[0.2pt](l){4-5}
& (1)& (2) & (3)& (4) \\   \addlinespace[0.15cm] \hline \addlinespace[0.15cm]
            Effect at t$\geq 0$ &       0.590\sym{***}&       0.588\sym{***}&                     &                     \\
$\times$ guerrilla attack&     (0.033)         &     (0.033)         &                     &                     \\
Effect at t=0       &                     &                     &       0.780\sym{***}&       0.788\sym{***}\\
\hspace{0.2cm} $\times$ guerrilla attack&                     &                     &     (0.049)         &     (0.049)         \\
Effect at t=+1      &                     &                     &       0.760\sym{***}&       0.760\sym{***}\\
\end{tabular}
}
\end{center}
\end{table}


\begin{table}[H]
\begin{center}
{
\renewcommand{\arraystretch}{0.8}
\setlength{\tabcolsep}{2pt}
\caption {Impact of first nC church on the prob. of a conflict-related attack: robustness to the  inclusion of additional controls}  \label{PGUEPANYPGUEPPARrobutscontrols_tab}
\small
\centering  \begin{tabular}{lccccccccc}
\hline\hline \addlinespace[0.15cm]
    & \multicolumn{9}{c}{Dep. variable = 1 if an attack by} \\\addlinespace[0.15cm]
    & \multicolumn{3}{c}{Any group} & \multicolumn{3}{c}{Guerrilla} & \multicolumn{3}{c}{Paramilitaries} \\\cmidrule[0.2pt](l){2-4}\cmidrule[0.2pt](l){5-7}\cmidrule[0.2pt](l){8-10}
& (1)& (2) & (3) & (4) & (5) & (6)& (7) & (8) & (9)   \\   \addlinespace[0.15cm] \hline \addlinespace[0.15cm]
            \multicolumn{1}{l}{\emph{\underline{Panel A}:}}            & \multicolumn{9}{c}{\emph{Dynamic} specification}\\\addlinespace[0.15cm]
 Effect at t=-3      &       0.041         &       0.043         &       0.033         &      -0.003         &      -0.005         &       0.005         &       0.048         &       0.053         &       0.031         \\
                    &     (0.045)         &     (0.045)         &     (0.038)         &     (0.040)         &     (0.041)         &     (0.035)         &     (0.036)         &     (0.036)         &     (0.032)         \\
Effect at t=-2      &      -0.003         &       0.001         &      -0.022         &      -0.014         &      -0.013         &      -0.017         &       0.020         &       0.024         &       0.005         \\
                    &     (0.044)         &     (0.044)         &     (0.039)         &     (0.039)         &     (0.039)         &     (0.036)         &     (0.036)         &     (0.036)         &     (0.033)         \\
\end{tabular}
}
\end{center}
\end{table}


\begin{table}[H]
\begin{center}
{
\renewcommand{\arraystretch}{0.8}
\setlength{\tabcolsep}{7pt}
\caption {Impact of first nC church on the prob. of a conflict-related attack:  robustness to clustering by department}  \label{PGUEPANYPGUEPPARclusterdepto_tab}
\small
\centering  \begin{tabular}{lcccccc}
\hline\hline \addlinespace[0.15cm]
    & \multicolumn{6}{c}{Dep. variable = 1 if an attack by} \\\addlinespace[0.15cm]
    & \multicolumn{2}{c}{Any group} & \multicolumn{2}{c}{Guerrilla} & \multicolumn{2}{c}{Paramilitaries} \\\cmidrule[0.2pt](l){2-3}\cmidrule[0.2pt](l){4-5}\cmidrule[0.2pt](l){6-7}
& (1)& (2) & (3) & (4) & (5) & (6)  \\   \addlinespace[0.15cm] \hline \addlinespace[0.15cm]
            \multicolumn{1}{l}{\emph{\underline{Panel A}:}}            & \multicolumn{6}{c}{\emph{Dynamic} specification}\\\addlinespace[0.15cm]
Effect at t=-3      &       0.041         &       0.042         &       0.004         &      -0.002         &       0.049         &       0.050         \\
                    &     (0.039)         &     (0.039)         &     (0.040)         &     (0.042)         &     (0.032)         &     (0.031)         \\
Effect at t=-2      &      -0.011         &      -0.004         &      -0.018         &      -0.016         &       0.021         &       0.021         \\
                    &     (0.039)         &     (0.037)         &     (0.031)         &     (0.032)         &     (0.034)         &     (0.035)         \\
\end{tabular}
}
\end{center}
\end{table}


\begin{table}[H]
\begin{center}
{
\renewcommand{\arraystretch}{0.8}
\setlength{\tabcolsep}{2pt}
\caption {Determinants of first non-Catholic church (NCHM data)}  \label{dfirstiglesiasncNCHMyearly_tab}
\small
\centering  \begin{tabular}{lccccc}
\hline\hline \addlinespace[0.15cm]
    & \multicolumn{5}{c}{Dep. variable = 1 when first} \\
    & \multicolumn{5}{c}{ non-Catholic church is established} \\\cmidrule[0.2pt](l){2-6}
& (1)& (2) & (3)  & (4) & (5) \\    \addlinespace[0.15cm] \hline \addlinespace[0.15cm]
Assassination by guerrilla (prob.) at t-1&      -0.004         &      -0.004         &       0.000         &       0.000         &      -0.000         \\
                    &     (0.007)         &     (0.007)         &     (0.007)         &     (0.007)         &     (0.007)         \\
Assassination by paramilitaries (prob.) at t-1&      -0.005         &      -0.001         &      -0.008         &       0.000         &       0.001         \\
                    &     (0.008)         &     (0.009)         &     (0.008)         &     (0.008)         &     (0.008)         \\
Assassination by unknown (prob.) at t-1&       0.000         &      -0.000         &       0.008         &       0.006         &       0.007         \\
\end{tabular}
}
\end{center}
\end{table}


\begin{table}[H]
\begin{center}
{
\renewcommand{\arraystretch}{0.65}
\setlength{\tabcolsep}{5pt}
\caption {Effect of first non-Catholic church on the prob. of a conflict-related attack: alternative estimators}  \label{PANYPGUEPPARiwaands_tab}
\small
%
\\
\addlinespace[0.1cm]\hline\hline

\multicolumn{4}{p{14cm}}{\footnotesizes{\textbf{Notes:} Columns (1) and (2) report the two-way fixed effects difference-in-differences estimates from Eq. (\ref{baseline})  without trends or controls. Columns (3) and (4) report \cite{AbrahamSun2018}'s interaction-weighted estimator (AS). Columns (5) and (6) report \cite{Tchuente_Wind2019}'s estimator (TW). Both AS and TW estimators are implemented following the procedures described in Section \ref{Robustness}, where $L=K=5$ for the TW estimator and $L=K=13$ for the AS estimator.} }
\end{tabular}
}
\end{center}
\end{table}


\begin{table}[H]
\begin{center}
{
\renewcommand{\arraystretch}{0.8}
\setlength{\tabcolsep}{1.5pt}
\caption {Determinants of first non-Catholic church the probability of a conflict-related assissantion: monthly data}  \label{dfirstiglesiasncNCHMmonthly_tab}
\small
\centering  \begin{tabular}{lccccccc}
\hline\hline \addlinespace[0.15cm]
    & \multicolumn{7}{c}{Dep. variable = 1 when first} \\
    & \multicolumn{7}{c}{ non-Catholic church is established} \\\cmidrule[0.2pt](l){2-8}
& (1)& (2) & (3)  & (4) & (5) & (6) & (7) \\    \addlinespace[0.15cm] \hline \addlinespace[0.15cm]
Assassination by guerrilla (prob.) at t-1&      -0.000         &      -0.000         &      -0.000         &      -0.000         &      -0.000         &       0.000         &       0.000         \\
                    &     (0.001)         &     (0.001)         &     (0.001)         &     (0.001)         &     (0.001)         &     (0.001)         &     (0.001)         \\
Assassination by paramilitaries (prob.) at t-1&      -0.000         &      -0.000         &       0.000         &       0.000         &       0.000         &       0.000         &      -0.000         \\
                    &     (0.001)         &     (0.001)         &     (0.001)         &     (0.001)         &     (0.001)         &     (0.001)         &     (0.001)         \\
Assassination by unknown (prob.) at t-1&       0.000         &       0.000         &       0.000         &       0.000         &       0.000         &       0.001         &       0.001         \\
                    &     (0.001)         &     (0.001)         &     (0.001)         &     (0.001)         &     (0.001)         &     (0.001)         &     (0.001)         \\
Assassination by guerrilla (prob.) at t-2&                     &       0.000         &                     &       0.000         &                     &                     &       0.001         \\
                    &                     &     (0.001)         &                     &     (0.001)         &                     &                     &     (0.001)         \\
Assassination by paramilitaries (prob.) at t-2&                     &      -0.001         &                     &      -0.001         &                     &                     &      -0.000         \\
                    &                     &     (0.001)         &                     &     (0.001)         &                     &                     &     (0.001)         \\
Assassination by unknown (prob.) at t-2&                     &       0.000         &                     &       0.000         &                     &                     &       0.000         \\
                    &                     &     (0.001)         &                     &     (0.001)         &                     &                     &     (0.001)         \\
Internally displaced pop. (outflow, rate) at t-1&                     &                     &                     &                     &       0.000         &       0.000         &       0.000         \\
                    &                     &                     &                     &                     &     (0.000)         &     (0.000)         &     (0.000)         \\
Internally displaced pop. (inflow, rate) at t-1&                     &                     &                     &                     &       0.000         &       0.000         &       0.000         \\
                    &                     &                     &                     &                     &     (0.000)         &     (0.000)         &     (0.000)         \\
Desertion from the guerrilla (prob.) at t-1&                     &                     &                     &                     &      -0.000         &      -0.000         &      -0.000         \\
                    &                     &                     &                     &                     &     (0.001)         &     (0.001)         &     (0.001)         \\
Forced recruitment (prob.) at t-1&                     &                     &                     &                     &       0.000         &       0.000         &       0.000         \\
                    &                     &                     &                     &                     &     (0.001)         &     (0.001)         &     (0.001)         \\
Log of total population at t-1&                     &                     &      -0.009         &      -0.009         &      -0.021\sym{**} &      -0.017\sym{*}  &      -0.017\sym{*}  \\
                    &                     &                     &     (0.007)         &     (0.007)         &     (0.009)         &     (0.010)         &     (0.010)         \\
Prop. of rural population at t-1&                     &                     &      -0.001         &      -0.001         &       0.002         &       0.002         &       0.002         \\
                    &                     &                     &     (0.002)         &     (0.002)         &     (0.002)         &     (0.002)         &     (0.002)         \\
Unsatisfied Basic Needs at t-1&                     &                     &       0.000         &       0.000         &       0.000         &      -0.000         &      -0.000         \\
                    &                     &                     &     (0.000)         &     (0.000)         &     (0.000)         &     (0.000)         &     (0.000)         \\
Ethinic minority population at t-1&                     &                     &       0.005         &       0.005         &      -0.002         &                     &                     \\
                    &                     &                     &     (0.003)         &     (0.003)         &     (0.003)         &                     &                     \\
Homicide (rate) at t-1&                     &                     &      -0.000         &      -0.000         &       0.000         &       0.000         &       0.000         \\
                    &                     &                     &     (0.000)         &     (0.000)         &     (0.000)         &     (0.000)         &     (0.000)         \\
Coffee int. x log coffee price at t-1&                     &                     &                     &                     &                     &       0.001         &       0.001         \\
                    &                     &                     &                     &                     &                     &     (0.001)         &     (0.001)         \\
Vote share for the Liberal party at t-1&                     &                     &                     &                     &                     &       0.001         &       0.001         \\
                    &                     &                     &                     &                     &                     &     (0.001)         &     (0.001)         \\
Vote share for the Conservative party at t-1&                     &                     &                     &                     &                     &       0.001         &       0.001         \\
                    &                     &                     &                     &                     &                     &     (0.002)         &     (0.002)         \\
Vote share for leftwing parties at t-1&                     &                     &                     &                     &                     &       0.004         &       0.004         \\
                    &                     &                     &                     &                     &                     &     (0.004)         &     (0.004)         \\
R-sq                &       0.085         &       0.085         &       0.082         &       0.082         &       0.097         &       0.088         &       0.088         \\
Observations        &      103411         &      103409         &      101393         &      101393         &       63211         &       56093         &       56093         \\

\addlinespace[0.15cm]\hline\hline
\multicolumn{8}{p{16.5cm}}{\footnotesizes{\textbf{Notes:} All models include municipality, year and department $\times$ year fixed effects, as well as  municipality-specific linear trends.  Samples for regression models include data from 1996 onwards.   Robust standard errors (in parentheses) are clustered by municipality. * denotes statistically significant estimates at 10\%, ** denotes significant at 5\% and *** denotes significant at 1\%.} }
\end{tabular}
}
\end{center}
\end{table}


\begin{table}[H]
\begin{center}
{
\renewcommand{\arraystretch}{0.9}
\setlength{\tabcolsep}{5pt}
\caption {Impact of first non-Catholic church on the prob. of a conflict-related attack: role of month of establishment at t=0}  \label{PGUEPANYPGUEPPARmonthofest_tab}
\small
\centering  \begin{tabular}{lcccccc}
\hline\hline \addlinespace[0.15cm]
    & \multicolumn{6}{c}{Dep. variable = 1 if an attack by} \\\addlinespace[0.15cm]
    & \multicolumn{2}{c}{Any group} & \multicolumn{2}{c}{Guerrilla} & \multicolumn{2}{c}{Paramilitaries} \\\cmidrule[0.2pt](l){2-3}\cmidrule[0.2pt](l){4-5}\cmidrule[0.2pt](l){6-7}
& (1)& (2) & (3) & (4) & (5) & (6)  \\   \addlinespace[0.15cm] \hline \addlinespace[0.15cm]
  Effect at t=0       &       0.553\sym{***}&       0.518\sym{***}&       0.377\sym{**} &       0.359\sym{**} &       0.225\sym{*}  &       0.218\sym{**} \\
                    &     (0.185)         &     (0.180)         &     (0.172)         &     (0.174)         &     (0.117)         &     (0.108)         \\
Effect at t=0       &      -0.395\sym{*}  &      -0.369\sym{*}  &      -0.239         &      -0.208         &      -0.312\sym{*}  &      -0.330\sym{*}  \\
\hspace{0.2cm} $\times$ church established in Feb.&     (0.209)         &     (0.206)         &     (0.192)         &     (0.197)         &     (0.188)         &     (0.180)         \\
Effect at t=0       &      -0.514\sym{*}  &      -0.488\sym{*}  &      -0.323         &      -0.325         &      -0.061         &      -0.052         \\
\hspace{0.2cm} $\times$ church established in Mar.&     (0.268)         &     (0.276)         &     (0.281)         &     (0.293)         &     (0.194)         &     (0.188)         \\
Effect at t=0       &      -0.887\sym{***}&      -0.841\sym{***}&      -0.854\sym{***}&      -0.818\sym{***}&      -0.183         &      -0.177         \\
\hspace{0.2cm} $\times$ church established in Apr.&     (0.244)         &     (0.242)         &     (0.234)         &     (0.238)         &     (0.189)         &     (0.186)         \\
Effect at t=0       &      -0.414\sym{*}  &      -0.385         &      -0.204         &      -0.197         &      -0.275         &      -0.261         \\
\hspace{0.2cm} $\times$ church established in May&     (0.235)         &     (0.239)         &     (0.216)         &     (0.223)         &     (0.198)         &     (0.192)         \\
Effect at t=0       &      -0.444\sym{*}  &      -0.429\sym{*}  &      -0.363\sym{*}  &      -0.309         &      -0.099         &      -0.152         \\
\hspace{0.2cm} $\times$ church established in Jun.&     (0.228)         &     (0.223)         &     (0.217)         &     (0.216)         &     (0.175)         &     (0.159)         \\
Effect at t=0       &      -0.499\sym{**} &      -0.482\sym{*}  &      -0.367\sym{*}  &      -0.345         &      -0.194         &      -0.197         \\
\hspace{0.2cm} $\times$ church established in Jul.&     (0.243)         &     (0.252)         &     (0.213)         &     (0.222)         &     (0.152)         &     (0.157)         \\
Effect at t=0       &      -0.509\sym{*}  &      -0.499\sym{*}  &      -0.326         &      -0.346         &      -0.318\sym{*}  &      -0.311\sym{*}  \\
\hspace{0.2cm} $\times$ church established in Aug.&     (0.279)         &     (0.276)         &     (0.255)         &     (0.262)         &     (0.176)         &     (0.169)         \\
Effect at t=0       &      -0.432\sym{*}  &      -0.395\sym{*}  &      -0.242         &      -0.221         &      -0.144         &      -0.136         \\
\hspace{0.2cm} $\times$ church established in Sept.&     (0.235)         &     (0.231)         &     (0.218)         &     (0.221)         &     (0.143)         &     (0.134)         \\
Effect at t=0       &      -0.605\sym{***}&      -0.541\sym{***}&      -0.463\sym{**} &      -0.424\sym{**} &      -0.272         &      -0.257         \\
\hspace{0.2cm} $\times$ church established in Oct.&     (0.208)         &     (0.206)         &     (0.192)         &     (0.199)         &     (0.197)         &     (0.197)         \\
Effect at t=0       &      -0.314         &      -0.298         &      -0.128         &      -0.113         &      -0.304\sym{**} &      -0.330\sym{**} \\
\hspace{0.2cm} $\times$ church established in Nov.&     (0.271)         &     (0.275)         &     (0.261)         &     (0.269)         &     (0.151)         &     (0.148)         \\
Effect at t=0       &      -0.350         &      -0.321         &      -0.205         &      -0.192         &      -0.256         &      -0.281         \\
\hspace{0.2cm} $\times$ church established in Dec.&     (0.220)         &     (0.219)         &     (0.184)         &     (0.187)         &     (0.192)         &     (0.196)         \\
R-sq                &       0.602         &       0.608         &       0.618         &       0.626         &       0.498         &       0.505         \\
Observations        &        5530         &        5357         &        5530         &        5357         &        5530         &        5357         \\

\addlinespace[0.15cm]\hline\addlinespace[0.15cm]\
Controls & No & Yes& No& Yes& No& Yes \\
\addlinespace[0.15cm]\hline\hline
\multicolumn{7}{p{15.5cm}}{\footnotesizes{\textbf{Notes:} All columns in Panel A report the estimates from Eq. (\ref{baseline}) and all columns in Panel B report the estimates from Eq. (\ref{didbaseline}).  All models include municipality and year fixed effects, department $\times$ year fixed effects and municipality-specific trends.  The models in Panel B include 13 lags and 13 leads, normalized to the period prior to treatment. The models with controls (columns (2), (4) and (6)) include the following (lagged) covariates:  log of total population, proportion of the population living in a rural area, proportion of the population with unsatisfied basic needs, proportion of ethnic minority population, the homicide rate and the share of vote for liberals, conservatives and the left. Samples for regression models use data from 1996 to 2009. Robust standard errors (in parentheses) are clustered by municipality. * denotes statistically significant estimates at 10\%, ** denotes significant at 5\% and *** denotes significant at 1\%.} }
\end{tabular}
}
\end{center}
\end{table}


\begin{table}[H]
\begin{center}
{
\renewcommand{\arraystretch}{0.7}
\setlength{\tabcolsep}{5pt}
\caption {Impact of first non-Catholic church on the prob. of a conflict-related assassination: monthly data}  \label{NCHMmonhtly_tab}
\vspace{-0.3cm}
\small
\centering  \begin{tabular}{lcccccccc}
\hline\hline \addlinespace[0.15cm]
    & \multicolumn{8}{c}{Dep. variable = 1 if an assassination by} \\\addlinespace[0.15cm]
    & \multicolumn{2}{c}{Any group} & \multicolumn{2}{c}{Guerrilla} & \multicolumn{2}{c}{Paramilitaries} & \multicolumn{2}{c}{Unknown}  \\\cmidrule[0.2pt](l){2-3}\cmidrule[0.2pt](l){4-5}\cmidrule[0.2pt](l){6-7}\cmidrule[0.2pt](l){8-9}
& (1)& (2) & (3) & (4) & (5) & (6)  & (7) & (8)  \\   \addlinespace[0.15cm] \hline \addlinespace[0.15cm]
            \multicolumn{1}{l}{\emph{\underline{Panel A}:}}            & \multicolumn{8}{c}{\emph{Dynamic} specification}\\\addlinespace[0.15cm]
Effect at t=-3      &       0.006         &       0.009         &       0.018         &       0.018         &      -0.002         &       0.002         &       0.002         &       0.003         \\
                    &     (0.018)         &     (0.018)         &     (0.011)         &     (0.011)         &     (0.011)         &     (0.011)         &     (0.015)         &     (0.015)         \\
Effect at t=-2      &       0.002         &       0.005         &       0.009         &       0.010         &       0.010         &       0.015         &       0.002         &       0.003         \\
                    &     (0.017)         &     (0.017)         &     (0.009)         &     (0.009)         &     (0.011)         &     (0.011)         &     (0.015)         &     (0.015)         \\
Effect at t=-1      &       0.000         &       0.000         &       0.000         &       0.000         &       0.000         &       0.000         &       0.000         &       0.000         \\
                    &         (.)         &         (.)         &         (.)         &         (.)         &         (.)         &         (.)         &         (.)         &         (.)         \\
Effect at t=0       &       0.018         &       0.022         &       0.020\sym{*}  &       0.021\sym{*}  &       0.003         &       0.008         &       0.005         &       0.008         \\
                    &     (0.018)         &     (0.018)         &     (0.011)         &     (0.011)         &     (0.011)         &     (0.011)         &     (0.015)         &     (0.016)         \\
Effect at t=+1      &       0.006         &       0.008         &       0.012         &       0.012         &       0.008         &       0.013         &      -0.008         &      -0.010         \\
                    &     (0.017)         &     (0.017)         &     (0.010)         &     (0.010)         &     (0.012)         &     (0.012)         &     (0.016)         &     (0.016)         \\
Effect at t=+2      &       0.005         &       0.007         &       0.012         &       0.012         &       0.001         &       0.004         &       0.021         &       0.021         \\
                    &     (0.017)         &     (0.017)         &     (0.011)         &     (0.011)         &     (0.011)         &     (0.011)         &     (0.016)         &     (0.016)         \\
Effect at t=+3      &      -0.005         &      -0.001         &       0.014         &       0.014         &       0.020\sym{*}  &       0.024\sym{**} &      -0.023\sym{*}  &      -0.020         \\
                    &     (0.016)         &     (0.017)         &     (0.011)         &     (0.011)         &     (0.011)         &     (0.012)         &     (0.014)         &     (0.014)         \\
Effect at t=+4      &       0.006         &       0.011         &       0.003         &       0.003         &       0.007         &       0.011         &       0.013         &       0.015         \\
                    &     (0.017)         &     (0.017)         &     (0.010)         &     (0.011)         &     (0.011)         &     (0.011)         &     (0.015)         &     (0.015)         \\
Effect at t=+5      &      -0.000         &       0.005         &       0.009         &       0.009         &       0.007         &       0.011         &      -0.001         &       0.001         \\
                    &     (0.017)         &     (0.017)         &     (0.010)         &     (0.010)         &     (0.010)         &     (0.011)         &     (0.015)         &     (0.015)         \\
Effect at t=+6      &      -0.014         &      -0.010         &       0.004         &       0.004         &       0.009         &       0.014         &      -0.010         &      -0.010         \\
                    &     (0.016)         &     (0.016)         &     (0.010)         &     (0.010)         &     (0.011)         &     (0.011)         &     (0.014)         &     (0.014)         \\
Effect at t=+7      &      -0.003         &       0.000         &      -0.003         &      -0.003         &       0.011         &       0.016         &      -0.015         &      -0.014         \\
                    &     (0.017)         &     (0.017)         &     (0.009)         &     (0.009)         &     (0.011)         &     (0.011)         &     (0.015)         &     (0.015)         \\
Effect at t=+8      &      -0.001         &      -0.002         &       0.009         &       0.009         &      -0.011         &      -0.007         &       0.003         &       0.002         \\
                    &     (0.017)         &     (0.017)         &     (0.011)         &     (0.011)         &     (0.010)         &     (0.011)         &     (0.014)         &     (0.014)         \\
Effect at t=+9      &       0.016         &       0.018         &      -0.010         &      -0.011         &       0.017         &       0.021\sym{*}  &       0.014         &       0.015         \\
                    &     (0.017)         &     (0.017)         &     (0.008)         &     (0.008)         &     (0.011)         &     (0.011)         &     (0.016)         &     (0.016)         \\
Effect at t=+10     &       0.019         &       0.020         &       0.015         &       0.015         &       0.027\sym{**} &       0.030\sym{**} &       0.004         &       0.005         \\
                    &     (0.017)         &     (0.017)         &     (0.011)         &     (0.011)         &     (0.012)         &     (0.012)         &     (0.016)         &     (0.017)         \\
Effect at t=+11     &       0.006         &       0.006         &       0.014         &       0.014         &       0.001         &       0.005         &      -0.001         &      -0.003         \\
                    &     (0.017)         &     (0.018)         &     (0.010)         &     (0.010)         &     (0.012)         &     (0.012)         &     (0.015)         &     (0.015)         \\
Effect at t=+12     &       0.003         &       0.006         &       0.018\sym{*}  &       0.019\sym{*}  &      -0.006         &      -0.002         &      -0.008         &      -0.007         \\
                    &     (0.016)         &     (0.016)         &     (0.010)         &     (0.011)         &     (0.012)         &     (0.012)         &     (0.014)         &     (0.014)         \\
R-sq                &       0.504         &       0.506         &       0.291         &       0.293         &       0.502         &       0.505         &       0.390         &       0.392         \\
Observations        &      104845         &      102807         &      104845         &      102807         &      104845         &      102807         &      104845         &      102807         \\

\addlinespace[0.15cm]\hline\addlinespace[0.15cm]
            \multicolumn{1}{l}{\emph{\underline{Panel B}:}}            & \multicolumn{8}{c}{\emph{Static} specification (long-term effect)}\\\addlinespace[0.15cm]
Effect at t$\geq 0$ &       0.032\sym{***}&       0.030\sym{***}&       0.006\sym{*}  &       0.005         &       0.003         &       0.001         &       0.028\sym{***}&       0.027\sym{***}\\
                    &     (0.008)         &     (0.008)         &     (0.003)         &     (0.003)         &     (0.007)         &     (0.007)         &     (0.007)         &     (0.007)         \\
R-sq                &       0.502         &       0.504         &       0.290         &       0.292         &       0.499         &       0.503         &       0.388         &       0.390         \\
Observations        &      103413         &      101393         &      103413         &      101393         &      103413         &      101393         &      103413         &      101393         \\

 \addlinespace[0.15cm]\hline \addlinespace[0.15cm]
Baseline controls & No & Yes& No& Yes& No& Yes & No& Yes \\
\addlinespace[0.1cm]\hline\hline
\multicolumn{9}{p{16cm}}{\footnotesizes{\textbf{Notes:} All columns in Panel A report the estimates from Eq. (\ref{baseline}) and all columns in Panel B report the estimates from Eq. (\ref{didbaseline}).  All models include municipality and year fixed effects, department $\times$ year fixed effects and municipality-specific trends.  The models in Panel A include 13 lags and 13 leads, normalized to the period prior to treatment. The models with Baseline controls  include the following (lagged) covariates:  log of total population, proportion of the population living in a rural area, proportion of the population with unsatisfied basic needs, proportion of ethnic minority population and the homicide rate. Samples for regression models use data from 1996 to 2009. Robust standard errors (in parentheses) are clustered by municipality. * denotes statistically significant estimates at 10\%, ** denotes significant at 5\% and *** denotes significant at 1\%.} }
\end{tabular}
}
\end{center}
\end{table}


\subsection{Additional evidence on the mechanisms}\label{approbustnessmech}


\begin{table}[H]
\begin{center}
{
\renewcommand{\arraystretch}{0.8}
\setlength{\tabcolsep}{6pt}
\caption {Effect of first nC church on the prob. of a conflict-related attack: heterogeneous effects by occurrence of a forced recruitment before 1996}  \label{PANYPGUEPPARrecruit_tab}
\small
\vspace{-0.3cm}\centering  \begin{tabular}{lcccccc}
\hline\hline \addlinespace[0.15cm]
    & \multicolumn{6}{c}{Dep. variable = 1 if an attack by a non-state armed group in} \\ \cmidrule[0.2pt](l){2-7}    
                 & \multicolumn{3}{c}{Municipalities with at least one}& \multicolumn{3}{c}{Municipalities without any} \\
                                  & \multicolumn{3}{c}{case of forced recruitment bf 1996}& \multicolumn{3}{c}{case of forced recruitment bf 1996}\\ \cmidrule[0.2pt](l){2-4}\cmidrule[0.2pt](l){5-7}
& \multicolumn{1}{c}{any group}& \multicolumn{1}{c}{guerrilla} & \multicolumn{1}{c}{paramilitaries} & \multicolumn{1}{c}{any group}& \multicolumn{1}{c}{guerrilla} & \multicolumn{1}{c}{paramilitaries}  \\\cmidrule[0.2pt](l){2-2} \cmidrule[0.2pt](l){3-3} \cmidrule[0.2pt](l){4-4} \cmidrule[0.2pt](l){5-5} \cmidrule[0.2pt](l){6-6}  \cmidrule[0.2pt](l){7-7}      
    & (1)& (2) & (3)& (4) & (5)& (6)\\   \addlinespace[0.15cm]\hline\addlinespace[0.15cm]                                          
Effect at t=-3      &       0.066         &      -0.158         &       0.139         &       0.024         &       0.007         &       0.023         \\
                    &     (0.119)         &     (0.119)         &     (0.130)         &     (0.042)         &     (0.038)         &     (0.034)         \\
Effect at t=-2      &       0.036         &       0.015         &       0.109         &      -0.043         &      -0.037         &      -0.005         \\
                    &     (0.114)         &     (0.103)         &     (0.098)         &     (0.042)         &     (0.039)         &     (0.035)         \\
Effect at t=-1      &       0.000         &       0.000         &       0.000         &       0.000         &       0.000         &       0.000         \\
                    &         (.)         &         (.)         &         (.)         &         (.)         &         (.)         &         (.)         \\
Effect at t=0       &       0.065         &       0.167\sym{*}  &       0.057         &       0.071         &       0.069         &      -0.003         \\
                    &     (0.096)         &     (0.085)         &     (0.150)         &     (0.049)         &     (0.046)         &     (0.036)         \\
Effect at t=+1      &       0.186\sym{*}  &       0.200\sym{*}  &       0.004         &       0.005         &       0.000         &       0.031         \\
                    &     (0.106)         &     (0.102)         &     (0.145)         &     (0.046)         &     (0.045)         &     (0.040)         \\
Effect at t=+2      &       0.130         &       0.195         &       0.094         &       0.023         &       0.041         &       0.023         \\
                    &     (0.123)         &     (0.137)         &     (0.146)         &     (0.057)         &     (0.055)         &     (0.041)         \\
Effect at t=+3      &      -0.042         &      -0.118         &      -0.045         &      -0.011         &       0.013         &       0.070         \\
                    &     (0.167)         &     (0.181)         &     (0.205)         &     (0.059)         &     (0.058)         &     (0.049)         \\
Effect at t=+4      &       0.211         &       0.190         &       0.106         &       0.026         &       0.066         &       0.122\sym{*}  \\
                    &     (0.151)         &     (0.162)         &     (0.185)         &     (0.066)         &     (0.064)         &     (0.065)         \\
Effect at t=+5      &       0.179         &       0.166         &       0.202         &       0.084         &       0.122\sym{*}  &       0.029         \\
                    &     (0.135)         &     (0.141)         &     (0.181)         &     (0.069)         &     (0.066)         &     (0.059)         \\
R-sq                &       0.684         &       0.722         &       0.562         &       0.506         &       0.532         &       0.388         \\
Observations        &         635         &         635         &         635         &        4615         &        4615         &        4615         \\

\addlinespace[0.15cm]\addlinespace[0.15cm]\hline\hline
\multicolumn{7}{p{16cm}}{\footnotesizes{\textbf{Notes:} All columns report the estimates from Eq. (\ref{baseline}).  All models include municipality and year fixed effects, department $\times$ year fixed effects, and the following (lagged) covariates:  log of total population, proportion of the population living in a rural area, proportion of the population with unsatisfied basic needs, proportion of ethnic minority population and the homicide rate. Samples for regression models use data from 1996 to 2009. Robust standard errors (in parentheses) are clustered by municipality.} }
\end{tabular}
}
\end{center}
\end{table}


\begin{table}[H]
\begin{center}
{
\renewcommand{\arraystretch}{0.8}
\setlength{\tabcolsep}{6pt}
\caption {Effect of first nC church on the prob. of a conflict-related attack: heterogeneous effects by presence of coca crops in 2000}  \label{DIDPANYPGUEPPARcocain2000_tab}
\small
\vspace{-0.3cm}\centering  \begin{tabular}{lcccccc}
\hline\hline \addlinespace[0.15cm]
    & \multicolumn{6}{c}{Dep. variable = 1 if an attack by a non-state armed group in} \\ \cmidrule[0.2pt](l){2-7}    
                 & \multicolumn{3}{c}{Municipalities with presence}& \multicolumn{3}{c}{Municipalities without presence} \\
                                  & \multicolumn{3}{c}{of coca crops in 2000}& \multicolumn{3}{c}{of coca crops in 2000}\\ \cmidrule[0.2pt](l){2-4}\cmidrule[0.2pt](l){5-7}
& \multicolumn{1}{c}{any group}& \multicolumn{1}{c}{guerrilla} & \multicolumn{1}{c}{paramilitaries} & \multicolumn{1}{c}{any group}& \multicolumn{1}{c}{guerrilla} & \multicolumn{1}{c}{paramilitaries}  \\\cmidrule[0.2pt](l){2-2} \cmidrule[0.2pt](l){3-3} \cmidrule[0.2pt](l){4-4} \cmidrule[0.2pt](l){5-5} \cmidrule[0.2pt](l){6-6}  \cmidrule[0.2pt](l){7-7}  
    & (1)& (2) & (3)& (4) & (5)& (6)\\   \addlinespace[0.15cm]\hline\addlinespace[0.15cm]                                              
Effect at t=-3      &      -0.016         &       0.033         &      -0.084         &       0.002         &      -0.011         &      -0.022         \\
                    &     (0.074)         &     (0.085)         &     (0.106)         &     (0.040)         &     (0.037)         &     (0.035)         \\
Effect at t=-2      &      -0.037         &       0.018         &      -0.028         &      -0.055         &      -0.014         &      -0.079\sym{**} \\
                    &     (0.122)         &     (0.113)         &     (0.111)         &     (0.041)         &     (0.037)         &     (0.033)         \\
Effect at t=-1      &       0.000         &       0.000         &       0.000         &       0.000         &       0.000         &       0.000         \\
                    &         (.)         &         (.)         &         (.)         &         (.)         &         (.)         &         (.)         \\
Effect at t=0       &       0.045         &       0.175         &      -0.052         &       0.022         &       0.033         &      -0.080\sym{*}  \\
                    &     (0.131)         &     (0.122)         &     (0.104)         &     (0.071)         &     (0.065)         &     (0.049)         \\
Effect at t=+1      &       0.218\sym{*}  &       0.157\sym{*}  &       0.173         &      -0.038         &       0.015         &      -0.067         \\
                    &     (0.125)         &     (0.086)         &     (0.198)         &     (0.061)         &     (0.062)         &     (0.046)         \\
Effect at t=+2      &      -0.029         &       0.038         &       0.022         &      -0.106         &      -0.055         &      -0.080         \\
                    &     (0.144)         &     (0.142)         &     (0.209)         &     (0.068)         &     (0.065)         &     (0.056)         \\
Effect at t=+3      &       0.220\sym{*}  &       0.121         &       0.128         &      -0.056         &      -0.038         &      -0.059         \\
                    &     (0.131)         &     (0.149)         &     (0.163)         &     (0.071)         &     (0.063)         &     (0.066)         \\
Effect at t=+4      &       0.104         &       0.192         &      -0.084         &       0.045         &       0.112         &       0.040         \\
                    &     (0.126)         &     (0.139)         &     (0.208)         &     (0.067)         &     (0.075)         &     (0.060)         \\
Effect at t=+5      &      -0.036         &       0.022         &      -0.362\sym{***}&       0.046         &       0.151\sym{***}&      -0.069         \\
                    &     (0.085)         &     (0.102)         &     (0.135)         &     (0.056)         &     (0.055)         &     (0.053)         \\
R-sq                &       0.638         &       0.688         &       0.541         &       0.533         &       0.562         &       0.426         \\
Observations        &         547         &         547         &         547         &        2866         &        2866         &        2866         \\

\addlinespace[0.15cm]
\addlinespace[0.15cm]\hline\hline
\multicolumn{7}{p{16cm}}{\footnotesizes{\textbf{Notes:} All columns report the estimates from Eq. (\ref{baseline}).  All models include municipality and year fixed effects and department $\times$ year fixed effects, and  the following (lagged) covariates:  log of total population, proportion of the population living in a rural area, proportion of the population with unsatisfied basic needs, proportion of ethnic minority population and the homicide rate.  All models include 9 lags and 9 leads, normalized to the period prior to treatment. Samples for regression models use data from 2001 to 2009. Robust standard errors (in parentheses) are clustered by municipality.} }
\end{tabular}
}
\end{center}
\end{table}


\begin{figure}[H]
             \caption{Effect of first nC church on the prob. of a conflict-related assassination: heterogeneous effects by existence of cases of forced recruitment and by presence of coca crops}
        \label{dvsvictkillanyyearhetero_fig}
\begin{subfigure}{0.5\textwidth}
\caption{Forced recruitment before 1996} \label{dvsvictkillanyyearheteroA_fig}
\includegraphics[width=\linewidth]{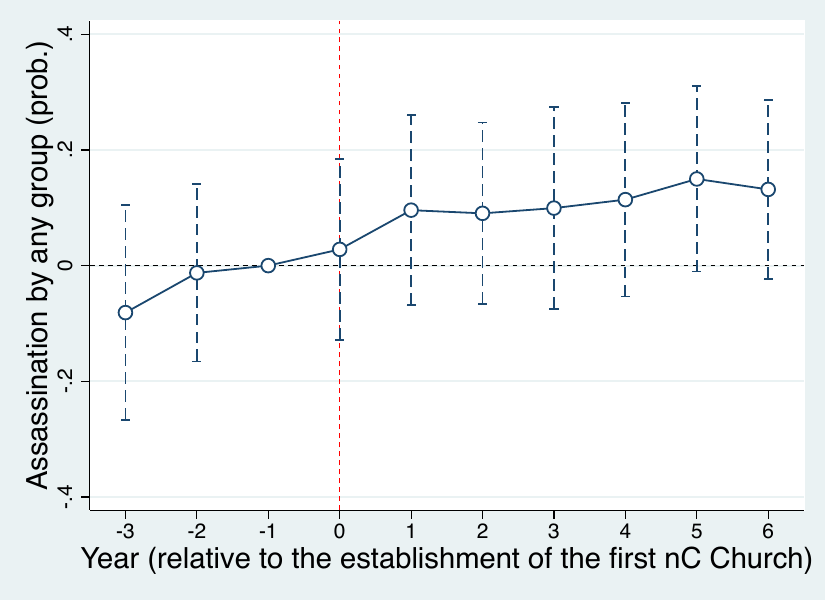}
\end{subfigure}\hspace*{\fill}
\begin{subfigure}{0.5\textwidth}
\caption{No forced recruitment before 1996} \label{dvsvictkillanyyearheteroB_fig}
\includegraphics[width=\linewidth]{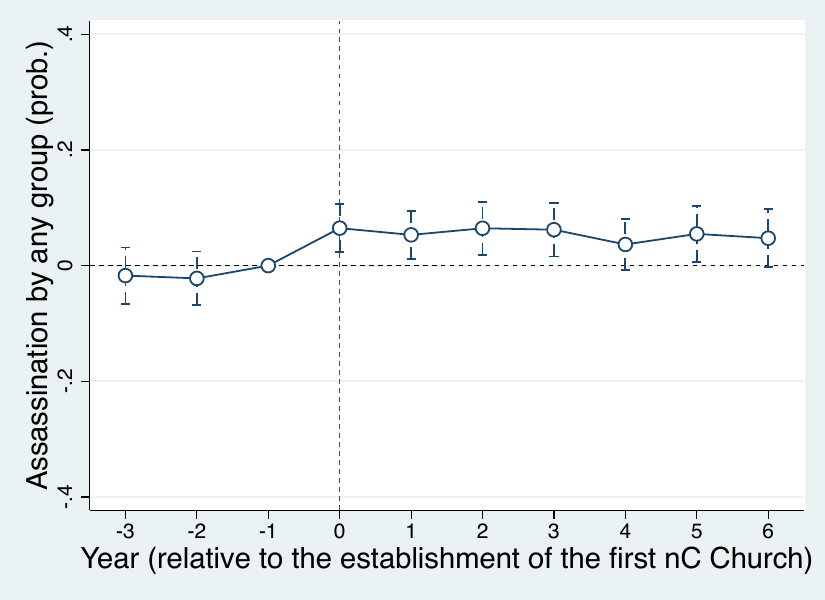}
\end{subfigure}

\medskip
\begin{subfigure}{0.5\textwidth}
\caption{Presence of coca crops in 2000} \label{dvsvictkillanyyearheteroC_fig}
\includegraphics[width=\linewidth]{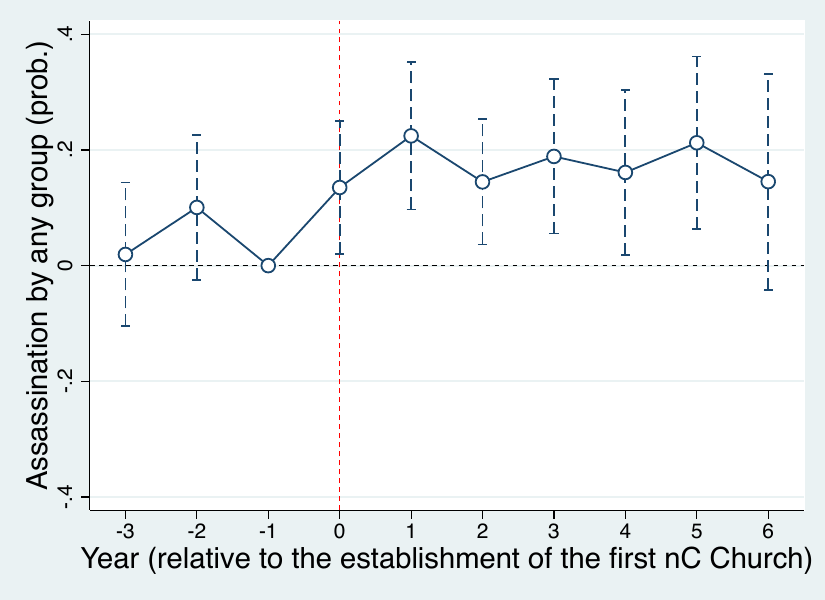}
\end{subfigure}\hspace*{\fill}
\begin{subfigure}{0.5\textwidth}
\caption{No presence of coca crops in 2000} \label{dvsvictkillanyyearheteroD_fig}
\includegraphics[width=\linewidth]{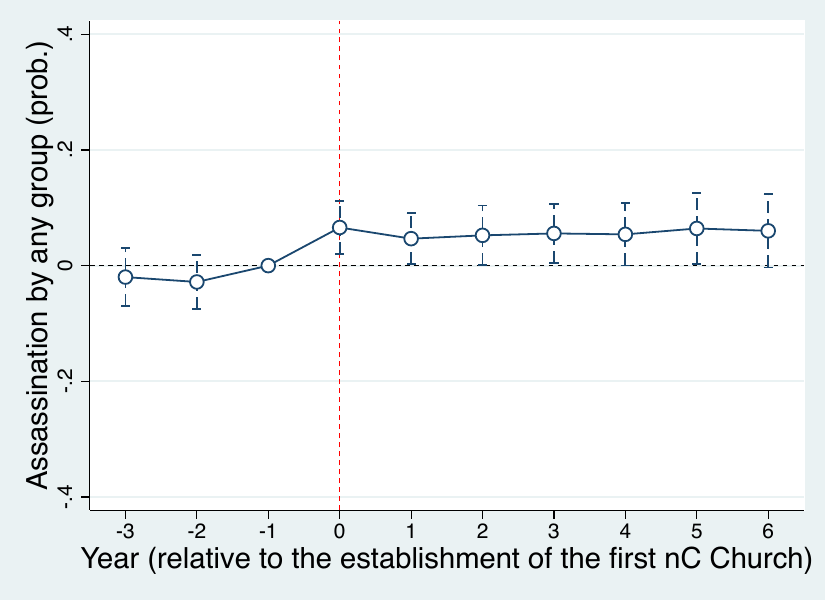}
\end{subfigure}
     \begin{minipage}{16cm} \footnotesizes These figures show the two-way fixed effects estimates from Eq. (\ref{baseline}), in a specification that includes municipality and year  fixed effects, department $\times$ year fixed effects, municipality-specific linear trends and the following (lagged) covariates:  log of total population, proportion of the population living in rural areas, proportion with unsatisfied basic needs (used as a proxy for poverty), proportion of ethnic minority population and homicide rate. Vertical lines indicate 90\% confidence intervals.
\end{minipage}
\end{figure}


\begin{table}[H]
\begin{center}
{
\renewcommand{\arraystretch}{0.8}
\setlength{\tabcolsep}{3pt}
\caption {Effect of first nC church on the prob. of a conflict-related assassination: heterogeneous effects by occurrence of a forced recruitment before 1996}  \label{DIDNCHMrecruit_tab}
\small
\vspace{-0.3cm}\centering  
\begin{tabular}{lcccccccc}
\hline\hline \addlinespace[0.15cm]
    & \multicolumn{8}{c}{Dep. variable = 1 if an assassination} \\ \cmidrule[0.2pt](l){2-9}    
    & (1)& (2) & (3)& (4) & (5)& (6)& (7)& (8)\\   \addlinespace[0.15cm]\hline\addlinespace[0.15cm]
               \multicolumn{1}{l}{\emph{\underline{Panel A}:}}    & \multicolumn{8}{c}{\emph{Dynamic} specification}\\\addlinespace[0.15cm]
                                & \multicolumn{4}{c}{Municipalities with at least one}& \multicolumn{4}{c}{Municipalities without any} \\
                                  & \multicolumn{4}{c}{case of forced recruitment bf 1996}& \multicolumn{4}{c}{case of forced recruitment bf 1996}\\ \cmidrule[0.2pt](l){2-5}\cmidrule[0.2pt](l){6-9}
 & \multicolumn{1}{c}{any}& \multicolumn{1}{c}{guerr.} & \multicolumn{1}{c}{param.}& \multicolumn{1}{c}{unknw.} & \multicolumn{1}{c}{any}& \multicolumn{1}{c}{guerr.} & \multicolumn{1}{c}{param.} & \multicolumn{1}{c}{unknw.}  \\\cmidrule[0.2pt](l){2-2} \cmidrule[0.2pt](l){3-3} \cmidrule[0.2pt](l){4-4} \cmidrule[0.2pt](l){5-5} \cmidrule[0.2pt](l){6-6}  \cmidrule[0.2pt](l){7-7}    \cmidrule[0.2pt](l){8-8}   \cmidrule[0.2pt](l){9-9}                                                
Effect at t=-3      &      -0.081         &      -0.087         &      -0.023         &      -0.057         &      -0.017         &      -0.013         &      -0.000         &      -0.043         \\
                    &     (0.110)         &     (0.101)         &     (0.048)         &     (0.107)         &     (0.029)         &     (0.024)         &     (0.020)         &     (0.030)         \\
\end{tabular}
}
\end{center}
\end{table}


\begin{table}[H]
\begin{center}
{
\renewcommand{\arraystretch}{0.8}
\setlength{\tabcolsep}{3pt}
\caption {Effect of first nC church on the prob. of a conflict-related assassination: heterogeneous effects by presence of coca crops in 2000}  \label{DIDNCHMcocain2000_tab}
\small
\vspace{-0.3cm}\centering  \begin{tabular}{lcccccccc}
\hline\hline \addlinespace[0.15cm]
    & \multicolumn{8}{c}{Dep. variable = 1 if an assassination} \\ \cmidrule[0.2pt](l){2-9}    
    & (1)& (2) & (3)& (4) & (5)& (6)& (7)& (8)\\   \addlinespace[0.15cm]\hline\addlinespace[0.15cm]
               \multicolumn{1}{l}{\emph{\underline{Panel A}:}}    & \multicolumn{8}{c}{\emph{Dynamic} specification}\\\addlinespace[0.15cm]
                  & \multicolumn{4}{c}{Municipalities with presence}& \multicolumn{4}{c}{Municipalities without presence} \\
                                  & \multicolumn{4}{c}{of coca crops in 2000}& \multicolumn{4}{c}{of coca crops in 2000}\\ \cmidrule[0.2pt](l){2-5}\cmidrule[0.2pt](l){6-9}
 & \multicolumn{1}{c}{any}& \multicolumn{1}{c}{guerr.} & \multicolumn{1}{c}{param.}& \multicolumn{1}{c}{unknw.} & \multicolumn{1}{c}{any}& \multicolumn{1}{c}{guerr.} & \multicolumn{1}{c}{param.} & \multicolumn{1}{c}{unknw.}  \\\cmidrule[0.2pt](l){2-2} \cmidrule[0.2pt](l){3-3} \cmidrule[0.2pt](l){4-4} \cmidrule[0.2pt](l){5-5} \cmidrule[0.2pt](l){6-6}  \cmidrule[0.2pt](l){7-7}    \cmidrule[0.2pt](l){8-8}   \cmidrule[0.2pt](l){9-9}                                                 
Effect at t=-3      &       0.019         &       0.021         &      -0.009         &       0.030         &      -0.020         &       0.010         &       0.004         &      -0.036         \\
                    &     (0.074)         &     (0.079)         &     (0.043)         &     (0.063)         &     (0.030)         &     (0.025)         &     (0.017)         &     (0.032)         \\
\end{tabular}
}
\end{center}
\end{table}


\begin{table}[H]
\begin{center}
{
\renewcommand{\arraystretch}{0.8}
\setlength{\tabcolsep}{6pt}
\caption {Effect of first non-Catholic church on  the prob. of a conflict-related attack: heterogeneous effects by the number of Catholic churches in 1995}  \label{DIDPANYPGUEPPARdpeinglecFS_tab}
\small
\vspace{-0.3cm}\centering  \begin{tabular}{lcccccc}
\hline\hline \addlinespace[0.15cm]
    & \multicolumn{6}{c}{Dep. variable = 1 if an attack by a non-state armed group} \\ \cmidrule[0.2pt](l){2-7}    
    & (1)& (2) & (3)& (4) & (5)& (6)\\   \addlinespace[0.15cm]\hline\addlinespace[0.15cm]
                 & \multicolumn{3}{c}{Municipalities with}& \multicolumn{3}{c}{Municipalities with} \\     
                 & \multicolumn{3}{c}{above-median number of per}& \multicolumn{3}{c}{below-median number of per} \\
                                  & \multicolumn{3}{c}{capita Catholic churches in 1995}& \multicolumn{3}{c}{capita Catholic churches in 1995}\\ \cmidrule[0.2pt](l){2-4}\cmidrule[0.2pt](l){5-7}
& \multicolumn{1}{c}{any group}& \multicolumn{1}{c}{guerrilla} & \multicolumn{1}{c}{paramilitaries} & \multicolumn{1}{c}{any group}& \multicolumn{1}{c}{guerrilla} & \multicolumn{1}{c}{paramilitaries}  \\\cmidrule[0.2pt](l){2-2} \cmidrule[0.2pt](l){3-3} \cmidrule[0.2pt](l){4-4} \cmidrule[0.2pt](l){5-5} \cmidrule[0.2pt](l){6-6}  \cmidrule[0.2pt](l){7-7}                                                
Effect at t=-3      &       0.060         &       0.051         &       0.036         &       0.008         &      -0.058         &       0.045         \\
                    &     (0.065)         &     (0.054)         &     (0.050)         &     (0.065)         &     (0.062)         &     (0.055)         \\
\end{tabular}
}
\end{center}
\end{table}


\begin{table}[H]
\begin{center}
{
\renewcommand{\arraystretch}{0.8}
\setlength{\tabcolsep}{6pt}
\caption {Effect of first non-Catholic church on the prob. of a conflict-related attack: heterogeneous effects by Conservative vote share before 1996}  \label{DIDPANYPGUEPPARdpropconvote_tab}
\small
\vspace{-0.3cm}\centering  \begin{tabular}{lcccccc}
\hline\hline \addlinespace[0.15cm]
    & \multicolumn{6}{c}{Dep. variable = 1 if an attack by a non-state armed group} \\ \cmidrule[0.2pt](l){2-7}    
    & (1)& (2) & (3)& (4) & (5)& (6)\\   \addlinespace[0.15cm]\hline\addlinespace[0.15cm]
    \addlinespace[0.15cm]
                 & \multicolumn{3}{c}{Municipalities with}& \multicolumn{3}{c}{Municipalities with} \\     
                 & \multicolumn{3}{c}{above-median vote share}& \multicolumn{3}{c}{below-median vote share} \\
                                  & \multicolumn{3}{c}{for Conservative party bf 1996}& \multicolumn{3}{c}{for Conservative party bf 1996}\\ \cmidrule[0.2pt](l){2-4}\cmidrule[0.2pt](l){5-7}
& \multicolumn{1}{c}{any group}& \multicolumn{1}{c}{guerrilla} & \multicolumn{1}{c}{paramilitaries} & \multicolumn{1}{c}{any group}& \multicolumn{1}{c}{guerrilla} & \multicolumn{1}{c}{paramilitaries}  \\\cmidrule[0.2pt](l){2-2} \cmidrule[0.2pt](l){3-3} \cmidrule[0.2pt](l){4-4} \cmidrule[0.2pt](l){5-5} \cmidrule[0.2pt](l){6-6}  \cmidrule[0.2pt](l){7-7}                                                
Effect at t=-3      &       0.142\sym{*}  &       0.073         &       0.086         &      -0.012         &      -0.050         &       0.012         \\
                    &     (0.076)         &     (0.073)         &     (0.061)         &     (0.058)         &     (0.051)         &     (0.047)         \\
\end{tabular}
}
\end{center}
\end{table}


\begin{table}[H]
\begin{center}
\renewcommand{\arraystretch}{0.7}
\setlength{\tabcolsep}{10pt}
\small
\caption{Left-wing political parties}\label{tableleftwingparties}
\vspace{-0.2cm}
\begin{tabular}{|c|c|c|c|}
\hline\hline\addlinespace[0.15cm]
1&\emph{Union Patriotica} &9&\emph{Frente Social y Politico} \\
2&\emph{Polo Democratico Alternativo} &10&\emph{Movimiento 19 de abril} \\
3&\emph{Polo Democratico Independiente} &11&\emph{Socialdemocrata Colombiano} \\
4&\emph{Partido Comunista} &12&\emph{Independiente Frente de Esperanza} \\
5&\emph{Alianza Nacional Popular ANAPO} &13&\emph{Movimiento Ciudadano} \\
6&\emph{Alianza Democratica M19} &14&\emph{Alternativa Democratica} \\
7&\emph{MOIR} &15&\emph{Unidad Democratica} \\
8&\emph{Frente Social y Politico} &16& \emph{Vamos Ipiales}\\
 \addlinespace[0.15cm]
\hline   \hline
\multicolumn{4}{p{16cm}}{\footnotesizes \textbf{Notes}: The classification of left-wing parties follows \cite{FergussonQuerubinVargas2017},  and adds the \emph{Partido Comunista} and excludes the \emph{Autoridades Indigenas de Colombia}, which may be described as an ethnic party \cite[see][]{Laurent2010}.}
\end{tabular}
\end{center}
\end{table}


\begin{table}[H]
\begin{center}
{
\renewcommand{\arraystretch}{0.7}
\setlength{\tabcolsep}{6pt}
\caption {Effect of first non-Catholic church on prob. of an attack:  heterogeneous effects by left-wing vote share before 1996}  \label{DIDPANYPGUEPPARdpropleftvote_tab}
\small
\vspace{-0.3cm}\centering  \begin{tabular}{lcccccc}
\hline\hline \addlinespace[0.15cm]
    & \multicolumn{6}{c}{Dep. variable = 1 if an attack by a non-state armed group} \\ \cmidrule[0.2pt](l){2-7}    
    & (1)& (2) & (3)& (4) & (5)& (6)\\   \addlinespace[0.15cm]\hline\addlinespace[0.15cm]
           \multicolumn{1}{l}{\emph{\underline{Panel A}:}}    & \multicolumn{6}{c}{\emph{Dynamic} specification}\\\addlinespace[0.15cm]
                 & \multicolumn{3}{c}{Municipalities with}& \multicolumn{3}{c}{Municipalities with} \\     
                 & \multicolumn{3}{c}{above-median vote share}& \multicolumn{3}{c}{below-median vote share} \\
                                  & \multicolumn{3}{c}{for a left-wing party bf 1996}& \multicolumn{3}{c}{for a left-wing party bf 1996}\\ \cmidrule[0.2pt](l){2-4}\cmidrule[0.2pt](l){5-7}
& \multicolumn{1}{c}{any group}& \multicolumn{1}{c}{guerrilla} & \multicolumn{1}{c}{paramilitaries} & \multicolumn{1}{c}{any group}& \multicolumn{1}{c}{guerrilla} & \multicolumn{1}{c}{paramilitaries}  \\\cmidrule[0.2pt](l){2-2} \cmidrule[0.2pt](l){3-3} \cmidrule[0.2pt](l){4-4} \cmidrule[0.2pt](l){5-5} \cmidrule[0.2pt](l){6-6}  \cmidrule[0.2pt](l){7-7}                                                
Effect at t=-3      &       0.272         &       0.094         &       0.207         &       0.034         &      -0.013         &       0.044         \\
\end{tabular}
}
\end{center}
\end{table}


\begin{table}[H]
\renewcommand{\arraystretch}{0.7}
\setlength{\tabcolsep}{15pt}
\caption {Effect of first non-Catholic church on election outcomes}  \label{nextelectoral_tab}
\small
\centering  \begin{tabular}{lccc}
\hline\hline \addlinespace[0.15cm]
    & \multicolumn{3}{c}{Dep. variable = vote share in next election for}  \\\cmidrule[0.2pt](l){2-4}
                 & \multicolumn{1}{c}{Liberal party}& \multicolumn{1}{c}{Conservative party}& \multicolumn{1}{c}{Leftwing party}    \\\cmidrule[0.2pt](l){2-2}\cmidrule[0.2pt](l){3-3}\cmidrule[0.2pt](l){4-4}
& (1)& (2) & (3)\\   \addlinespace[0.15cm] \hline \addlinespace[0.15cm]
            \multicolumn{1}{l}{\emph{\underline{Panel A}:}}            & \multicolumn{3}{c}{\emph{Dynamic} specification}\\\addlinespace[0.15cm]
Effect at t=-3      &      -0.008         &      -0.010         &      -0.006         \\
                    &     (0.021)         &     (0.016)         &     (0.008)         \\
Effect at t=-2      &       0.004         &       0.011         &       0.003         \\
                    &     (0.016)         &     (0.013)         &     (0.006)         \\
\end{tabular}
\end{table}


\begin{table}[H]
\begin{center}
{
\renewcommand{\arraystretch}{0.8}
\setlength{\tabcolsep}{6pt}
\caption {Effect of first nC church on the prob. of a guerrilla attack: heterogeneous effects by ethnic minority population in 1993}  \label{PANYPGUEPPARindigblacktotprop_tab}
\small
\vspace{-0.3cm}\centering  \begin{tabular}{lcccccc}
\hline\hline \addlinespace[0.15cm]
    & \multicolumn{6}{c}{Dep. variable = 1 if an attack by a non-state armed group} \\ \cmidrule[0.2pt](l){2-7}    
    & (1)& (2) & (3)& (4) & (5)& (6)\\   \addlinespace[0.15cm]\hline\addlinespace[0.15cm]
                            & \multicolumn{3}{c}{Municipalities}& \multicolumn{3}{c}{Municipalities} \\
                 & \multicolumn{3}{c}{with above-median}& \multicolumn{3}{c}{with below-median} \\
                 
                 & \multicolumn{3}{c}{share of ethnic minorities}& \multicolumn{3}{c}{share of ethnic minorities} \\ \cmidrule[0.2pt](l){2-4}\cmidrule[0.2pt](l){5-7}
& \multicolumn{1}{c}{any group}& \multicolumn{1}{c}{guerrilla} & \multicolumn{1}{c}{paramilitaries} & \multicolumn{1}{c}{any group}& \multicolumn{1}{c}{guerrilla} & \multicolumn{1}{c}{paramilitaries}  \\\cmidrule[0.2pt](l){2-2} \cmidrule[0.2pt](l){3-3} \cmidrule[0.2pt](l){4-4} \cmidrule[0.2pt](l){5-5} \cmidrule[0.2pt](l){6-6}  \cmidrule[0.2pt](l){7-7}                                                
Effect at t=-3      &       0.017         &       0.007         &       0.019         &       0.029         &      -0.001         &       0.039         \\
                    &     (0.056)         &     (0.053)         &     (0.051)         &     (0.049)         &     (0.045)         &     (0.036)         \\
\end{tabular}
}
\end{center}
\end{table}


\begin{table}[H]
\begin{center}
{
\renewcommand{\arraystretch}{0.7}
\setlength{\tabcolsep}{7pt}
\caption {Effect of first nC church on the prob. of a guerrilla attack: non-affiliated nC churches}  \label{PGUEPANYPGUEPPARspe_tab}
\small
\centering  \begin{tabular}{lcccccc}
\hline\hline \addlinespace[0.15cm]
    & \multicolumn{6}{c}{Dep. variable = 1 if an attack by} \\\addlinespace[0.15cm]
    & \multicolumn{2}{c}{Any group} & \multicolumn{2}{c}{Guerrilla} & \multicolumn{2}{c}{Paramilitaries} \\\cmidrule[0.2pt](l){2-3}\cmidrule[0.2pt](l){4-5}\cmidrule[0.2pt](l){6-7}
& (1)& (2) & (3) & (4) & (5) & (6)  \\   \addlinespace[0.15cm] \hline \addlinespace[0.15cm]
            \multicolumn{1}{l}{\emph{\underline{Panel A}:}}            & \multicolumn{6}{c}{\emph{Dynamic} specification}\\\addlinespace[0.15cm]
Effect at t=-3      &       0.029         &       0.031         &       0.019         &       0.016         &       0.025         &       0.020         \\
                    &     (0.054)         &     (0.055)         &     (0.047)         &     (0.048)         &     (0.043)         &     (0.044)         \\
Effect at t=-2      &       0.010         &       0.008         &       0.004         &       0.001         &       0.009         &       0.002         \\
                    &     (0.053)         &     (0.055)         &     (0.047)         &     (0.049)         &     (0.046)         &     (0.048)         \\
\end{tabular}
}
\end{center}
\end{table}


\begin{table}[H]
\begin{center}
{
\renewcommand{\arraystretch}{0.7}
\setlength{\tabcolsep}{7pt}
\caption {Effect of first nC church on the prob. of a guerrilla attack: affiliated churches}  \label{PGUEPANYPGUEPPARext_tab}
\small
\centering  \begin{tabular}{lcccccc}
\hline\hline \addlinespace[0.15cm]
    & \multicolumn{6}{c}{Dep. variable = 1 if an attack by} \\\addlinespace[0.15cm]
    & \multicolumn{2}{c}{Any group} & \multicolumn{2}{c}{Guerrilla} & \multicolumn{2}{c}{Paramilitaries} \\\cmidrule[0.2pt](l){2-3}\cmidrule[0.2pt](l){4-5}\cmidrule[0.2pt](l){6-7}
& (1)& (2) & (3) & (4) & (5) & (6)  \\   \addlinespace[0.15cm] \hline \addlinespace[0.15cm]
            \multicolumn{1}{l}{\emph{\underline{Panel A}:}}            & \multicolumn{6}{c}{\emph{Dynamic} specification}\\\addlinespace[0.15cm]
 Effect at t=-3      &      -0.012         &      -0.018         &      -0.001         &      -0.008         &       0.026         &       0.029         \\
                    &     (0.058)         &     (0.058)         &     (0.055)         &     (0.055)         &     (0.062)         &     (0.063)         \\
Effect at t=-2      &      -0.049         &      -0.044         &      -0.026         &      -0.020         &       0.029         &       0.030         \\
                    &     (0.048)         &     (0.048)         &     (0.047)         &     (0.047)         &     (0.048)         &     (0.048)         \\
\end{tabular}
}
\end{center}
\end{table}


\hbox {}
\hbox {} \newpage
\subsection{Determinants of first non-Catholic church}\label{appdeterminants}
\small

As mentioned in footnote \ref{ftdetfirshistorical}, the results in Table  \ref{dfirstiglesiasnc_tab} do not shed any light on why a non-Catholic church establishes in a municipality. In this section we propose an explanation that is consistent with the results in Table  \ref{dfirstiglesiasnc_tab}, and provide evidence in its favor. Specifically, we hypothesize that urbanization is still relevant, but only in places where, for historical reasons, other channels for social (or political) action are not available.

To examine the plausibility of this hypothesis, we exploit a key characteristic of the Colombian party system:  its historical domination by two political parties (Liberal and Conservative), which maintained a duopoly of power at all levels of government for more than a century, and that only in the 2000s started to lose out to other contenders. In this context, we propose that during the period we focus on, i) in places where the two traditional parties had been strong, there were fewer alternatives for political action, and ii)  in places where the two traditional parties had been weak, there were more alternatives for political action. We claim that in the first group of municipalities, non-Catholic churches were expected to be more successful in exploiting new desires for hope, and in the second group of municipalities, non-Catholic churches were expected to use other strategies (such as focusing on other more traditional challenges that may also result in hope, e.g. poverty).

Table \ref{dfirstiglesiasnchist_tab} below examines the plausibility of this hypothesis by using a specification similar to that in column (5) of Table \ref{dfirstiglesiasnc_tab}  (which we repeat for the sake of comparison in column (1)), but distinguishes between those municipalities where the historical level of support for the two traditional parties  (as measured by their vote share in mayoral elections before 1995) is either above the median (column (2)) or below the median (column (3)). Consistent with our hypothesis, the estimates for column (2) show that in municipalities with high historical support for the two traditional parties, a lower proportion of the population living in rural areas makes the establishment of the first non-Catholic church more likely. In addition, the estimates in column (3) show that in municipalities with low historical support for the two traditional parties, the level of poverty (and rurality) is positively correlated with the establishment of the first non-Catholic church. Finally, note that column (3) also shows a negative correlation between the occurrence of a previous guerrilla attack and the establishment of the first non-Catholic church; an explanation may be that in municipalities with low historical support for the two traditional parties, a new religious organization may also compete with other new (political) organizations, and this may discourage religious leaders from establishing roots there, since they may expect a stronger reaction from the guerrillas who were already present in the municipality.

The results in Table \ref{dfirstiglesiasnchist_tab} allow us to argue that in our main specification, we account for the main factors that plausibly explain the success of non-Catholic churches in Colombia: a historical characteristic, captured by municipality fixed effects and whose evolution may be captured by the municipality-specific trends, and the proportion of the population living in rural areas, the level of poverty and attacks by guerrillas, which we include as controls.

\begin{table}[H]
\begin{center}
{
\renewcommand{\arraystretch}{1}
\setlength{\tabcolsep}{3pt}
\caption {Determinants of first non-Catholic church (by historical vote for Liberal and Conservative parties (i.e. traditional parties)}  \label{dfirstiglesiasnchist_tab}
\small
\centering  \begin{tabular}{lcccccc}
\hline\hline \addlinespace[0.15cm]
    & \multicolumn{3}{c}{Dep. variable = 1 when first} \\
    & \multicolumn{3}{c}{non-Catholic church is established in} \\\cmidrule[0.2pt](l){2-4}
    & \multicolumn{1}{c}{} & \multicolumn{1}{c}{Municipalities} & \multicolumn{1}{c}{Municipalities}\\
            & \multicolumn{1}{c}{} & \multicolumn{1}{c}{with historically} & \multicolumn{1}{c}{with historically}\\
        & \multicolumn{1}{c}{Any} & \multicolumn{1}{c}{high support for} & \multicolumn{1}{c}{low support for}\\
       & \multicolumn{1}{c}{municipality} & \multicolumn{1}{c}{traditional parties} & \multicolumn{1}{c}{traditional parties}\\ \cmidrule[0.2pt](l){2-2}\cmidrule[0.2pt](l){3-3}\cmidrule[0.2pt](l){4-4}
& (1)& (2) & (3)   \\    \addlinespace[0.15cm] \hline \addlinespace[0.15cm]
Attack by guerrilla (prob.) at t-1&      -0.001         &       0.009         &      -0.014         \\
                    &     (0.008)         &     (0.011)         &     (0.011)         \\
Attack by paramilitaries (prob.) at t-1&      -0.003         &      -0.014         &       0.012         \\
                    &     (0.007)         &     (0.010)         &     (0.010)         \\
Internally displaced pop. (outflow, rate) at t-1&       0.000         &       0.000\sym{**} &      -0.000         \\
                    &     (0.000)         &     (0.000)         &     (0.000)         \\
Internally displaced pop. (inflow, rate) at t-1&       0.000         &      -0.000         &       0.000         \\
                    &     (0.000)         &     (0.000)         &     (0.000)         \\
Desertion from the guerrilla (prob.) at t-1&      -0.005         &       0.010         &      -0.015         \\
                    &     (0.008)         &     (0.013)         &     (0.011)         \\
Forced recruitment (prob.) at t-1&       0.003         &      -0.000         &       0.008         \\
                    &     (0.009)         &     (0.013)         &     (0.012)         \\
Log of total population at t-1&      -0.203\sym{*}  &       0.004         &      -0.386\sym{**} \\
                    &     (0.118)         &     (0.162)         &     (0.163)         \\
Prop. of rural population at t-1&       0.012         &      -0.100\sym{**} &       0.058\sym{*}  \\
                    &     (0.026)         &     (0.050)         &     (0.030)         \\
Unsatisfied Basic Needs at t-1&       0.000         &      -0.000         &       0.002\sym{**} \\
                    &     (0.001)         &     (0.001)         &     (0.001)         \\
Ethinic minority population at t-1&      -0.020         &       0.002         &      -0.013         \\
                    &     (0.034)         &     (0.045)         &     (0.057)         \\
Homicide (rate) at t-1&       0.000         &       0.000         &      -0.000         \\
                    &     (0.000)         &     (0.000)         &     (0.000)         \\
R-sq                &       0.258         &       0.326         &       0.282         \\
Observations        &        5313         &        2677         &        2537         \\

\addlinespace[0.15cm]\hline\hline
\multicolumn{4}{p{16cm}}{\footnotesize{\textbf{Notes:} All models include municipality, year and department $\times$ year fixed effects, as well as  municipality-specific linear trends.  Samples for regression models include data from 1996 to 2009.   Robust standard errors (in parentheses) are clustered by municipality. * denotes statistically significant estimates at 10\%, ** denotes significant at 5\% and *** denotes significant at 1\%.} }
\end{tabular}
}
\end{center}
\end{table}


\hbox {}
\hbox {} \newpage

\bibliographystyle{ecca}
\bibliography{bibrelandconf}

\begin{thebibliography}{72}
\providecommand{\natexlab}[1]{#1}

\bibitem[{{Abraham} and {Sun}(2018)}]{AbrahamSun2018}
\textsc{{Abraham}, S.} and \textsc{{Sun}, L.} (2018). {Estimating Dynamic
  Treatment Effects in Event Studies with Heterogeneous Treatment Effects}.
  \textit{ArXiv e-prints}.

\bibitem[{Abraham and Sun(2019)}]{abraham2019estimating}
\textsc{Abraham, S.} and \textsc{Sun, L.} (2019). Estimating dynamic treatment
  effects in event studies with heterogeneous treatment effects. \textit{MIT
  working paper}.

\bibitem[{Appleby(2000)}]{Appleby2000}
\textsc{Appleby, R.~S.} (2000). \textit{The Ambivalence of the Sacred:
  Religion, Violence, and Reconciliation}. Rowman and Littlefield: Lahman, MD.

\bibitem[{Barros and Garoupa(2002)}]{BarrosGaroupa2002}
\textsc{Barros, P.~P.} and \textsc{Garoupa, N.} (2002). An economic theory of
  church strictness. \textit{The Economic Journal}, \textbf{112}.

\bibitem[{Basedau(2011)}]{Basedau2011}
\textsc{Basedau, G. V. J. W.~T., Matthias;~Strüver} (2011). Do religious
  factors impact armed conflict? empirical evidence from sub-saharan africa.
  \textit{Terrorism and Political Violence}, \textbf{23}.

\bibitem[{Bastian(2005)}]{Bastian2005}
\textsc{Bastian, J.~P.} (2005). Pentecostalismos lationoamericanos. In A.~M.
  Bidega\'in and J.~Demera (eds.), \textit{Globalizaci\'on y diversidad
  religiosa en Colombia}, Universidad Nacional de Colombia.

\bibitem[{Becker and Pascali(2019)}]{BeckerPascali2019}
\textsc{Becker, S.~O.} and \textsc{Pascali, L.} (2019). Religion, division of
  labor, and conflict: Anti-semitism in germany over 600 years.
  \textit{American Economic Review}, \textbf{109}~(5), 1764--1804.

\bibitem[{Beltr\'an(2013)}]{Beltran2013}
\textsc{Beltr\'an, W.} (2013). \textit{Del monopolio cat\'olico a la
  explosi\'on pentecostal Pluralizaci\'on religiosa, secularizaci\'on y cambio
  social en Colombia}. Universidad Nacional de Colombia.

\bibitem[{Blattman and Miguel(2010)}]{BlattmanMiguel2010}
\textsc{Blattman, C.} and \textsc{Miguel, E.} (2010). Civil war.
  \textit{Journal of Economic Literature}, \textbf{48}~(1), 3--57.

\bibitem[{Bormann \textit{et~al.}(2015)Bormann, Cederman and
  Vogt}]{BormannCedermanVogt2015}
\textsc{Bormann, N.}, \textsc{Cederman, L.} and \textsc{Vogt, M.} (2015).
  Language, religion, and ethnic civil war. \textit{Journal of Conflict
  Resolution}.

\bibitem[{Borusyak and Jaravel(2017)}]{BorusyakJaravel2017}
\textsc{Borusyak, K.} and \textsc{Jaravel, X.} (2017). Revisiting event study
  designs. Working Paper, Harvard University, May 2017.

\bibitem[{{Callaway} and {Sant'Anna}(2018)}]{CallawaySantAnna2018}
\textsc{{Callaway}, B.} and \textsc{{Sant'Anna}, P.~H.~C.} (2018).
  {Difference-in-Differences with Multiple Time Periods and an Application on
  the Minimum Wage and Employment}. \textit{ArXiv e-prints}.

\bibitem[{Campante and Yanagizawa-Drott(2015)}]{YanagizawaDrotCampante2015}
\textsc{Campante, F.} and \textsc{Yanagizawa-Drott, D.} (2015). { Does Religion
  Affect Economic Growth and Happiness? Evidence from Ramadan *}. \textit{The
  Quarterly Journal of Economics}, \textbf{130}~(2), 615--658.

\bibitem[{Cantoni(2015)}]{Cantoni2015}
\textsc{Cantoni, D.} (2015). {The Economic Effects Of The Protestant
  Reformation: Testing The Weber Hypothesis In The German Lands}.
  \textit{Journal of the European Economic Association}, \textbf{13}~(4),
  561--598.

\bibitem[{Cantoni \textit{et~al.}(2018)Cantoni, Dittmar and
  Yuchtman}]{CantoniDittmarYuchtman2018}
\textsc{---}, \textsc{Dittmar, J.} and \textsc{Yuchtman, N.} (2018). {Religious
  Competition and Reallocation: the Political Economy of Secularization in the
  Protestant Reformation*}. \textit{The Quarterly Journal of Economics},
  \textbf{133}~(4), 2037--2096.

\bibitem[{{de Chaisemartin} and
  {D'Haultf{\oe}uille}(2018)}]{deChaisemartinDHaultfoeuille2018}
\textsc{{de Chaisemartin}, C.} and \textsc{{D'Haultf{\oe}uille}, X.} (2018).
  {Two-way fixed effects estimators with heterogeneous treatment effects}.
  \textit{ArXiv e-prints}.

\bibitem[{Deas(2015)}]{Deas2015}
\textsc{Deas, M.} (2015). The colombian conflict: A historical perspective. In
  B.~M. Bagley and J.~D. Rosen (eds.), \textit{Colombia's Political Economy at
  the Outset of the Twenty-First Century: From Uribe to Santos and Beyond},
  Lexington Books.

\bibitem[{Dube and Vargas(2013)}]{DubeVargas2013}
\textsc{Dube, O.} and \textsc{Vargas, J.~F.} (2013). {Commodity Price Shocks
  and Civil Conflict: Evidence from Colombia}. \textit{Review of Economic
  Studies}, \textbf{80}~(4), 1384--1421.

\bibitem[{{El tiempo}(2009)}]{eltiempoJUL082019}
\textsc{{El tiempo}} (2009). Las farc aumentan el reclutamiento de menores para
  sustituir desertores.
  \url{https://www.eltiempo.com/archivo/documento/CMS-5599854}, online;
  consulted 17 Jan 2019.

\bibitem[{Esteban \textit{et~al.}(2012)Esteban, Mayoral and
  Ray}]{EstebanMayoralRay2012AER}
\textsc{Esteban, J.}, \textsc{Mayoral, L.} and \textsc{Ray, D.} (2012).
  Ethnicity and conflict: An empirical study. \textit{American Economic
  Review}, \textbf{102}~(4), 1310\'{\i}1342.

\bibitem[{Eswaran(2011)}]{Eswaran2011}
\textsc{Eswaran, M.} (2011). Competition and performance in the marketplace for
  religion: A theoretical perspective. \textit{The B E Journal of Economic
  Analysis and Policy}, \textbf{11}.

\bibitem[{Fearon and Laitin(2003)}]{FearonLaitin2003}
\textsc{Fearon, J.~D.} and \textsc{Laitin, D.~D.} (2003). Ethnicity,
  insurgency, and civil war. \textit{The American Political Science Review},
  \textbf{97}~(1), pp. 75--90.

\bibitem[{Fergusson \textit{et~al.}(2017)Fergusson, Querubín, Ruiz and
  Vargas}]{FergussonQuerubinVargas2017}
\textsc{Fergusson, L.}, \textsc{Querubín, P.}, \textsc{Ruiz, N.~A.} and
  \textsc{Vargas, J.~F.} (2017). \textit{{The Real Winner's Curse}}. Documentos
  CEDE 015279, Universidad de los Andes - CEDE.

\bibitem[{Fernandez and Fogli(2009)}]{FernandezFogli2009}
\textsc{Fernandez, R.} and \textsc{Fogli, A.} (2009). {Culture: An Empirical
  Investigation of Beliefs, Work, and Fertility}. \textit{American Economic
  Journal: Macroeconomics}, \textbf{1}~(1), 146--177.

\bibitem[{Fox(1999)}]{Fox1999}
\textsc{Fox, J.} (1999). Do religious institutions support violence or the
  status quo? \textit{Studies in Conflict \& Terrorism}, \textbf{22}~(2),
  119--139.

\bibitem[{Fox(2001)}]{Fox2001}
\textsc{---} (2001). {Are Middle East Conflicts More Religious?} \textit{Middle
  East Quarterly}, \textbf{8}~(4), 31--40.

\bibitem[{Galtung(1996)}]{Galtung1996}
\textsc{Galtung, J.} (1996). \textit{Peace by Peaceful Means: Peace and
  Conflict, Development and Civilization}.

\bibitem[{Gill(1998)}]{Gill1998}
\textsc{Gill, A.} (1998). \textit{Rendering Unto Caesar: The Catholic Church
  and the State in Latin America}. Chicago, The University of Chicago Press.

\bibitem[{Giuliano(2007)}]{Giuliano2007}
\textsc{Giuliano, P.} (2007). Living arrangements in western europe: Does
  cultural origin matter? \textit{Journal of the European Economic
  Association}, \textbf{5}~(5), 927--952.

\bibitem[{GMH(2013{\natexlab{a}})}]{GMH2013}
\textsc{GMH} (2013{\natexlab{a}}). \textit{Basta ya! Colombia: memorias de
  guerra y dignidad. Informe General}. Grupo de Memoria Historica; Imprenta
  Nacional.

\bibitem[{GMH(2013{\natexlab{b}})}]{GMHFARC2013}
\textsc{---} (2013{\natexlab{b}}). \textit{Guerrilla y Poblaci\'on Civil.
  Trayectoria de las FARC 1949-2013}. Grupo de Memoria Historica, Imprenta
  Nacional.

\bibitem[{Gonzalez(2005)}]{Gonzalez2005}
\textsc{Gonzalez, F.} (2005). Iglesia cat\'olica y conflicto en colombia: De la
  lucha contra la modernidad a los di\'alogos de paz. \textit{Revista
  Controversia}, \textbf{184}~(71).

\bibitem[{Goodman-Bacon(2018)}]{GoodmanBacon2018}
\textsc{Goodman-Bacon, A.} (2018). \textit{Difference-in-Differences with
  Variation in Treatment Timing}. Working Paper 25018, National Bureau of
  Economic Research.

\bibitem[{Grosfeld \textit{et~al.}(2019)Grosfeld, Sakalli and
  Zhuravskaya}]{GrosfeldSakalliSeyhunZhuravskaya2019}
\textsc{Grosfeld, I.}, \textsc{Sakalli, S.~O.} and \textsc{Zhuravskaya, E.}
  (2019). {Middleman Minorities and Ethnic Violence: Anti-Jewish Pogroms in the
  Russian Empire}. \textit{The Review of Economic Studies}, \textbf{87}~(1),
  289--342.

\bibitem[{Gutierrez and Baron(2005)}]{GutierrezBaron2005}
\textsc{Gutierrez, F.} and \textsc{Baron, M.} (2005). \textit{Re-stating the
  State: Paramilitary territorial control and political order in Colombia,
  1978-2004}. Working paper 66, Development Research Center, LSE, [Online;
  accessed 7 July 2015].

\bibitem[{Haynes(2009)}]{Haynes2009}
\textsc{Haynes, J.} (2009). Conflict, conflict resolution and peace-building:
  The role of religion in mozambique, nigeria and cambodia.
  \textit{Commonwealth and Comparative Politics}, \textbf{47}.

\bibitem[{Isaacs(2017)}]{Isaacs2017}
\textsc{Isaacs, M.} (2017). Faith in contention: Explaining the salience of
  religion in ethnic conflict. \textit{Comparative Political Studies},
  \textbf{50}~(2), 200--231.

\bibitem[{Jedwab \textit{et~al.}(2019)Jedwab, Johnson and
  Koyama}]{JedwabJohnsonKoyama2019}
\textsc{Jedwab, R.}, \textsc{Johnson, N.~D.} and \textsc{Koyama, M.} (2019).
  Negative shocks and mass persecutions: evidence from the black death.
  \textit{Journal of Economic Growth}, \textbf{24}~(4), 345--395.

\bibitem[{Klaiber(1998)}]{Klaiber1998}
\textsc{Klaiber, J.} (1998). \textit{The Church, Dictatorships, and Democracy
  in Latin America}. Orbis Books.

\bibitem[{Laporte and Windmeijer(2005)}]{laporte2005estimation}
\textsc{Laporte, A.} and \textsc{Windmeijer, F.} (2005). Estimation of panel
  data models with binary indicators when treatment effects are not constant
  over time. \textit{Economics Letters}, \textbf{88}~(3), 389--396.

\bibitem[{LaRosa(2000)}]{LaRosa2000}
\textsc{LaRosa, M.} (2000). \textit{De la izquierda a la derecha. La Iglesia
  cat\'olica en la Colombia contempor\'anea}. Planeta.

\bibitem[{Laurent(2010)}]{Laurent2010}
\textsc{Laurent, V.} (2010). Con bastones de mando o en el tarjeton.
  movilizaciones politicas indigenas en colombia. \textit{Colombia
  Internacional}, ~(71).

\bibitem[{Lopez(2010)}]{Lopez2010cap1}
\textsc{Lopez, C.} (2010). La refundacion de la patria, de la teoria a la
  evidencia. In C.~Lopez (ed.), \textit{Y Refundaron la Patria.. De como
  mafiosos y politicos reconfiguraron el Estado Colombiano}, Debate.

\bibitem[{Mitra and Ray(2014)}]{MitraRay2014}
\textsc{Mitra, A.} and \textsc{Ray, D.} (2014). Implications of an economic
  theory of conflict: Hindu-muslim violence in india. \textit{Journal of
  Political Economy}, \textbf{122}~(4), 719--765.

\bibitem[{Montalvo and Reynal-Querol(2005)}]{MontalvoReynalQuerol2005}
\textsc{Montalvo, J.~G.} and \textsc{Reynal-Querol, M.} (2005). Ethnic
  polarization, potential conflict, and civil wars. \textit{American Economic
  Review}, \textbf{95}~(3), 796--816.

\bibitem[{Moscona \textit{et~al.}(2018)Moscona, Nunn and
  Robinson}]{MosconaNunnRobinson2018}
\textsc{Moscona, J.}, \textsc{Nunn, N.} and \textsc{Robinson, J.~A.} (2018).
  \textit{Social Structure and Conflict: Evidence from Sub-Saharan Africa}.
  Working Paper 24209, National Bureau of Economic Research.

\bibitem[{Nunn(2012)}]{Nunn2012}
\textsc{Nunn, N.} (2012). {Culture and the Historical Process}.
  \textit{Economic History of Developing Regions}, \textbf{27}~(S1), 108--126.

\bibitem[{Pe\~naranda(2015)}]{Penaranda2015}
\textsc{Pe\~naranda, D.} (2015). \textit{Conflictos armados y reconstrucci\'on
  identitaria en los Andes colombianos.} Bogota: Cnmh, Iepri.

\bibitem[{{Peace Direct}(2019)}]{Peacedirect2019}
\textsc{{Peace Direct}} (2019). \textit{Civil Society \& Inclusive Peace Key
  insights and lessons from a global consultation convened on Peace Insight}.
  Report, Peace Direct and Inclusive Peace \& Transition Initiative (IPTI).

\bibitem[{Perchoc(2016)}]{Perchoc2016}
\textsc{Perchoc, P.} (2016). \textit{Religious organisations and conflict
  resolution}. Briefing 593515, European Parliamentary Research Service.

\bibitem[{{Pew Research Center}(2014)}]{PewResearchCenter2014}
\textsc{{Pew Research Center}} (2014). \textit{Religion in Latin America.
  Widespread Change in a Historically Catholic Region}. Report, Pew Research
  Center.

\bibitem[{Restrepo \textit{et~al.}(2004)Restrepo, Spagat and
  Vargas}]{RestrepoSpagatVargas2004}
\textsc{Restrepo, J.}, \textsc{Spagat, M.} and \textsc{Vargas, J.} (2004). {The
  Dynamics of the Columbian Civil Conflict: A New Dataset}. \textit{Homo
  Oeconomicus}, \textbf{21}, 396--429.

\bibitem[{{Revista Semana}(2005)}]{SemanaMARCH202005}
\textsc{{Revista Semana}} (2005). Violento martirio.
  \url{https://www.semana.com/nacion/articulo/violento-martirio/71517-3},
  online; consulted 17 Jan 2019.

\bibitem[{Romero(2005)}]{Romero2005}
\textsc{Romero, M.} (2005). \textit{Paramilitares y Autodefensas, 1982-2003}.
  Instituto de Estudios Politicos y Relationes Internacionales.

\bibitem[{Sanchez and Palau(2006)}]{SanchezPalau2006}
\textsc{Sanchez, F.} and \textsc{Palau, M.} (2006). \textit{{Conflict,
  Decentralisation and Local Governance in Colombia, 1974-2004}}. HiCN Working
  Papers~14, Households in Conflict Network.

\bibitem[{Sanchez(2001)}]{Sanchez2001}
\textsc{Sanchez, G.} (2001). Problems of violence, prospects for peace. In
  C.~W. Bergquist, G.~Sanchez and R.~Penaranda (eds.), \textit{Violence in
  Colombia, 1990-2000: Waging War and Negotiating Peace}, Wilmington, Del.: SR
  Books.

\bibitem[{Seul(1999)}]{Seul1999}
\textsc{Seul, J.~R.} (1999). Ours is the way of god': Religion, identity, and
  intergroup conflict. \textit{Journal of Peace Research}, \textbf{36}~(5),
  553--569.

\bibitem[{Shore(2009)}]{Shore2009}
\textsc{Shore, M.} (2009). \textit{Religion and Conflict Resolution:
  Christianity and South Africa's Truth and Reconciliation Commission}.
  Burlington, VT: Ashgate.

\bibitem[{Somma \textit{et~al.}(2017)Somma, Bargsted and
  Valenzuela}]{SommaBargstedValenzuela2017}
\textsc{Somma, N.~M.}, \textsc{Bargsted, M.~A.} and \textsc{Valenzuela, E.}
  (2017). Mapping religious change in latin america. \textit{Latin American
  Politics and Society}, \textbf{59}~(1), 119--142.

\bibitem[{Spolaore and Wacziarg(2009)}]{SpolaoreWacziarg2009}
\textsc{Spolaore, E.} and \textsc{Wacziarg, R.} (2009). {The Diffusion of
  Development}. \textit{The Quarterly Journal of Economics}, \textbf{124}~(2),
  469--529.

\bibitem[{Spolaore and Wacziarg(2016)}]{SpolaoreWacziarg2016}
\textsc{---} and \textsc{---} (2016). War and relatedness. \textit{The Review
  of Economics and Statistics}, \textbf{98}~(5), 925--939.

\bibitem[{Svensson(2013)}]{Svensson2013}
\textsc{Svensson, I.} (2013). One god, many wars: Religious dimensions of armed
  conflict in the middle east and north africa. \textit{Civil Wars},
  \textbf{15}.

\bibitem[{Svensson and Nilsson(2017)}]{SvenssonNilsson2017}
\textsc{---} and \textsc{Nilsson, D.} (2017). Disputes over the divine.
  \textit{Journal of Conflict Resolution}.

\bibitem[{Tchuente and Windmeijer(2019)}]{Tchuente_Wind2019}
\textsc{Tchuente, G.} and \textsc{Windmeijer, F.} (2019). \textit{Dynamic
  Treatment Effects: Some Identification Results}. Report, Miemo.

\bibitem[{Tejeiro(2010)}]{Tejeiro2010}
\textsc{Tejeiro, C.} (2010). El pentecostalismo en el contexto del cambio
  social y religioso en am\'erica latina y colombia. In C.~Tejeiro (ed.),
  \textit{El pentecostalismo en Colombia. Pr\'acticas religiosas, liderazgo y
  participaci\'on pol\'itica}, Universidad Nacional de Colombia.

\bibitem[{Toft(2007)}]{Toft2007}
\textsc{Toft, M.~D.} (2007). Getting religion? the puzzling case of islam and
  civil war. \textit{International Security}, \textbf{31}.

\bibitem[{{USCIRF}(2005)}]{USCIRF2005}
\textsc{{USCIRF}} (2005). \textit{International Religious Freedom Report for
  2005}. Report, US Commission on International Religious Freedom.

\bibitem[{{USCIRF}(2009)}]{USCIRF2009}
\textsc{{USCIRF}} (2009). \textit{International Religious Freedom Report for
  2009}. Report, US Commission on International Religious Freedom.

\bibitem[{{USCIRF}(2010)}]{USCIRF2010}
\textsc{{USCIRF}} (2010). \textit{International Religious Freedom Report for
  2010}. Report, US Commission on International Religious Freedom.

\bibitem[{{USCIRF}(2011)}]{USCIRF2011}
\textsc{{USCIRF}} (2011). \textit{International Religious Freedom Report for
  2011}. Report, US Commission on International Religious Freedom.

\bibitem[{{Verdad Abierta}(2010)}]{verdadabiertaSEPT202010}
\textsc{{Verdad Abierta}} (2010). La decadencia de las farc.
  \url{https://verdadabierta.com/la-decadencia-de-las-farc/}, online; consulted
  17 Jan 2019.

\bibitem[{Wilson(2001)}]{Wilson2001}
\textsc{Wilson, R.~A.} (2001). \textit{The Politics of Truth and Reconciliation
  in South Africa: Legitimizing the Post-Apartheid State}. Cambridge Studies in
  Law and Society, Cambridge University Press.

\end{thebibliography}

\end{document}